\documentclass[aps,prd,superscriptaddress,showpacs,preprintnumbers,nofootinbib]{revtex4}
\usepackage{lineno,hyperref}
\usepackage{amssymb, graphics, setspace}
\usepackage{pgfplots}
\usepackage[abs]{overpic}
\usepackage{physics}
\usepackage{textcomp}
\usepackage[caption=false]{subfig}
\usepackage{amsmath}
\usepackage{commath}
\usepackage{graphicx,bm}
\usepackage{url}
\usepackage{float}
\usepackage{textcomp}
\usepackage{verbatim}
\begin{document}

\title{Elastic and diffractive scattering in the LHC era}

\author{L\'aszl\'o Jenkovszky}
\affiliation{Bogolyubov Institute for Theoretical Physics (BITP),
	Ukrainian National Academy of Sciences \\14-b, Metrologicheskaya str.,
	Kiev, 03680, UKRAINE \\ jenk@bitp.kiev.ua}

\author{Rainer Schicker}
\affiliation{Physikalisches Institut, Im Neuenheimer Feld 226, 
	Heidelberg University, 69120 Heidelberg, GERMANY \\ schicker@physi.uni-heidelberg.de}

\author{Istv\'an Szanyi}
\affiliation{Uzhgorod National University, \\14, Universytets'ka str.,  
	Uzhgorod, 88000, UKRAINE \\ sz.istvan03@gmail.com}

\begin{abstract}
We review elastic and diffractive scattering of protons (called also "forward physics") with emphasis on the LHC data, especially those deviating from the expectations based on extrapolations from earlier measurements at the ISR, Fermilab and thus triggering searches for new ideas, models and theories. We list these new data and provide a brief introduction of available theoretical approaches, mainly those based on analyticity, crossing symmetry and unitarity, particularly the Regge-pole model realizing these concepts. Fits to the data are presented and tensions between theoretical predictions and the data that may indicate the way to further progress are in the focus of our paper.    

\end{abstract}
\pacs{12.40.Nn, 11.55.Jy}
\maketitle

\section{Introduction} \label{s1}
The Regge-pole theory makes part of the ambitious $S$-matrix program, developed in the 60-ies By G. Chew, S.C Frautchi {\it et al.} (see Ref.~\cite{Collins} and references therein). The scattering matrix ($S$-matrix) is a linear operator transforming the initial state into the final state: $S\ket{i}=\ket{f}.$ The initial states final states are defined at the time $-\infty$ and $+\infty$, $i.e.$ the $S-$matrix does not know the detailed time evolution of the system between asymptotically free states.       

The linearity of $S$ reflects the superposition principle of quantum mechanics, so it is natural to require relativistic invariance of $S$. This can be provided by use of Lorenz-invariant combinations of the kinematic variables, used throughout this paper.
Apart from Lorenz invariance, the $S$ matrix is assumed to satisfy unitarity, analyticity and crossing symmetry. The above postulates were believed to constraint the $S-$matrix until the ultimate theory of the strong interaction. According to the analytic $S-$matrix theory, unitarity constrains the structure of the scattering amplitude (or $S-$matrix), resulting in their practical realizations, such as the Regge-pole model and its further extension in dual models. 

Towards the end of 60-ies and beginning of 70-ies of the past century this extreme point of view was replaced by another extreme, namely that quantum chromodynamics (QCD) is the only way, leaving little room for any alternative, especially for the "old fashion" analytic $S-$matrix theory. No doubt, jet production or asymptotic freedom is difficult to incorporate in the framework of the analytic $S-$matrix theory, as is difficult to construct a "soft" scattering amplitude in QCD. By this paper we try to balance between two extremes by preserving the advantages of the Regge-pole model in the "soft" region, inaccessible (at least for the moment) in QCD. We believe that the right way to progress is by complementarity and admission of possible alternatives.      

The aim of the present paper is revise the status of the Regge-pole theory in light of recent experimental results, coming mainly from the Large Hadron Collider (LHC), CERN and reduce the available freedom inherit in the model. 

To be more specific, these are: 

1.The nature of the leading Regge singularity, the pomeron. Is it a pole, simple or multiple, accumulation of poles as predicted by QCD or a more complicated singularity (cut), generated by unitarity? We do not anticipate an ultimate, definite answer, still with the new LHC data in hand, we hope to narrow the corridor of the above options. 

Unitarity is an indispensable ingredient of any theory, be it Regge or QCD or dual, however it does resolve our problems. We shall show that 
the unitarization procedure, whatever its nature is closely related to the input (Born) amplitude (see item 1, above).

2. Regge trajectories, in a sense are building blocks of the theory, they may be considered as dynamical "variables" replacing $s,\ t,\ u$. We stress their fundamental role (linear vs. non-linear?) in the construction of the theory.

3. The LHC data confirm part of the predictions of the Regge-pole model, but at the same time, they provided surprising in deviating from expected extrapolations from lower energies. Any conflict/disagreement between theory and experiment is a good chance to improve it or, at least constrain the available freedom of the $S-$ matrix theory, and Regge-pole models in particular. We scrutinize the recently observed discrepancies in the behavior of the forward slope $B(s,t)$, of the ratio $\rho=\frac{Re A(s,t)}{Im A(s,t)}.$          

4. Of special importance is the still awaited detailed fit of the differential cross sections of both $pp$ and $\bar{p}p$ scattering in a wide interval of $s$ and $t$. A relevant compilation of successful fits or failures (by COMPASS or any other collaboration) is a challenging but badly needed task to be done!

Let us briefly recapitulate the present state of the theory. The construction of the scattering amplitude implies two steps: choice of the input ("Born term") and subsequent unitarization. The better the input (the closer to the expected unitary output), the better are the chances of a successfully converging solution ({\it i.e.} the smaller are the unitarity corrections). The standard procedure is to use a simple Regge-pole amplitude as input with subsequent eikonalization. 

The common feature in all papers (see Refs.~\cite{DLl,Pancheri,Khoze1,Khoze2,Khoze3,Petrov,Alkin,Godizov,Paulo} is the use of a supercritical pomeron, $\alpha_P(0)>1,$ as input, motivated by the rise of the cross sections and by perturbative QCD calculations ("Lipatov pomeron") \cite{BFKL1,BFKL2}. The next indispensable step is unitarization, usually realized in the eikonal formalism. Unitarization is necessary at least for two reasons: to reconcile the rise of the total cross sections with the Froissart-Martin bound and to generate the diffractive dip-bump structures in the differential cross sections. The latter issue is critical for most of the theoretical constructions since the standard eikonalization procedure results in a sequence of secondary dips and bumps, while experimentally a single dip-bump is observed only, as confirmed by all measurements including those recent, at highest LHC energy. This deficiency is usually resolved {\it e.g.} by introducing so-called enhanced diagrams, introducing extra free parameters etc. Still, none of the above-mentioned models was able to reproduce the whole set of $pp$ and $\bar pp$ data data from the ISR to the LHC in the dip-bump region. This is a crucial test for all existing models. These models did not predict the unexpected rapid, as $ln^2 s$, rise of the forward slope revealed by TOTEM \cite{TOTEM_tot} or the drastic decrease of the ratio $\rho(s)$ reported in Ref.~\cite{TOTEM_rho}. The latter (a single data point) was fitted a post\'eriori \cite{MN} by an odderon, although in an earlier paper by one of the authors Ref.~\cite{Alkin}, the model predicted quite a different trend. To summarize, the existing Regge-eikonal models are compatible with the general trend of high-energy diffractive scattering, but many details, such as the dynamics of the dip-bump, the role of the odderon etc. still remain open and controversial. Moreover, as already mentioned, none of the listed papers predicted the new phenomena (rise of $B(s)$ or fall of $\rho(s)$), discovered recently at the LHC by TOTEM \cite{TOTEM_rho}. 

Model-independent fits to forward ($t=0$) observables, are produced by the COMPETE Collaboration \cite{COMPETE}. The LHC forward  data were recently scrutinized in the excellent paper by the Brazilian group \cite{Paulo}. A similar unbiased analysis including also non-forward data with a comparison of the relevant theoretical models is highly desirable. 

The energy dependence of the forward slope of the diffraction cone is a basic observable related to the interaction radius. While in Regge-pole models the rise of the  total cross sections is regulated by "hardness" of the Regge pole (here, the pomeron), the slope 
$B(s)$ in case of a single and simple Regge pole is always logarithmic. As we shall see in Sec. \ref{Sec:fits}, deviation (acceleration) may arise from more complicated Regge singularities, the odderon and, definitely from unitarity corrections.

The TOTEM Collaboration announced \cite{TOTEM_tot} new results of the measurements on the proton-proton elastic slope at $7$ and $8$ TeV, $B$(7~TeV)$=19.89\pm0.27$ and $B$(8~TeV)$=19.9\pm0.3$ GeV$^{-2}$, showing that its energy dependence exceeds the value of the logarithmic approximation from lower energies, compatible with a $\ln s$ behavior \cite{TOTEM_tot}. These data offer new information concerning the burning problem of the strong interaction dynamics, namely the onset of (or approach to) the asymptotic regime of the strong interaction. For similar studies see also Refs.~\cite{TT0,TT01,TT02,TT03,TT2}.  

A possible alternative to the simple Regge-pole model as input is a double pole (double pomeron pole, or simply dipole pomeron, DP) in the angular momentum $(j)$ plane. It has a number of advantages over the simple pomeron Regge pole. In particular, it produces logarithmically rising cross sections already at the "Born" level. 


\section{High-energy elastic $pp$ and $p\bar p$ scattering; experimental situation}\label{Sec:experiment}
In this section we present the most representative results obtained at the LHC at various energies and by different groups.

\begin{figure}[H]
	\centering
	\includegraphics[scale=0.9]{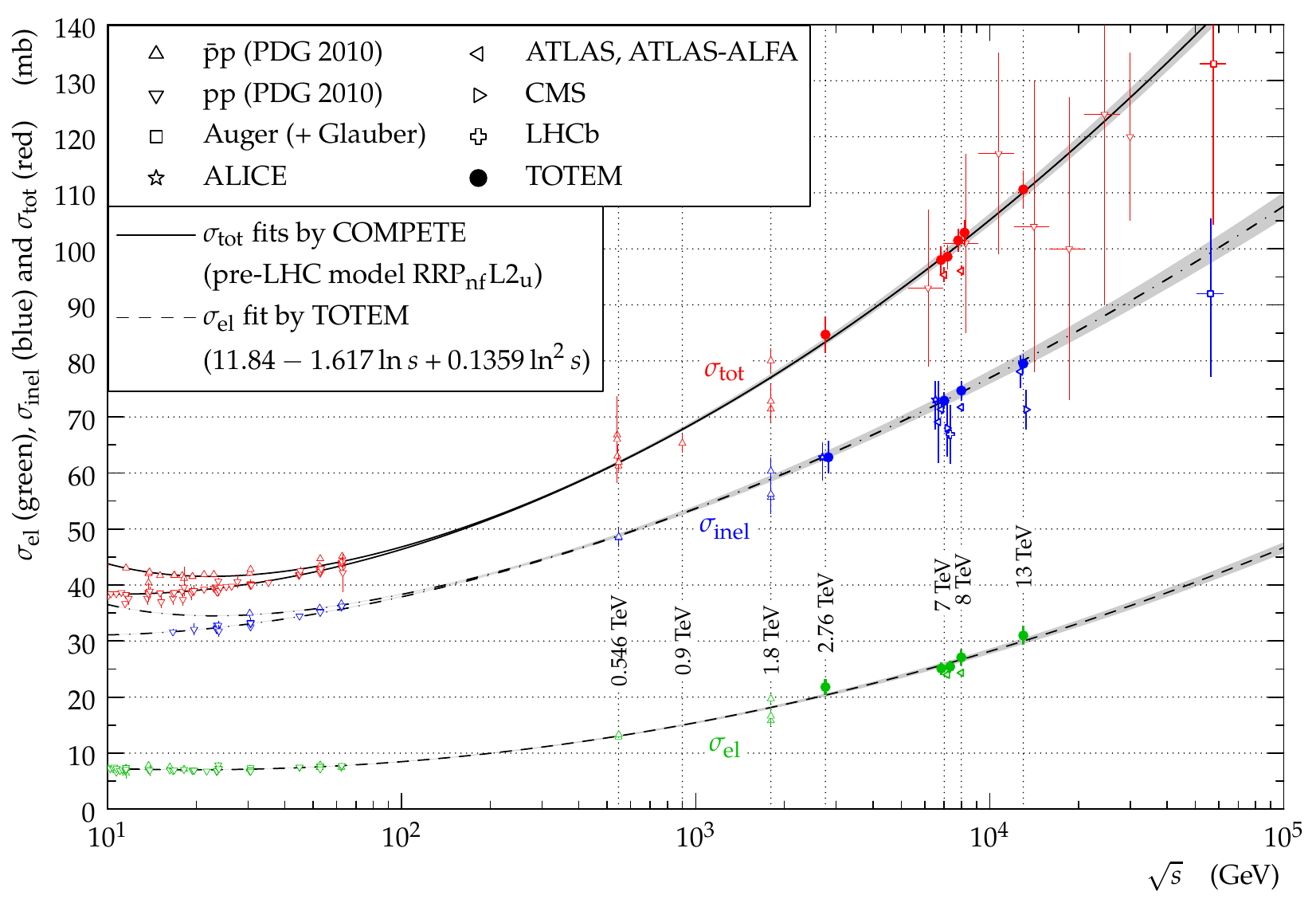}
	\caption{$pp$ and $p \bar p$ elastic, inelastic and total cross sections. Figure from Ref.~\cite{TOTEM_tot}}
	\label{Fig:sTOTEM}
\end{figure}

Total cross section at the LHC was measured at $2.76,\ 7,\ 8$ and $13$ TeV by TOTEM, ALICE, ATLAS, CMS, LHCb \cite{TOTEM_tot}. Fig.~\ref{Fig:sTOTEM} shows the obtained total, elastic and inelastic cross sections as well as a nice fit by a Brazilian group in Ref.~\cite{Paulo}. Note that while the TOTEM data on $\sigma_{tot}$ follow the conventional \cite{COMPETE} $\ln^2 s$ extrapolation from lower energies, ATLAS' data fall systematically below it. 

\begin{figure}[H]
	\centering
	\includegraphics[scale=1]{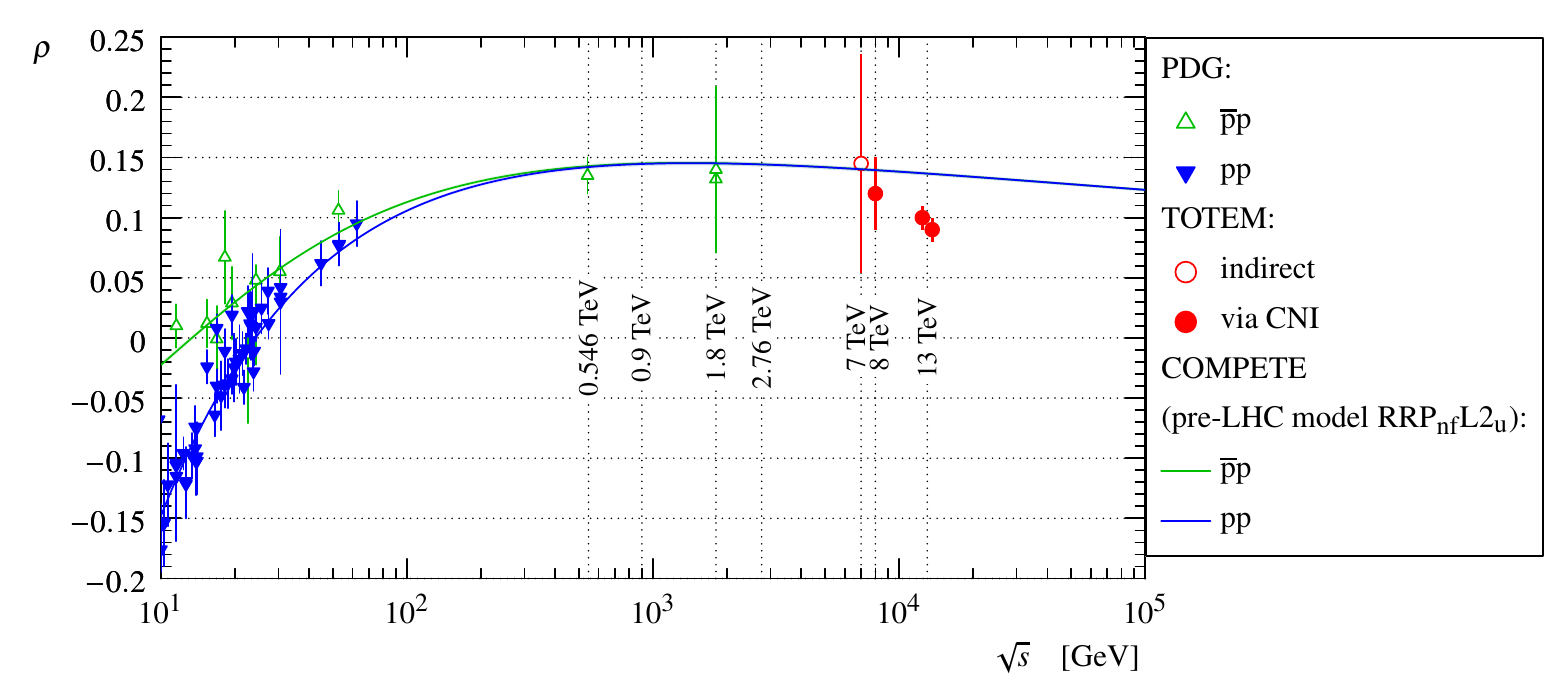}
	\caption{The $rho$-parameter in $pp$ and $p \bar p$ scattering, from Ref.~\cite{TOTEM_rho}}
	\label{Fig:rTOTEM}
\end{figure}

\begin{figure}[H]
	\centering
	\includegraphics[scale=1]{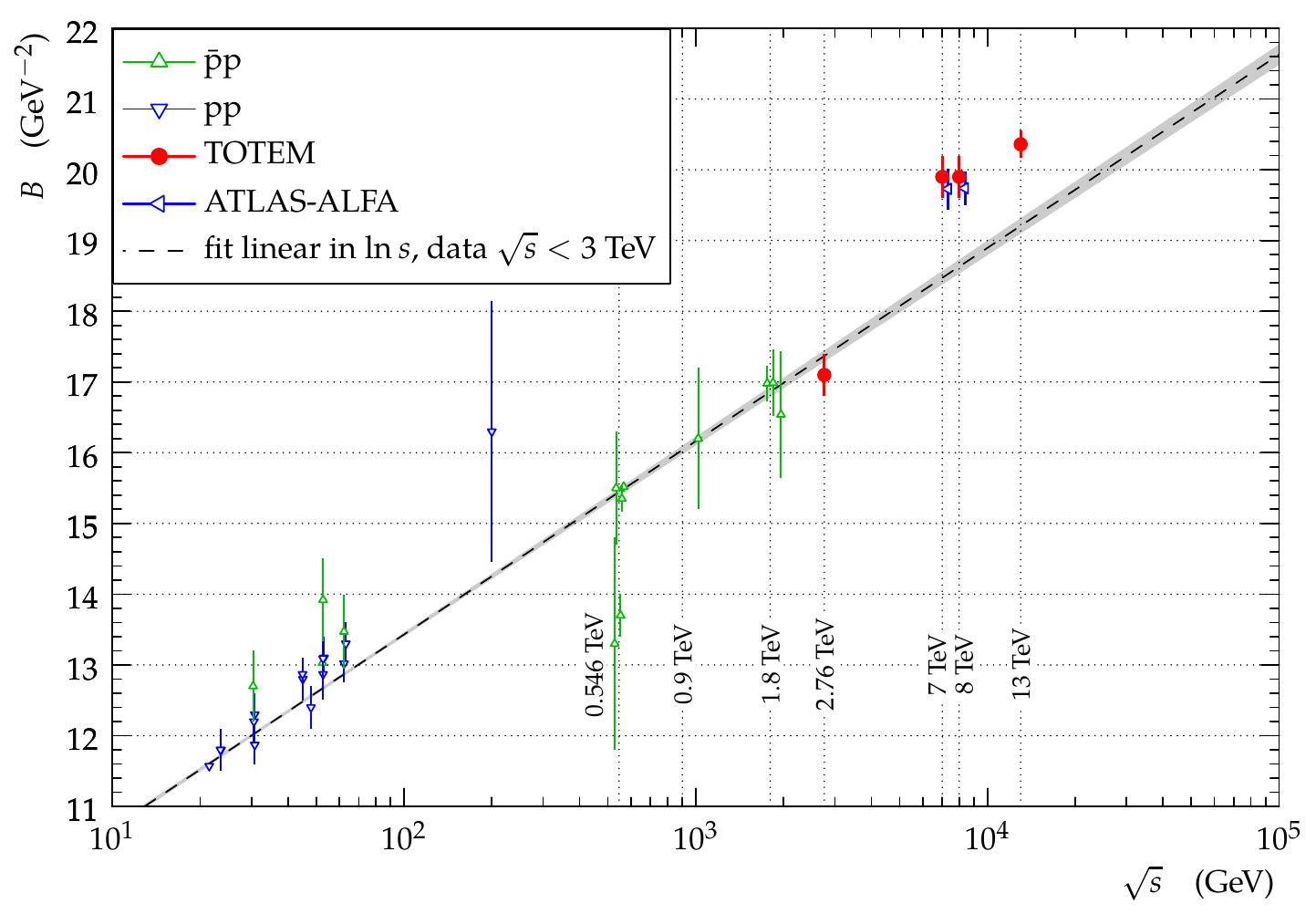}
	\caption{ $pp$ and $p \bar p$ elastic slope, from Ref.~\cite{TOTEM_tot}}
	\label{Fig:BTOTEM}
\end{figure}

Among many LHC discoveries of special interest is the unexpected decrease of the ratio of the real to imaginary part of the forward amplitude at $13$ TeV, shown in Fig.~\ref{Fig:rTOTEM} and the surprisingly rapid increase of the forward slope, Fig.~\ref{Fig:BTOTEM}.
They will be discussed below, in Sec. \ref{Sec:fits}.

Recall that the slope of the diffraction cone is  
\begin{equation}\label{Eq:slope}
B(s,t\rightarrow 0)=\frac{d}{dt}\Biggl(\ln\frac{d\sigma}{dt}\Biggr)\bigg|_{t=0},
\end{equation}
where $A(s,t)$ is the elastic scattering amplitude.
In the case of a single and simple Regge pole, the slope increases logarithmically with $s$:
\begin{equation}\label{Eq:tslope}
B(s,t)=B_0(t)+2\alpha'(t)\ln(s/s_0).
\end{equation}
The slope of the trajectory is not constant, therefore the forward cone in the nearly forward direction deviating from the exponential, see Refs.~\cite{JSZ2,JSZT,RPM} and earlier references therein.
The "forward" slope $B$ is extracted form the experimental data within finite bins in $t$ \cite{JL}, consequently it is function of both $s$ and $t$.

The slope of the cone $B(s,t)$ is not measured directly, it is deduced from the data on the directly measurable differential cross sections within certain bins in $t$. Therefore, the primary sources are the cross sections or the scattering amplitude fitted to these cross sections. To scrutinize the slope we first prepare the ground by constructing a model amplitude from which the slope can be calculated.

For example, at the SPS Collider the slope was measured at $\sqrt{s}=540$ GeV \cite{Burq} in two $t$-bins with the result
$$B=13.3\pm 1.5\ {\rm GeV}^{-2}\ \ \  (0.15<|t|<0.26)\ {\rm GeV}^2, $$
$$B=17.2\pm 1.0\ {\rm GeV}^{-2}\ \ \  (0.05<|t|<0.18)\ {\rm GeV}^2. $$ 

At energies below the Tevatron, including those of SPS and definitely so at the ISR, secondary trajectories contribute significantly. At the LHC and beyond, we are at the fortunate situation where these non-leading contribution may be neglected. Apart from the pomeron, the odderon may be present, although its contribution may be noticeable only away from the forward direction, {\it e.g.} at the dip \cite{JLL}. Recall also that a single Regge pole produces monotonic $\ln s$ rise of the slope, discarded by the recent LHC data \cite{TOTEM_tot}. 

\begin{figure}[H]
	\centering
	\includegraphics[scale=0.7]{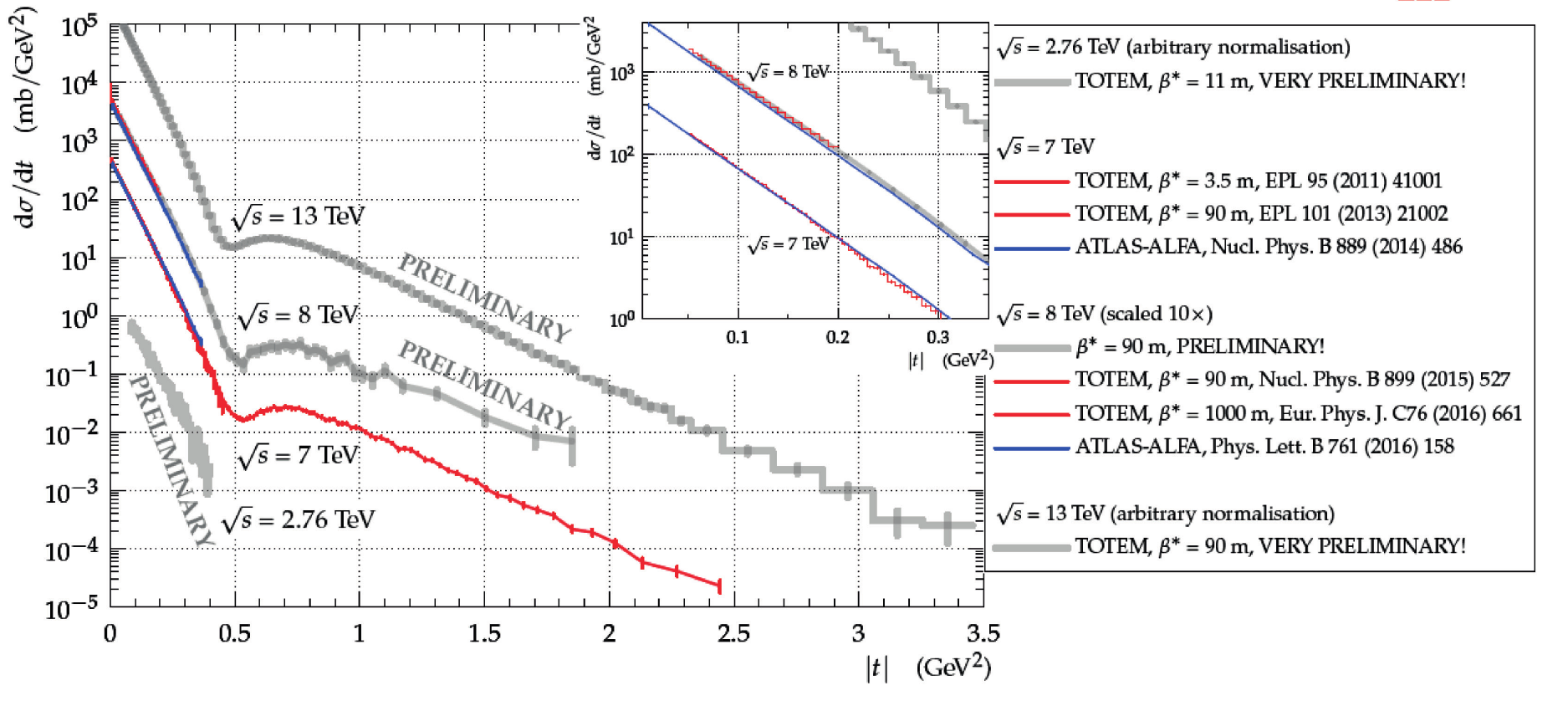}
	\caption{Resent LHC measurements on $pp$ elastic differential cross section. The figure is from Ref.~\cite{Deile}}
	\label{Fig:dsTOTEM}
\end{figure}

\begin{figure}[H]
	\centering
	\subfloat[\label{fig:dsigma8}]{%
		\includegraphics[scale=0.8]{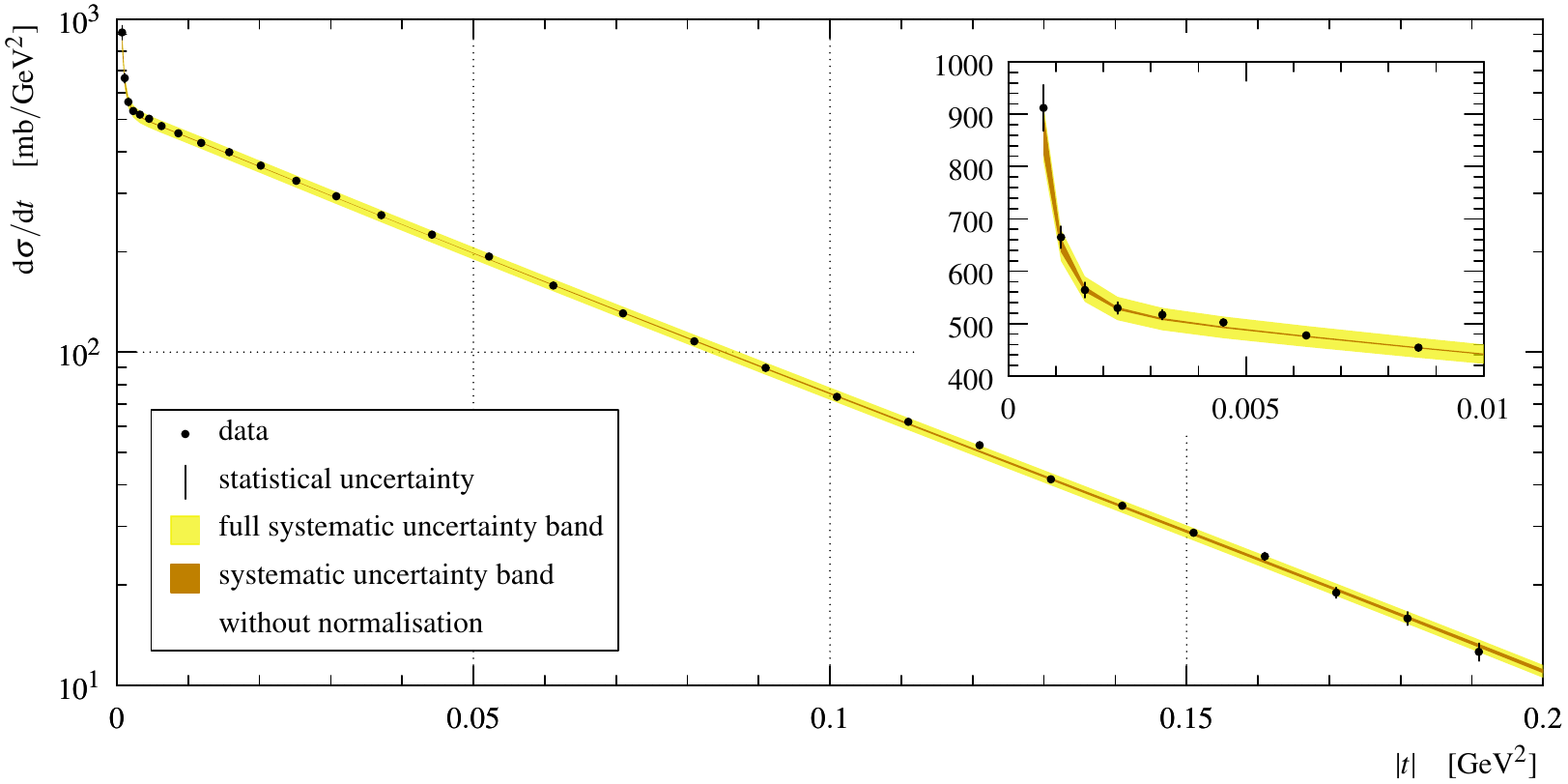}%
	}\hfil
	\subfloat[\label{fig:dsigma13}]{%
		\includegraphics[scale=0.8]{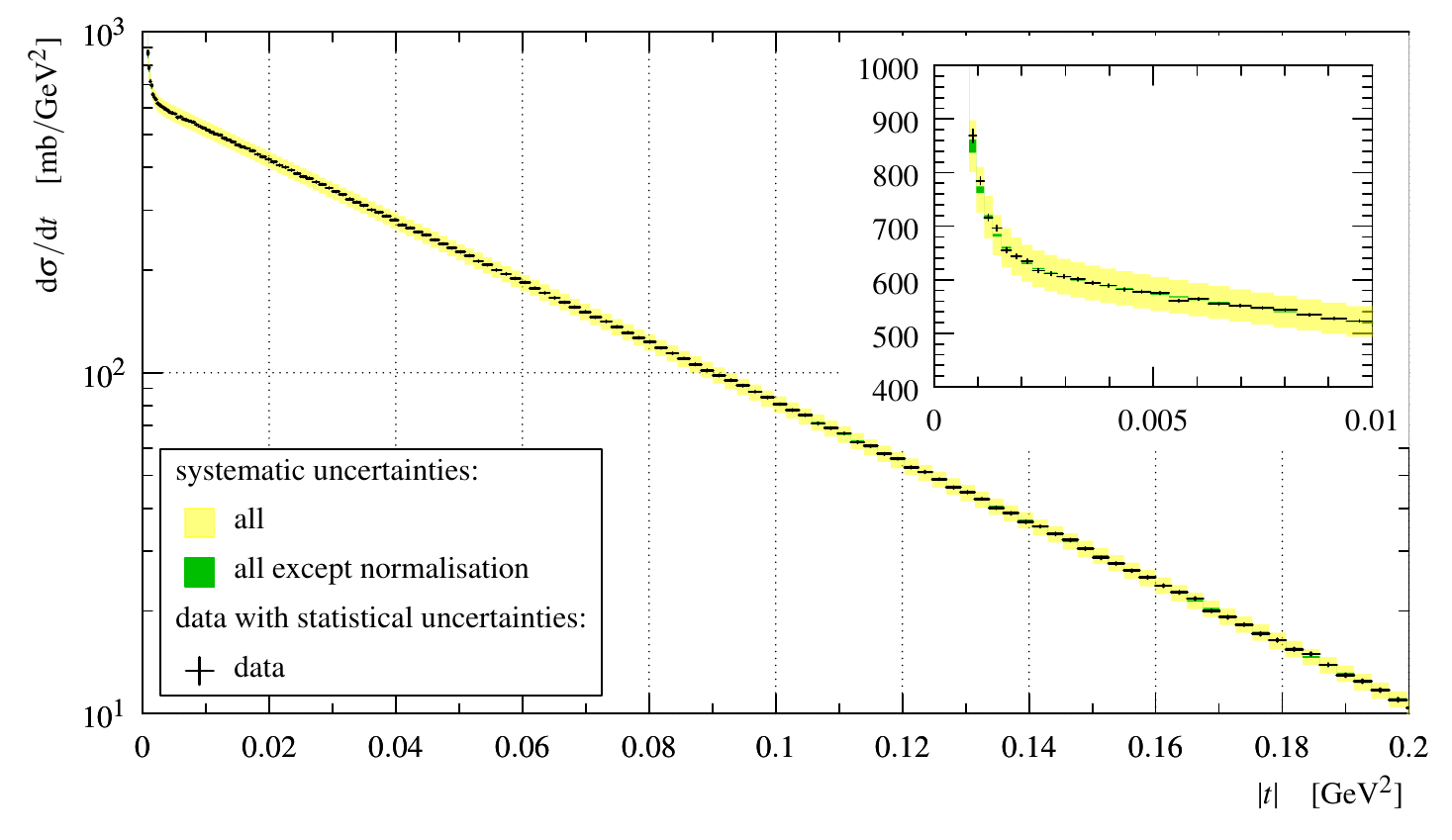}%
	}
	\caption{Resent TOTEM  measurements of $pp$ low-$|t|$ elastic differential cross section at 8 TeV (a) and 13 TeV (b), from Refs.~\cite{totem82,TOTEM_rho}}
	\label{Fig:ds_low_totem}
\end{figure}

\begin{figure}[H]
	\centering
	\includegraphics[scale=1.3]{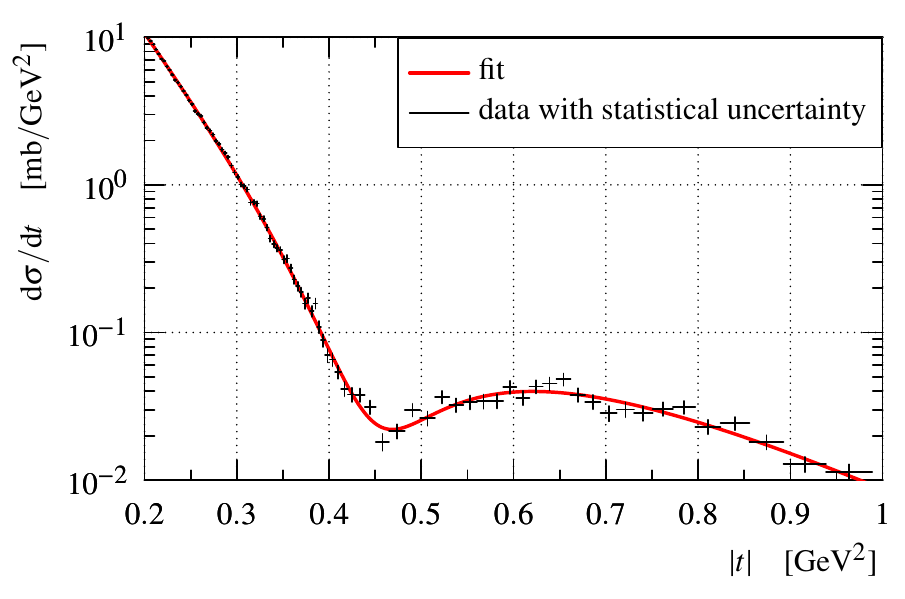}
	\caption{Resent LHC measurements of $pp$ elastic differential cross section in the dip-bump region at 13 TeV, from Ref.~\cite{TOTEM_rho}}
	\label{Fig:ds_db13_TOTEM}
\end{figure}

The resent elastic pp differential cross section measurements are shown in Figs.\ref{Fig:dsTOTEM}-\ref{Fig:ds_db13_TOTEM} 

\section{$S-$matrix theory, Regge-pole models}\label{Regge}
Regge-pole theory is the adequate tool to handle "soft" or "forward" physics. It is a successful example of the analytic $S$ matrix theory, based on analyticity, unitarity and crossing symmetry of the scattering amplitude. It was developed in the 60-ies \cite{Chew,ChF} of the past century, culminating in discovery of duality\cite{DHSch} and dual amplitudes\cite{Veneziano}, whereupon, in the 70-ies was overshadowed by local quantum field theories, more specifically by quantum chromodynamics (QCD), see Sec. 8 in \cite{BP} and references therein.   

The  rejection (in the 60-ies) of field-theoretical approaches was replaced by its extreme opposite, namely by complete neglect of the "old fashion" analytic $S-$matrix theory, including Regge poles, with one exception: attempts to derive the pomeron "from first principles", i.e. from perturbative QCD. 

A generation of physicists grew up lacking any knowledge of the analytic $S$ matrix theory and Regge-pole models. In this Section we try to fill this gap by introducing the basic notions of this efficient and indispensable tool of high-energy phenomenology. 

\subsection{Kinematics} \label{Sec:Kinematics}
Throughout this paper we use relativistic-invariant Mandelstam variables. It will be useful to recall the notation to be used throughout the paper. 

In a generic production process 
\begin{equation}
1+2\rightarrow 3+4++...+N
\end{equation}
the number of independent Lorenz-invariant variables reduces to $3N-10$
due to the conservation of four-momentum, $p_1+p_2=p_3+p_4+...p_N$ (four constraints), the mass shall condition $p_i^2=m_i^2,\ \ i=1,2,...N,$ ($N$ constraints) and the arbitrariness in fixing a four-dimensional reference frame (six constraints).

In this paper we shall be concerned with two-body exclusive scattering and diffraction  dissociation.

The kinematics of two-body reactions 
$$1+2\rightarrow 3+4$$
is described by two variables chose among three Mandelstam variables, defined as 
$$s=(p_1+p_2)^2=(p_3+p_4)^2,$$
$$t=(p_1-p_3)^2=(p_2-p_4)^2,$$
$$u=(p_1-p_4)^2=(p_2-p_3)^2,$$
obeying the  identity
$$s+t+u=\sum_{i=1}^4m_i^4,$$
by which only two, generally $s$ and $t,$ are taken independent. 

\subsection{Regge poles and trajectories; factorization}
Below we introduce the Regge-pole model with emphases on its practical applications. Its derivation from potential scattering, the Schr\"odinger equation and relation to quantum mechanics can be found in many textbooks, e.g. \cite{Collins,BP,DDLN}.

To start with, let us recall that in non-relativistic quantum mechanics bound states appear as poles of the partial wave amplitude $a_l(k)$ (or the $S$ matrix) for integer values of the angular momentum $l$. Regge's original ideas \cite{Regge1,Regge2} was to continue $a_l(k)$ to complex values of $l$, resulting in an interpolating function $a(l,k)$, which reduces to $a_l(k)$ for integer $l=0,1,2,...$. For a Yukawa potential the singularity of $a(l,k)$ appear to be simple moving poles located at values defined by $l=\alpha(k)$, a function of energy, called Regge trajectory. 

In relativistic $S$-matrix theory we do not have a Schr\"odinger equation, and the existence of Regge poles is conjectured by analogy with quantum mechanics. The use of the complex angular moments results (for details see Refs. \cite{Collins,BP,DDLN}) in a representation for the amplitude 
$$A(s,t)=\beta(t)\xi(t)s^{\alpha(t)}, $$
valid in all channels, where $\beta(t)$ is the residue and
$$\xi(t)=-\frac{1\pm e^{-i\pi\alpha(t)}}{\sin\pi\alpha(t)}$$   
is the signature factor. 

\begin{figure}[h]
	\centering
	\includegraphics[scale=0.4]{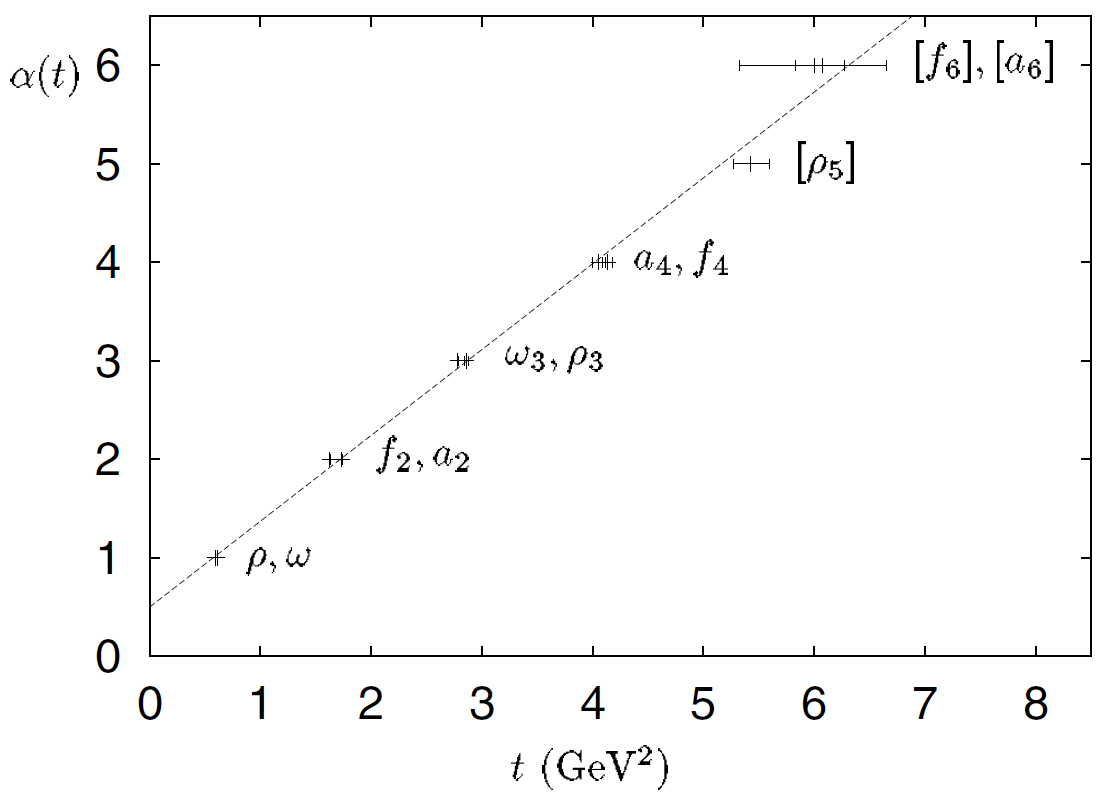}
	\caption{Linear mesonic Chew-Frautschi plot (spin vs. squared masses, $Re\alpha(t)$).}
	\label{Fig:ReggeCF}
\end{figure}

Baryon and meson trajectories are nearly linear functions in a limited range of their variables. This is suggested by the (nearly) exponential shape of forward cone in elastic scattering and by the meson and baryon spectrum. In Fig.~\ref{Fig:ReggeCF} a typical Chew-Frautschi plot is shown. Similar nearly linear plots are known for other mesons and baryons, see \cite{Collins,BP,DDLN}. Whatever appealing, this simplicity is only an approximation to reality: analyticity and unitarity as well as the finiteness of resonances' width require (see below) Regge trajectories to be non-linear complex functions \cite{Barut,Jenk2,Fiore2}.        

Let us reiterate that Regge trajectories are building blocks of the scattering amplitude. In dual models (see below) they appear as the only variables. By crossing symmetry, they connect (smoothly interpolate between) resonance formation (positive $x=s$ or $t$) with  scattering (negative $x$), thus anticipating duality.     

A basic prediction in favouring Regge-pole models and confirmed by all previous experiments is the logarithmic shrinkage of the diffraction cone. Surprisingly, this regularity seems to be broken by recent measurements by TOTEM at the LHC, see Secs. \ref{Sec:experiment} and \ref{Sec:fits}.

\begin{figure}[H]
	\centering
	\includegraphics[scale=0.45]{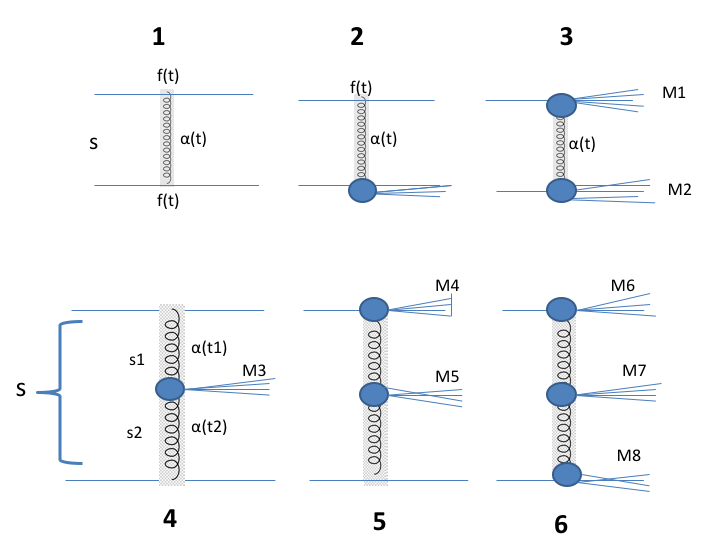}
	\caption{Regge-pole factorization.}
	\label{Fig:Factor}
\end{figure}

Factorization (Fig.~\ref{Fig:Factor}) of the Regge residue $\beta(t)$ and the "propagator" $(s/s_0)^{\alpha_P(t)-1}$ is a basic property of the theory, see Ref.~\cite{BP}. As mentioned, at the LHC for the first time,
we have the opportunity to test directly Regge-factorization  in diffraction, since the scattering amplitude here is dominated by a pomeron exchange, identical in elastic and inelastic diffraction.
Simple factorization relations between elastic scattering ($\frac{d\sigma_{el}}{dt}$), single diffraction (SD) ($\frac{d\sigma_{SD}}{dt})$ and double ($\frac{d^3\sigma_{DD}}{dtdM_1^2dM_2^2})$ DD are known from the literature \cite{Goulianos}.
By writing the scattering amplitude as a product of vertices, elastic $f$ and inelastic $F$, multiplied by the (universal) propagator (pomeron exchange), $f^2s^{\alpha} \ \ fFs^{\alpha},\ \ F^2s^{\alpha}$ for elastic scattering, single (SD) and double (DD) diffraction dissociation, one gets
\begin{equation}\label{factor1}
\frac{d^3\sigma_{DD}}{dtdM_1^2dM_2^2}=\frac{d^2\sigma_{SD1}}{dtdM_1^2}\frac{d^2\sigma_{SD2}}{dtdM_2^2}/\frac{d\sigma_{el}}{dt}.
\end{equation}

Assuming exponential residua $e^{Bt}$ for both SD and elastic scattering and integrating over $t$ one gets:
\begin{equation}\label{factor2}
\frac{d^3\sigma_{DD}}{dM_1^2dM_2^2}=k\frac{d^2\sigma_{SD1}}{dM_1^2}\frac{d^2\sigma_{SD2}}{dM_2^2}/\sigma_{el}.
\end{equation}
where $k=r^2/(2r-1),\ \ r=B_{SD}/B_{el}$.

For $pp$ interactions at the ISR, $r=2/3$ and hence $k=4/3$. Taking the value $r=2/3$, consistent with the experimental results at Fermilab and ISR, one obtains
$\sigma_{DD}=\frac{4\sigma^2_{SD}}{3\sigma_{el}}$. Notice that for $r=1/2,\ \ k \rightarrow\infty.$  Thus $k$ is very sensitive to the ratio $r$, which shows that direct measurements of the slopes at the LHC are important. Interestingly, relation (\ref{factor2}) can be used in different ways, e.g. to cross check any among the four inputs.

To summarize this discussion, we emphasize the important role of the ratio between the inelastic and elastic slope, which at the LHC is close to its
critical value $B_{SD}/B_{el}=0.5$, which means a very sensitive correlation between these two quantities. The right balance may require a
correlated study of the two by keeping the ratio above $0.5$. This constrain may guide future experiments on elastic and inelastic diffraction.

A simple Regge pole is not unique, a dipole being an exclusive alternative. The unique properties of the dipole pomeron (DP) will be discussed and used below, in Sec. \ref{Sec:fits}.

\subsection{Pomeron and odderon}
Regge trajectories (reggeons), introduced in the 60-ies of the past century corresponded to a family of mesons or baryons sitting on the real part of the trajectories -- the co-called Chew-Frautschi plot, to which their parameters (intercept and slope) are adjusted. There are two exceptions, namely the pomeron and odderon. The pomeron was introduced by I.Ya. Pomeranchuk \cite{Pomeranchuk} as a fictive trajectory with postulated unit intercept to provide for non-decreasing asymptotic total cross-sections. In those days, the common belief was that asymptotically the cross sections tend to a constant limit. This has changed after the rise of cross sections was discovered at the ISR. The new,  fictitious trajectory accommodates the asymptotically constant or rising cross section provided its intercept is respectively one or bigger, 
$\alpha_P(0)\ge 1$. The so-called supercritical pomeron, typically with $\alpha_P(0)\approx 1.1$ violates the Froissart bound (and unitarity) at very high energies, beyond any credible extrapolation. Nevertheless, formally and for aesthetic reasons, the input amplitude should be subjected to unitarization. The simple pole input has an interesting alternative, a dipole, to be introduced and used in Sec. \ref{Sec:fits}. Here we only mention, that the dipole pomeron (DP) produces rising cross sections at unit intercept, moreover it has the property of reproducing itself in $s$-channel unitarization \cite{Fort}, i.e. the unitarization procedure converges better than in the case of a simple pole. 

Unlike the case of ordinary (called also secondary or sub-leading) reggeons the pomeron trajectory was not connected to any observable particle. This changed in the 70-ies with the advent of the quark model and QCD. Now the pomeron trajectory has its own Chew-Frautschi plot with glueballs, bound states of gluons, eventually mixed with quarks, forming "hybrids" this making difficult their experimental identification. 

The existence of the pomeron makes plausible the existence of its odd-$C$ counterpart - the odderon. While the pomeron is made of an even number of gluons, the odderon is a bound state of odd number of gluons. The pomeron is "seen" as the imaginary part of the forward amplitude (total cross section), the identification of the odderon is not so unique. Most likely it plays on important role in filling the diffraction minimum in $\bar pp$ scattering, see Sec.\ref{Sec:fits}. 

What is the pomeron (more generally, a reggeon)? It is a virtual particle with continuously varying spin and virtuality, lying on the relevant trajectory. At certain fixed values of virtuality (squared transferred momentum $t$) they correspond to  real mesons or baryons. More on glueball production identification see in Sec. \ref{Sec:CED}.

\subsection{Duality}\label{Ssec:Dual}
\begin{figure} 	
	\centering
	\includegraphics[scale=0.4]{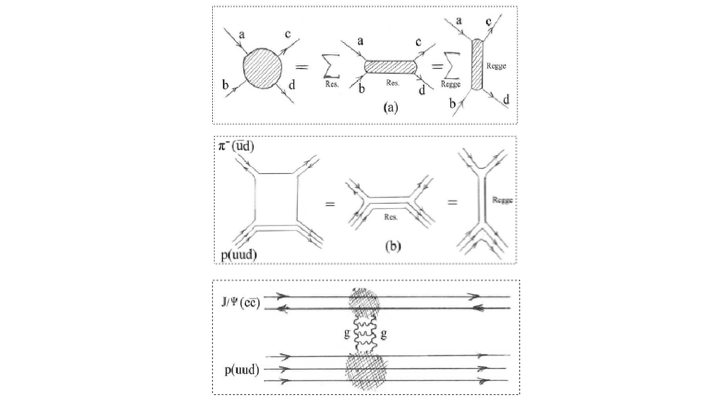}
	\caption{Resonance-Regge duality}
	\label{Fig:dual}
\end{figure}

The notion of duality has many facets values. Here we deal with resonance-Regge duality (Fig.~\ref{Fig:dual}), discovered \cite{DHSch} by saturating the so-called finite energy sum rules.  Their analysis showed that, contrary to expectations, the proper sum or resonances produces smooth Regge behavior and vice versa, their sum producing double counting. As a next step, an explicit dual amplitude was constructed \cite{Veneziano} It is an Euler $B$ function
\begin{equation}\label{Eq:Venezianoo}
A(s,t)=\int_0^1dxx^{-\alpha(s)-1}(1-x)^{\alpha(t)-1}=B\Bigl(-\alpha(s),-\alpha(t)\Bigr)=\frac{\Gamma\bigl(-\alpha(s)\bigr)\Gamma\bigl(-\alpha(t)\bigr)}{\Gamma\bigl(-\alpha(s)-\alpha(t)\bigr)}.
\end{equation}   

The Veneziano amplitude has several remarkable properties: it is crossing symmetric by construction, can be expanded in a pole series (resonance poles) in the $s$ and $t$ channel, and at large $s$, by the Stirling formula it is Regge-behaved, thus explicitly showing resonance-Regge duality. 

At the same time, the model is not free from difficulties, limitations:
it is valid for real and linear trajectories only, that means that finite widths of resonances cannot be incorporated (so-called narrow-resonance approximation) and unitarity is violated. After may attempts a solution was find in dual amplitudes with Mandelstam analyticity (DAMA) \cite{DAMA}, replacing Eq.(\ref{Eq:Venezianoo}) with: 
\begin{equation} \label{Eq:DAMA}
A(s,t)=\int_0^1dx(x/g)^{-\alpha(s,x)-1}((1-x)/g)^{\alpha(t,1-x)-1},
\end{equation}
where $g$ is a parameter, $g>1$.
The functions $\alpha(s,x)$ and $\alpha(t,1-x)$, in (\ref{Eq:DAMA}) called {\it homotopies} \cite{DAMA} are physical trajectories on and of integration,
$\alpha(s,0)=\alpha(s),\  \  \alpha(t,0)=\alpha(t)$ and are linear functions on the other one, $\alpha(s,1)=\alpha_0(s)=a+bs,\ \  \alpha(t,1)=\alpha_0(t),\ \ b\geq 0,$ i.e. the homotopies map the physical trajectories onto linear ones (in the Veneziano model (\ref{Eq:Venezianoo}) the homotopy $\alpha(s,x)$ is an identity mapping, $\alpha(s)=\alpha_0(s).$

DAMA (\ref{Eq:DAMA}) results in a number of dramatic changes with respect to (\ref{Eq:Venezianoo}). It not only allows for non-linear complex trajectories, but rather requires their use, providing for analytic properties of the amplitude required by unitarity. It producing finite-width resonances and incorporate fixed-angle scaling behavior typical of the the quark model. 

\begin{figure}\label{Fig:FMSR.pdf}
	\centering
	\includegraphics[scale=0.4]{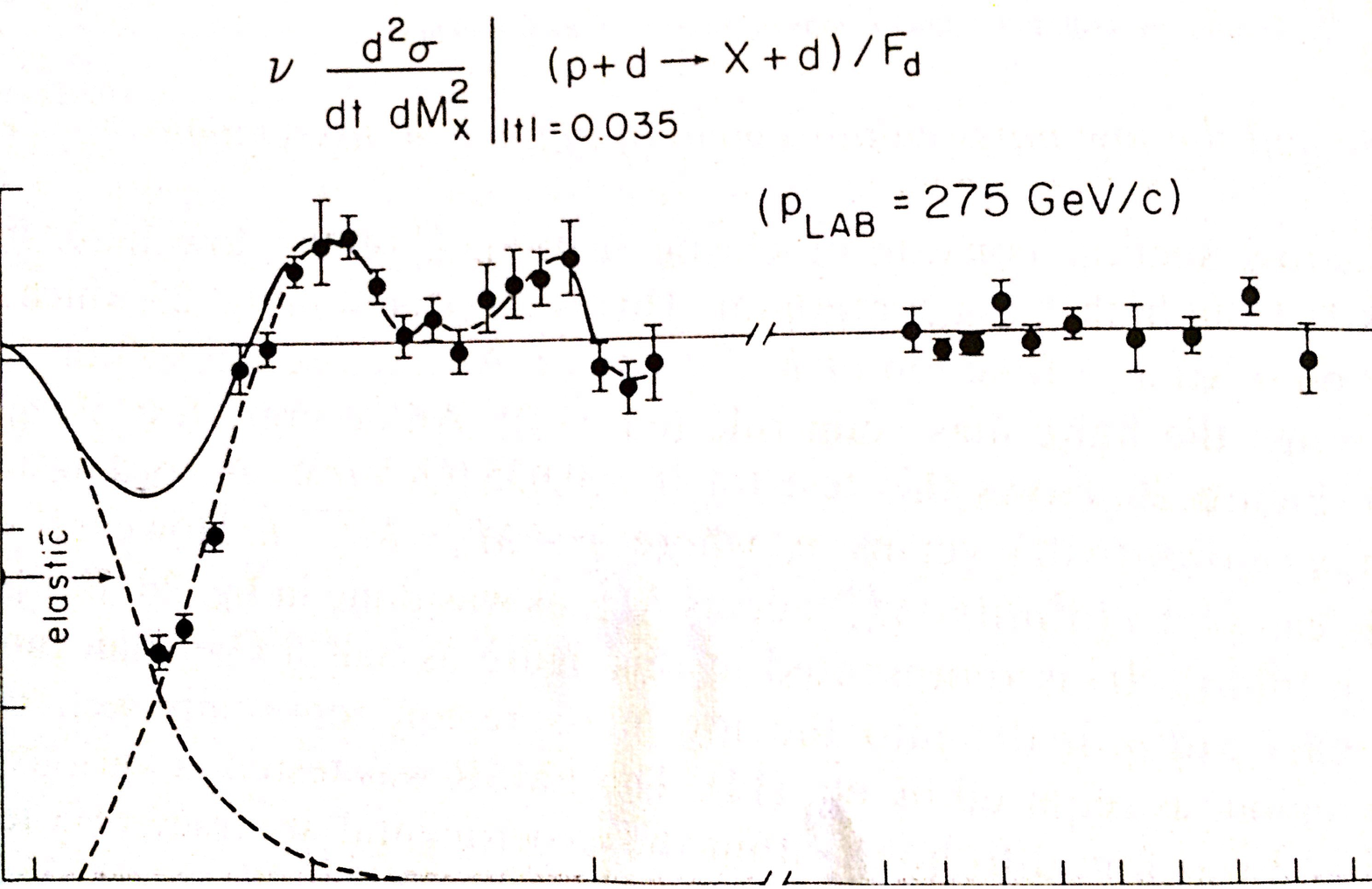}
	\caption{Finite-mass sum rules (FMSR) relate low missing-mass resonances with high mass production in diffraction dissociation.}
	\label{Fig:dual2}
\end{figure}

Its low-energy pole decomposition has the form \cite{DAMA} 
\begin{equation}\label{Eq:Pole}
g^{n+1}\sum{l+1}^n[-s\alpha(s)]^l\frac{C_{n-l}(t)}{[n-\alpha(s)]^{l+1}},
\end{equation}   
where $C(t)$ is a polynomial, see Ref.~\cite{DAMA}. The pole decomposition of the dual model (\ref{Eq:Pole}) reproduces the Breit-Wigner resonance formula (see Appendix in Ref.~\cite{Complete}. Moreover, contrary to the conventional Breit-Wigner model, it comprises a tower of direct-channel resonances, rather than a single Breit-Wigner one. In Sec. \ref{Sec:DD} full use of this property will be made to predict direct-channel, missing-mass resonances produced in diffraction dissociation (DD).

Resonance-Regge duality is applicable also in relating resonances in the missing mass of DD to the high-mass smooth asymptotics, as shown in Fig. \ref{Fig:dual2}.      

Finally, we mention parton-hadron, or Bloom-Gilman duality, relating resonance production in deep-inelastic scattering to the smooth scaling behavior of the structure functions, may be a clue to the confinement problem! 

\subsection{Unitarity, geometry and the black disc limit}
The unitarity condition is simple in the impact parameters representation of the scattering amplitude: 
\begin{equation}\label{eq:impact ampl}
h(s,b)=\frac{1}{4s}\int \frac{d^{2}\vec k}{(2\pi)^{2}}e^{i\vec k\vec b}A(s,t)=
\frac{1}{8\pi s}\int \limits_{0}^{\infty} dk k J_{0}(b\sqrt{-t})A(s,t)
\end{equation}
with the inverse transformation
\begin{equation}\label{eq:impact inv}
A(s,t)=16\pi^{2}s\int \frac{d^{2}\vec b}{(2\pi)^{2}}e^{-i\vec k\vec b}h(s,b)=
8\pi s\int \limits_{0}^{\infty} db b J_{0}(b\sqrt{-t})h(s,b)
\end{equation}
where $A(s,t)$ is the elastic amplitude, $J_{0}(z)$ is the Bessel function; $\vec k$ a two-dimensional vector and  $t\approx -k^{2}_{\bot}$ and $b$ is the impact parameter. 

Unitarity in the impact parameter representation is a simple algebraic equation \cite{Collins}) (no integration in $t$) 
\begin{equation}\label{eq:u.c.impact gen}
Im  h(s,b)=|h(s,b)|^{2}+G_{inel}(s,b),
\end{equation}
where $G_{inel}(s,b)>0$  takes into account inelastic
intermediate states from the original unitarity condition for $A(s,t)$. Therefore
\begin{equation}\label{eq:impact bound-1}
Im h(s,b)>0.
\end{equation}
and from unitarity 
\begin{equation}\label{eq:impact bound-2}
|h(s,b)|\leq 1
\end{equation}
follows. 

Unitarity has two solutions. One is the eikonal one, commonly known and used, by which 
\begin{equation}
h(s,b)=(e^{2i\chi(s,b)}-1)/i,
\end{equation} 
where the input "Born term" (here the eikonal) is usually of Regge-pole form. There is a less familiar alternative to the eikonal, developed mainly in Protvino \cite{TT01, TT02, TT03}, and called "$u$-matrix". In this approach, the unitarized amplitude is obtained from the input "Born" term by the following rational transformation:
\begin{equation}
h(s,b)=\frac{u(s,b)}{1-iu(s,b)},
\end{equation}
where $u(s,b)$ is the "Born term", similar to the eikonal $\chi(s,b),$ usually (but not necessarily) chosen to be of Regge-pole form. 

For the observables the following expressions hold:
\begin{equation}\label{eq:sigt}
\sigma_{t}(s)=\frac{1}{s}Im A(s,0)=8\pi\int\limits_{0}^{\infty}db
b Im h(s,b) 
\end{equation}
\begin{equation}\label{eq:sigel}
\sigma_{el}(s)=\frac{1}{16\pi s^{2}}\int\limits_{-\infty}^{0}dt
|A(s,t)|^{2}=8\pi \int\limits_{0}^{\infty} db
b|h(s,b)|^{2} 
\end{equation}
\begin{equation}\label{eq:siginel}
\sigma_{inel}(s)=8\pi \int\limits_{0}^{\infty} db
b (Im h(s,b)-|h(s,b)|^{2}).
\end{equation}

These two (eikonal and $u$ matrix) approaches differ dramatically concerting the "black disc limit", absolute in the eikonal model, but merely transitory for the $u$ matrix. A large number of paper appeared dramatizing the "dangerous" vicinity of the black disc limit $Im h(s,b)=1/2$, reached or even crossed at the LHC. The transformation of the experimental data, including the differential cross sections measured at the LHC can be always questioned because of the real part of the amplitude (or the phase) is not measured directly\footnote{Lacking direct data on $Re A(s,t)$ away from $t=0,$ one relies on dispersion relations or model amplitudes.} 

Contrary to the eikonal, in the $u$ matrix approach the black disc is not an absolute limit. Having reached $Im h(s,0)=1/2$, the nucleon will tend more transparent, see Ref.~\cite{DJS}. This phenomenon was discussed in a number of papers by S.M. Troshin and N.E. Tyurin, see Ref.~\cite{TT01,TT02,TT03} and references therein.

Another recent development, triggered by the LHC data, is the possible "hollowness" in nucleon's profile function \cite{AB}, i.e. the decrease (gap, "hollow") of the profile function at its center, i.e. $b=0$) with subsequent increase towards the periphery. 

\section{Analysis of the elastic data}\label{Sec:fits}

Regge-pole models were successful in quite a number of studies of high-energy hadron scattering. A partial list of recent papers on elastic proton proton and proton-antiproton scattering includes Refs.~\cite{DLl, Pancheri, Khoze1, Khoze2, Khoze3, Brazil1, Brazil2, Petrov, Alkin, Godizov, Paulo}. In these papers good fits are presented for a limited number of observables, ignoring others. The most critical and still open issue is the non-trivial dynamics and a detailed fit to both $pp$ and $\bar pp$ scattering in the dip-bump region  where high-precision data exist in a wide range of energies. Furthermore, recent measurements at the LHC by the TOTEM collaboration, namely the acceleration of $B(s)$ from $\ln s$ to $\ln^2 s$ \cite{TOTEM_tot} and  the unexpected low value of the ration $\rho(s)$ at 13 TeV \cite{TOTEM_rho} show that dynamics in the LHC energy region deviates from the what was expected from the standard Regge-pomeron models. 

The scattering amplitude is \cite{JLL}:
\begin{equation}\label{Eq:Amplitude}
A\left(s,t\right)_{pp}^{\bar pp}=A_P\left(s,t\right)+A_f\left(s,t\right)\pm\left[A_{\omega}\left(s,t\right)+A_O\left(s,t\right)\right].
\end{equation}

Secondary reggeons are parametrized in a standard way \cite{KKL, KKL1}, with linear Regge trajectories and exponential residua. The $f$ and $\omega$ reggeons are the principal non-leading contributions to $pp$ or $\bar p p$ scattering:
\begin{equation}\label{Reggeon1}
A_f\left(s,t\right)=a_f{\rm e}^{-i\pi\alpha_f\left(t\right)/2}{\rm e}
^{b_ft}\Bigl(s/s_0\Bigr)^{\alpha_f\left(t\right)},
\end{equation}
\begin{equation}\label{Reggeon2}
A_\omega\left(s,t\right)=ia_\omega{\rm e}^{-i\pi\alpha_\omega\left(t\right)/2}{\rm e}
^{b_\omega t}\Bigl(s/s_0\Bigr)^{\alpha_\omega\left(t\right)},
\end{equation}
with $\alpha_f\left(t\right)=0.703+0.84t$ and
$\alpha_{\omega}\left(t\right)=0.435+0.93t$. 

\subsection{Forward scattering; the ratio $\rho(s,0),$ forward slope $B(s,0)$}

Unlike most of the studies, where the pomeron is a simple pole, we use a dipole in the $j-$plane
\begin{eqnarray}\label{Pomeron}
& &A_P(s,t)={d\over{d\alpha_P}}\Bigl[{\rm e}^{-i\pi\alpha_P/2}G(\alpha_P)\Bigl(s/s_{0P}\Bigr)^{\alpha_P}\Bigr]= \\ \nonumber
& &{\rm e}^{-i\pi\alpha_P(t)/2}\Bigl(s/s_{0P}\Bigr)^{\alpha_P(t)}\Bigl[G'(\alpha_P)+\Bigl(L-i\pi
/2\Bigr)G(\alpha_P)\Bigr].
\end{eqnarray}
Since the first term in squared brackets determines the shape of the cone, one fixes
\begin{equation} \label{residue} G'(\alpha_P)=-a_P{\rm
	e}^{b_P[\alpha_P-1]},\end{equation} where $G(\alpha_P)$ is recovered
by integration. Consequently the pomeron amplitude Eq.(\ref{Pomeron}) may be rewritten in the following "geometrical" form (for details see Ref.~\cite{PEPAN} and references therein):
\begin{equation}\label{GP}
A_P(s,t)=i{a_P\ s\over{b_P\ s_{0P}}}[r_1^2(s){\rm e}^{r^2_1(s)[\alpha_P-1]}-\varepsilon_P r_2^2(s){\rm e}^{r^2_2(s)[\alpha_P-1]}],
\end{equation} 
where $r_1^2(s)=b_P+L-i\pi/2,\  r_2^2(s)=L-i\pi/2,\ L\equiv\ln(s/s_{0P}),\ \epsilon_P=1-b_P\lambda,\ \lambda=G(t=0)/G'(t=0)$, and the pomeron trajectory:
\begin{equation}\label{Ptray}
\alpha_P\equiv \alpha_P(t)= 1+\delta_P+\alpha_{1P}t - \alpha_{2P}(\sqrt{4m_{\pi}^2-t}-2m_{\pi}).
\end{equation}
The two-pion threshold in the trajectory accounts for the non-exponential behavior of the diffraction
cone at low $|t|$, see Refs.~\cite{RPM,JSZ2,JSZT}.

The odderon is assumed to be similar to the pomeron apart from its $C-$parity (signature) and different values of adjustable parameters (labeled by subscript ``$O$''): 
\begin{equation}\label{Odd}
A_O(s,t)={a_O\ s\over{b_O\ s_{0O}}}[r_{1O}^2(s){\rm e}^{r^2_{1O}(s)[\alpha_O-1]}-\varepsilon_O r_{2O}^2(s){\rm e}^{r^2_{2O}(s)[\alpha_O-1]}],
\end{equation}
where $r_{1O}^2(s)=b_O+L-i\pi/2$, $r_{2O}^2(s)=L-i\pi/2$, \mbox{$L\equiv\ln(s/s_{0O})$} and the trajectory
\begin{equation}\label{Eq:Otray}
\alpha_O\equiv \alpha_O(t) = 1+\delta_O+\alpha_{1O}t - \alpha_{2O}(\sqrt{4m_{\pi}^2-t}-2m_{\pi}).
\end{equation} 

In earlier versions of the DP, to avoid conflict with the Froissart bound, the intercept of the pomeron was fixed at $\alpha(0)=1$. However later it was realized that the logarithmic rise of the total cross sections provided by the DP may not be sufficient to meet the data, therefore a supercritical intercept was allowed for. From the fits to the data the value $\delta_P=\alpha(0)-1=0.04,$ half of Landshoff's value \cite{Land} follows. This is understandable: the DP promotes half of the rising dynamics, thus moderating the departure from unitarity at the "Born" level (smaller unitarity corrections). Unitarization of the Regge-pole input is indispensable in any case, and in the next section we proceed along these lines.

We use the norm where
\begin{eqnarray}\label{norm}
& & {d\sigma\over{dt}}(s,t)={\pi\over s^2}|A(s,t)|^2 \ \  {\rm and}\ \ \\ \nonumber
& &\sigma_{tot}(s)={4\pi\over s}Im  A(s,t)\Bigl.\Bigr|_{t=0}\ .
\end{eqnarray}

The free parameters of the model were simultaneously fitted to the data on elastic $pp$ and $p\bar p$ differential cross section in the region of the first cone, $|t|<1$ GeV$^2$ as well onto the data on total cross section and the ratio
\begin{equation}
\rho(s)=\frac{Re A(s,t=0)}{Im  A(s,t=0)}
\end{equation}
in the energy range between $5$ GeV and $57000$ GeV. Our fitting strategy is to keep control of those parameters that govern the behavior of the forward slope, neglecting details that are irrelevant to the forward slope, such is the diffraction minimum (here related to absorption corrections through the parameter $\epsilon$ in Eq. (\ref{GP})). 

\begin{figure}[h]
		\centering
		\includegraphics[scale=0.22]{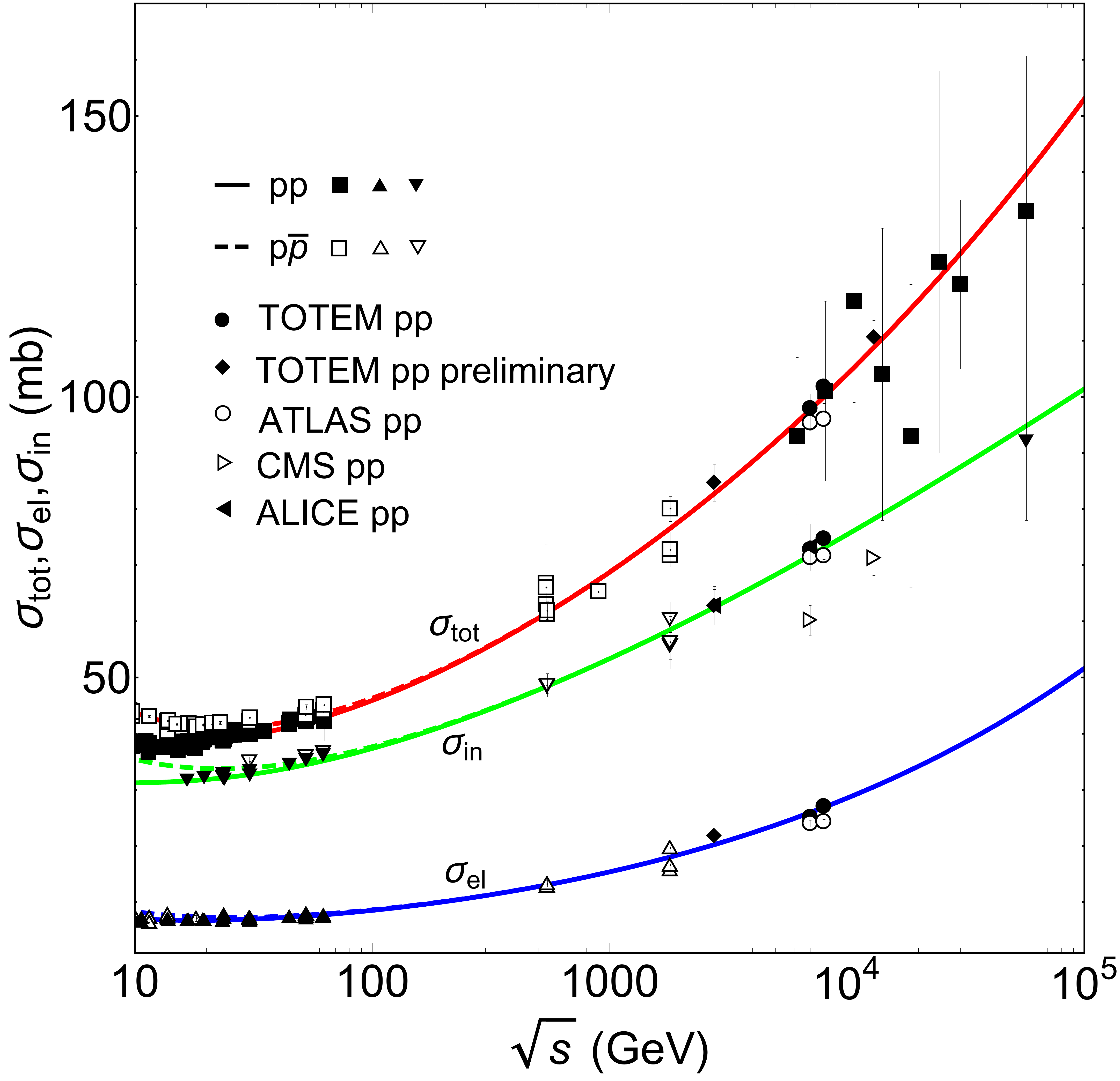}
		\caption{Fits to $pp$ and $\bar pp$ (a) total cross section data using Eqs.~(\ref{Eq:Amplitude}-\ref{Ptray}) without the odderon; calculated elastic and inelastic cross sections, Eqs.~(\ref{eq:el}-\ref{eq:inel}) are also shown.}
		\label{Fig:sigma}
	\end{figure}

\begin{figure}[h]
	\centering
	\includegraphics[scale=0.22]{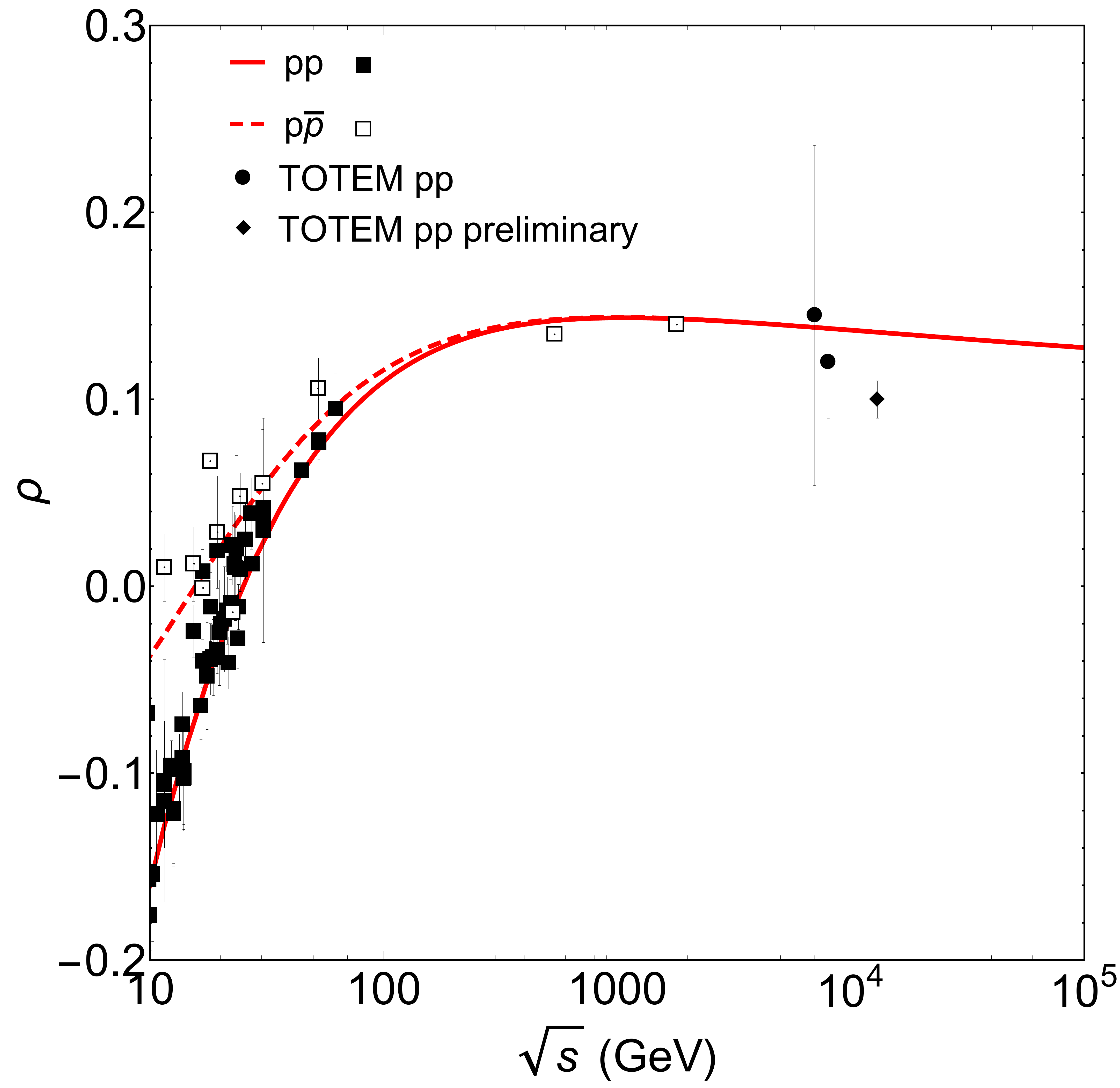}
	\caption{Fits to $pp$ and $\bar pp$ data on the ratio $\rho$ using Eqs.~(\ref{Eq:Amplitude}-\ref{Ptray}) without the odderon.}
	\label{fig:rho}
\end{figure}

\begin{figure}[H]
		\centering
			\includegraphics[scale=0.22]{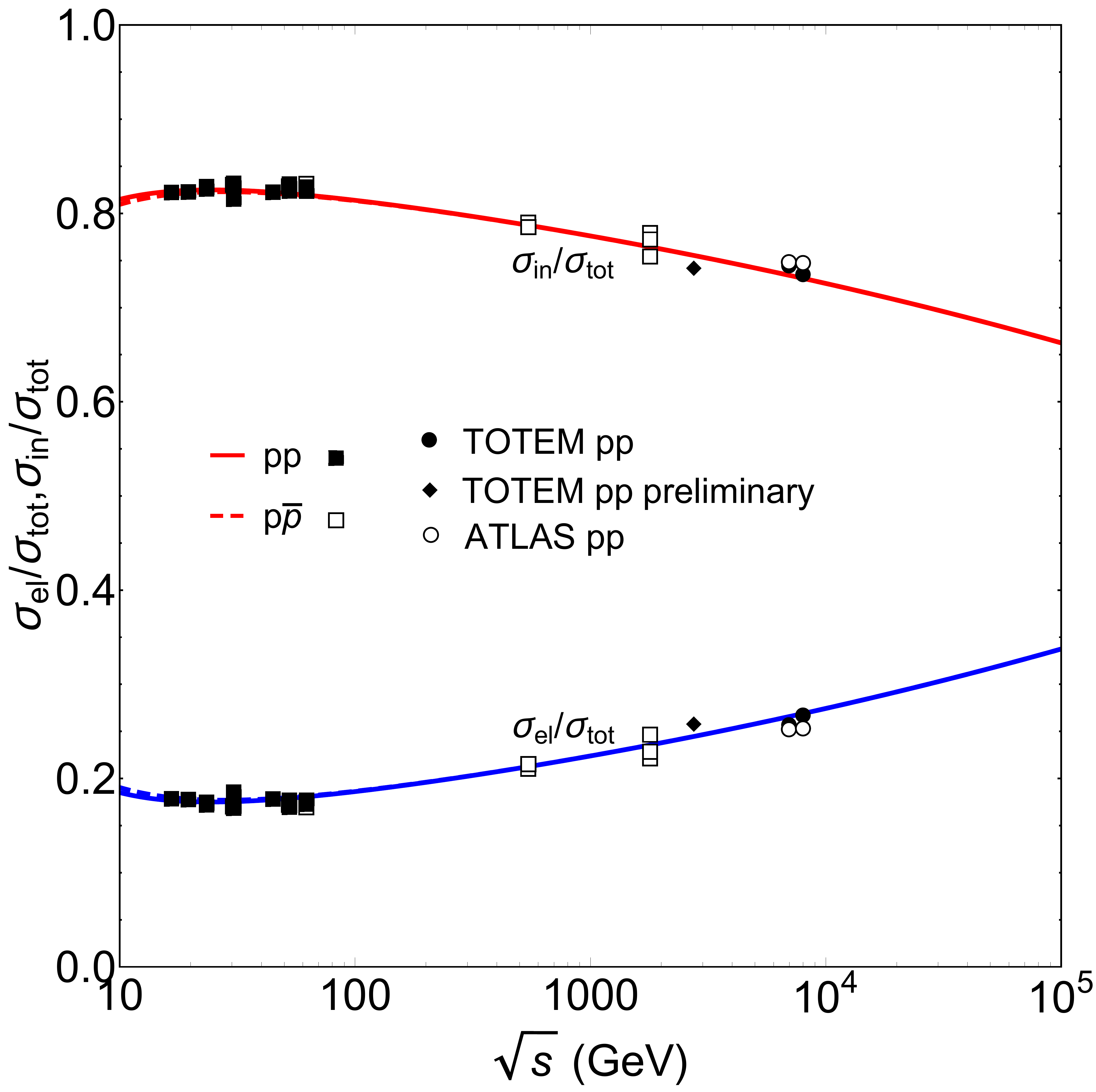}%
		\caption{$\sigma_{el}/\sigma_{tot}$ and $\sigma_{in}/\sigma_{tot}$ ratios calculated from Eqs.~(\ref{Eq:Amplitude}-\ref{Ptray}), without the odderon.}
		\label{Fig:sratios1}
\end{figure}

\begin{figure}[H]
	\centering
		\includegraphics[scale=0.22]{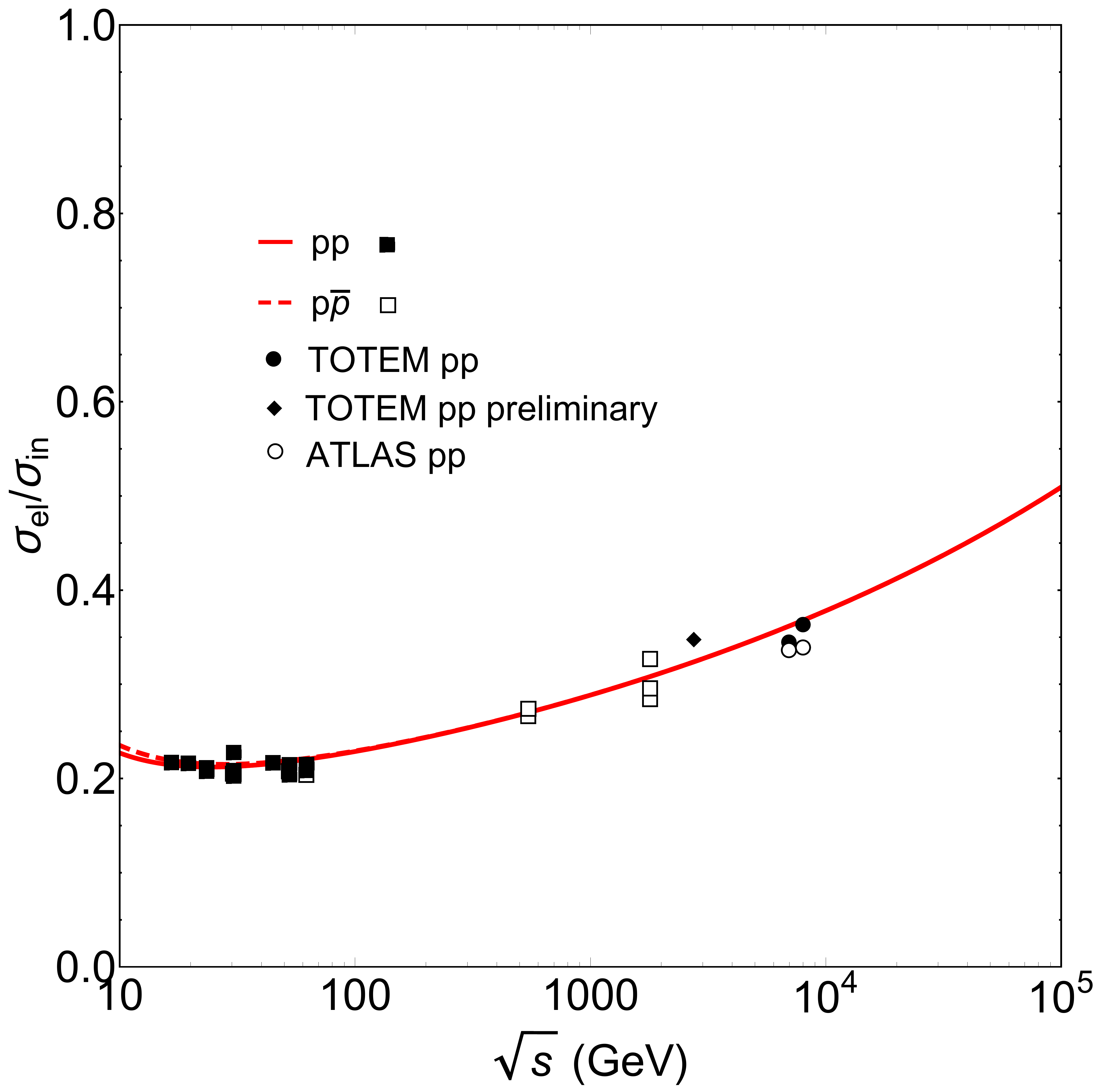}
	\caption{$\sigma_{el}/\sigma_{in}$ ratio calculated from Eqs.~(\ref{Eq:Amplitude}-\ref{Ptray}), without the odderon.}
	\label{Fig:sratios2}
\end{figure}

\begin{figure}[H]
		\centering
		\includegraphics[scale=0.22]{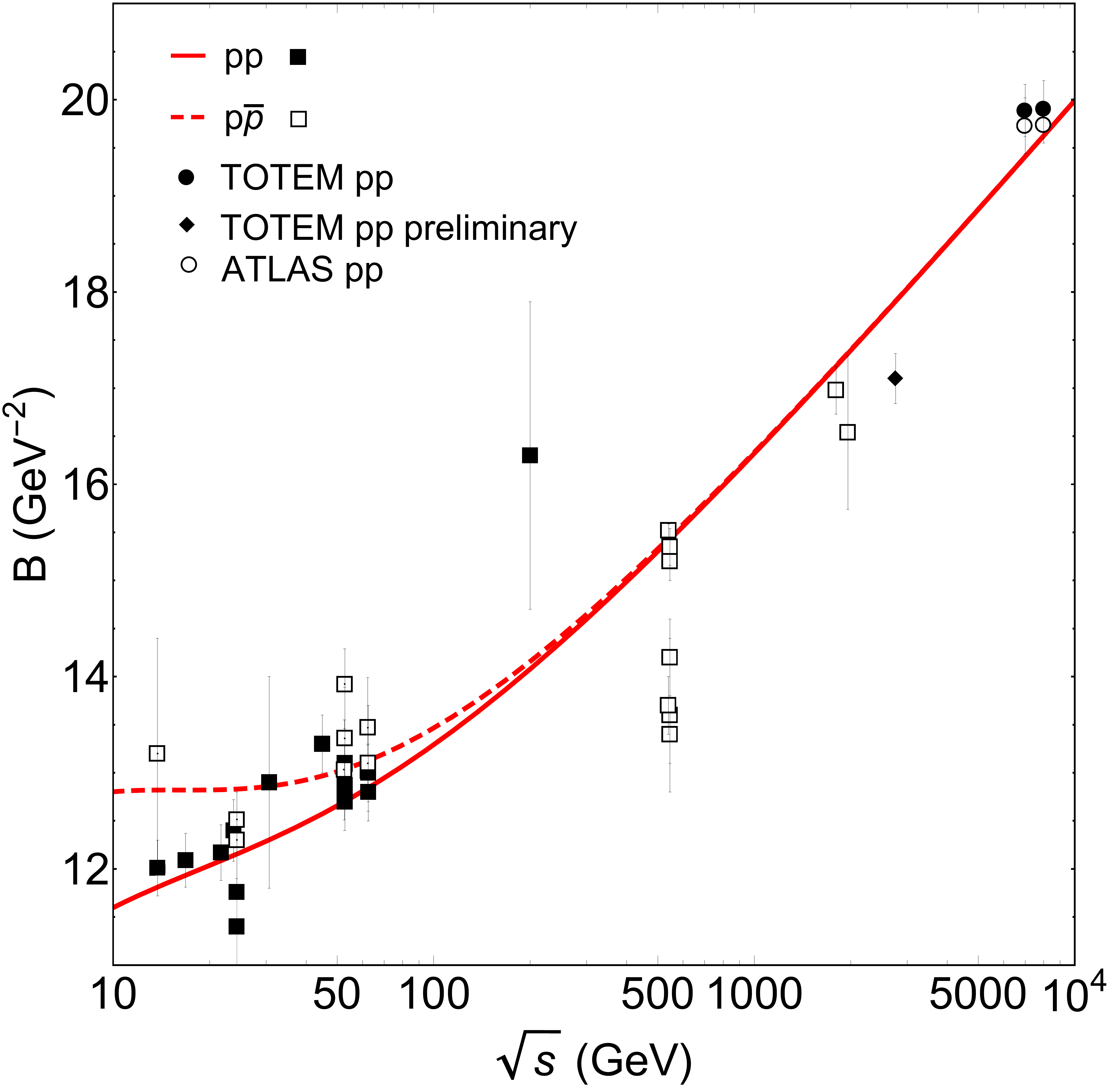}
		\caption{$pp$ and $\bar pp$ elastic slope $B(s,0)$ calculated from Eqs.~(\ref{Eq:Amplitude})-(\ref{Ptray}) (without the odderon).}
		\label{Fig:B1}
\end{figure}

\begin{figure}[H]
	\centering
	\includegraphics[scale=0.22]{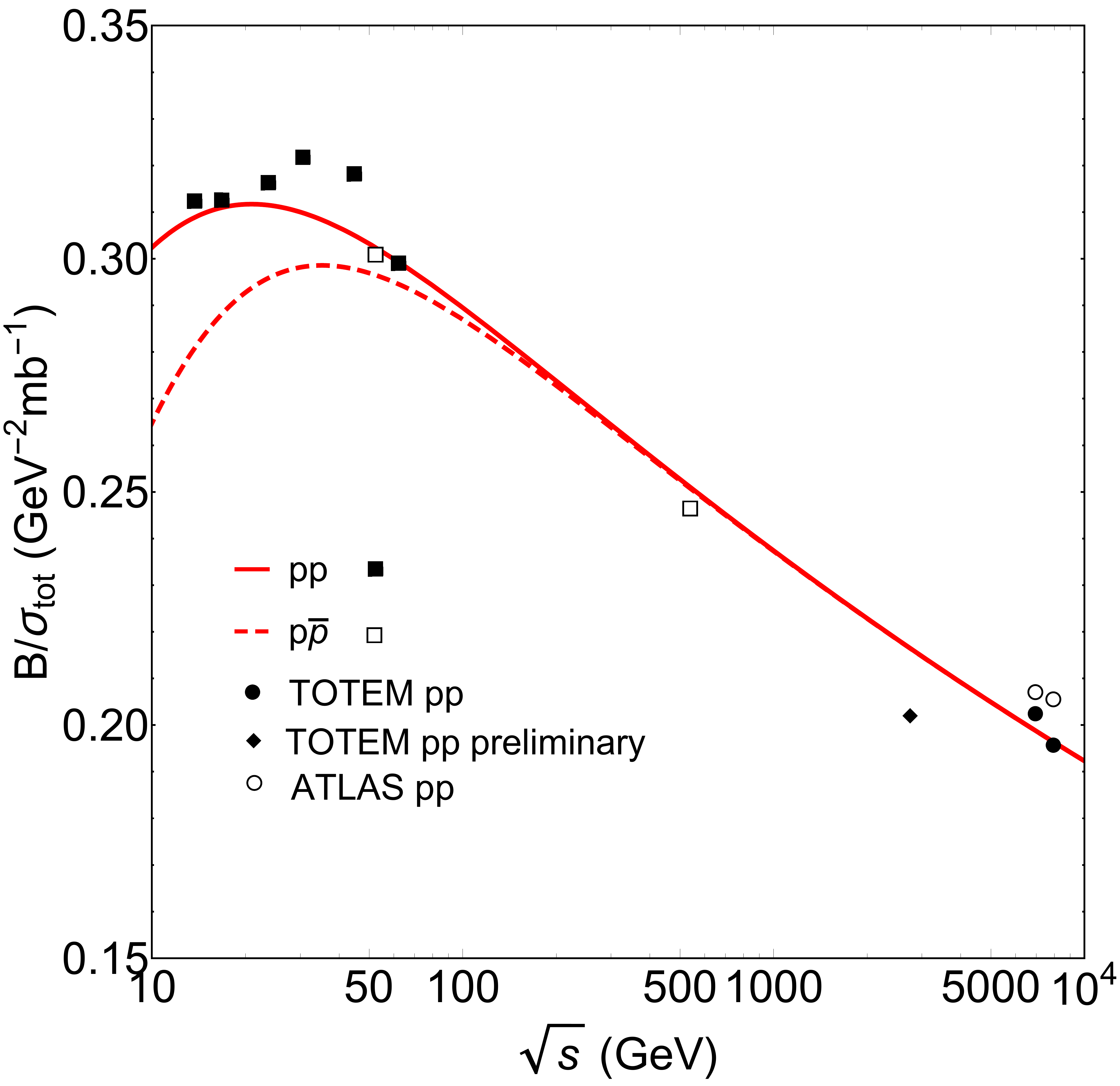}
	\caption{$B(s,t=0)/\sigma_{tot}$ ratio calculated from Eqs.~(\ref{Eq:Amplitude})-(\ref{Ptray}) (without the odderon).}
	\label{Fig:B2}
\end{figure}

\begin{figure}[H]
	\centering
	\includegraphics[scale=0.22]{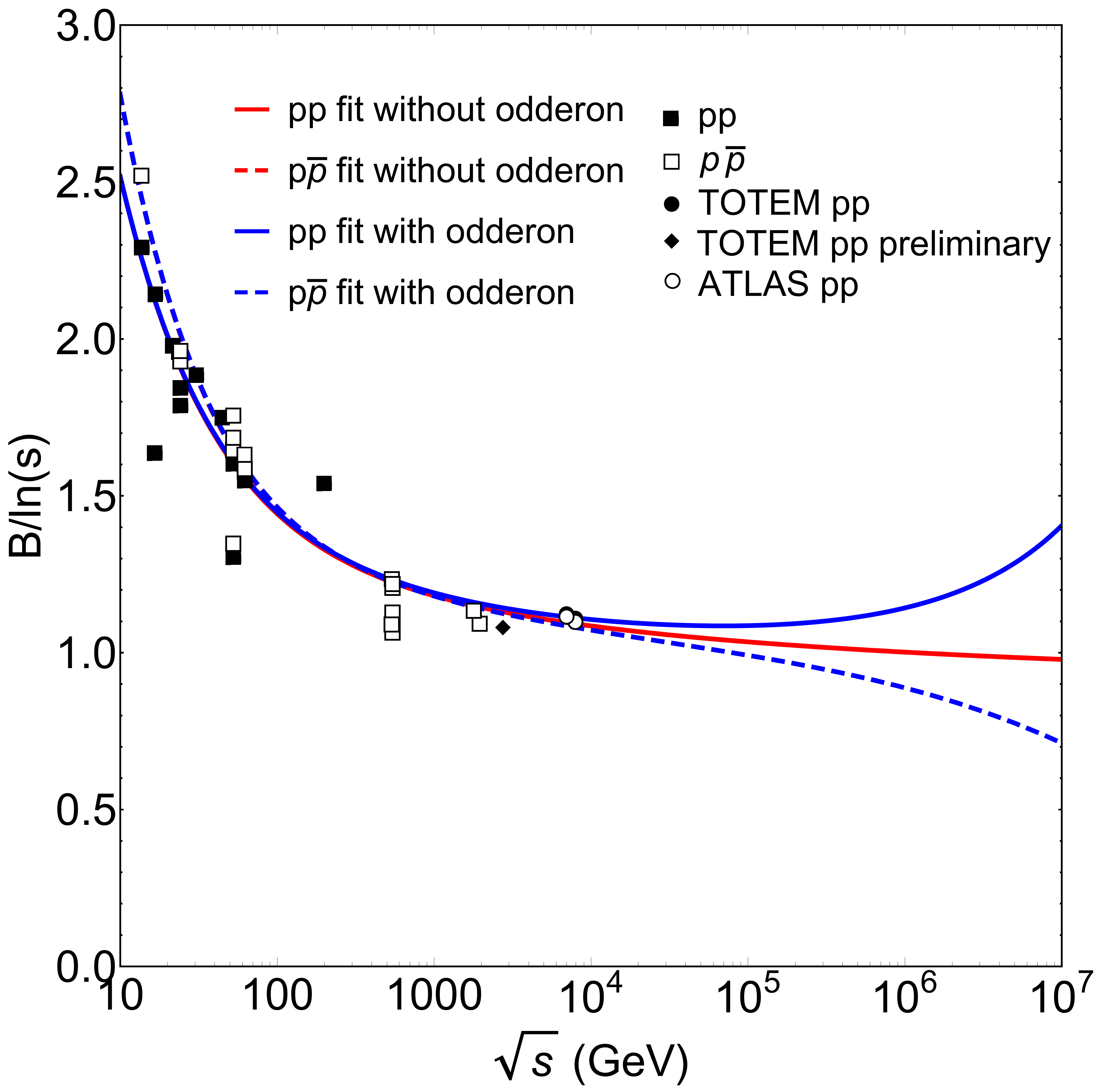}
	\caption{The ratio $B/ln(s)$ calculated from Eqs.~(\ref{Eq:Amplitude}-\ref{Eq:Otray}) with and without the odderon.}
	\label{Fig:Bo}
\end{figure}

Figs.~\ref{Fig:sigma}-\ref{Fig:Bo} and the parameters in Table~\ref{tab:parameters} are from Ref~\cite{BJSz} (published before the recent TOTEM 13 TeV measurements). Fig.~\ref{Fig:sigma} and Fig.~\ref{fig:rho} show the results of the fits to $pp$ and $\bar pp$ total cross section and the ratio $\rho(s,0)$ data.The data are from Refs.~\cite{TOTEM_tot,PDG,data,atlas7,atlas8,totem7,totem81,totem82,totem83,Auger}. The values of the fitted parameters and relevant values of $\chi^2/dof$ are presented in Table~\ref{tab:parameters}. 

Elastic cross section $\sigma_{el}(s)$ is calculated by integration
\begin{equation}\label{eq:el}
\sigma_{el}(s)=\int_{t_{min}}^{t_{max}}\frac{d\sigma}{dt}(s,t)\, dt,
\end{equation}
whereupon
\begin{equation}\label{eq:inel}
\sigma_{in}(s)=\sigma_{tot}(s)-\sigma_{el}(s). 
\end{equation}
Formally, $t_{min}=-s/2$ and $t_{max}=t_{threshold}$, however since the integral is saturated basically by the first cone, we set $t_{max}=0$ and $t_{min}=-1$ GeV$^2$. The results are shown in Fig.~\ref{Fig:sigma}.

The calculated ratios of $\sigma_{el}/\sigma_{tot}$, $\sigma_{in}/\sigma_{tot}$ and $\sigma_{el}/\sigma_{in}$ are shown in Figs. \ref{Fig:sratios1} and \ref{Fig:sratios2}. 
   
\begin{table}
	\caption{Fitted parameters to $pp$ and $p\bar p$ data on elastic differential cross section, total cross section and the parameter~$\rho$.}	
	\centering
	\begin{tabular}{cccccc}
		\multicolumn{1}{c}{}&\multicolumn{2}{c}{Pomeron}&\multicolumn{2}{c}{Reggeons}&\multicolumn{1}{c}{} \\
		\hline 
		&$a_P$ & $301\pm0.78$  & $a_f$ & $-16.4\pm0.061$& \\
		&$b_P$ [GeV$^{-2}$] & $9.91\pm0.049$  & $b_f$ [GeV$^{-2}$] & $4.22\pm0.055$ &         \\
		&$\delta_P$ & $0.0458\pm0.00011$ & $a_{\omega}$ &$9.71\pm0.093$& \\
		&$\alpha_{1P}$ [GeV$^{-2}$]& $0.394\pm0.002$  &$b_{\omega}$ [GeV$^{-2}$] &$8$ (fixed)& \\
		&$\alpha_{2P}$ [GeV$^{-1}$]  & $0.0148\pm0.00073$ &$s_0$ [GeV$^2$] &$1$& \\ \cline{4-6}
		&$\varepsilon_P$ & $0$  & \multicolumn{2}{c}{$\chi^2/dof=3.2$}&        \\ 
		&$s_{0P}$ [GeV$^2$] & $100$ &  & &\\ \hline
	\end{tabular}
	\label{tab:parameters}
\end{table}

With the model and its fitted parameters in hand, we now proceed to study the slope $B(s,t)$.

The calculated elastic slope $B(s,0)$ and the ratio $B(s,0)/\sigma_{tot}(s)$ are shown in Figs.~\ref{Fig:B1} and \ref{Fig:B2}. To see better the effect of the odderon, the deviation of $B(s)$ from its "canonical", logarithmic form, we show in Fig.\ref{Fig:Bo} its "normalized" shape, $B(s)/\ln(s)$. A similar approach was useful in studies Refs.~\cite{totem83, RPM,JSZ2,JSZT} of the fine structure (in $t$) of the diffraction cone.

The main conclusion from this section is that the dipole pomeron at the "Born" level, fitting data on elastic, inelastic and total cross section, does not reproduce the irregular behavior of the forward slope observed at the LHC.  Remarkably, the inclusion of the odderon produces a $ln^2(s)$ behavior of the elastic slope $B(s,0)$ beyond the LHC energy region, Fig.~\ref{Fig:B1}, although this fit needs further checks since the parameters of the odderon are sensitive to the data away from $t=0$, particularly to those in the dip-bump region $t\approx -1$ GeV$^2$.

Recent TOTEM data \cite{TOTEM_rho}, Fig.~\ref{fig:rho}, showing a dramatic decrease of the the ratio $\rho$ at 13 TeV offer new opportunities to test the existing models.  Preliminary comments on the possible connection between the odderon and TOTEM's new data \cite{TOTEM_rho} already appeared in the literature \cite{Khoze1, Khoze2, Khoze3, MN}. The above {\it ad hoc} fits however cannot resolve the problem that requires a global analysis of all observables in a wide kinematic region.

Irrespective of the role played by the odderon in the behavior of the slope $B(s)$ and the ratio $\rho(s)$, unitarity corrections are indispensable and they are discussed in the next section.

\subsection{The (exponential?) diffraction cone; structures: the “break”, dip and bump}

\begin{figure}[h]
	\center{
		\includegraphics[width=0.35\linewidth]{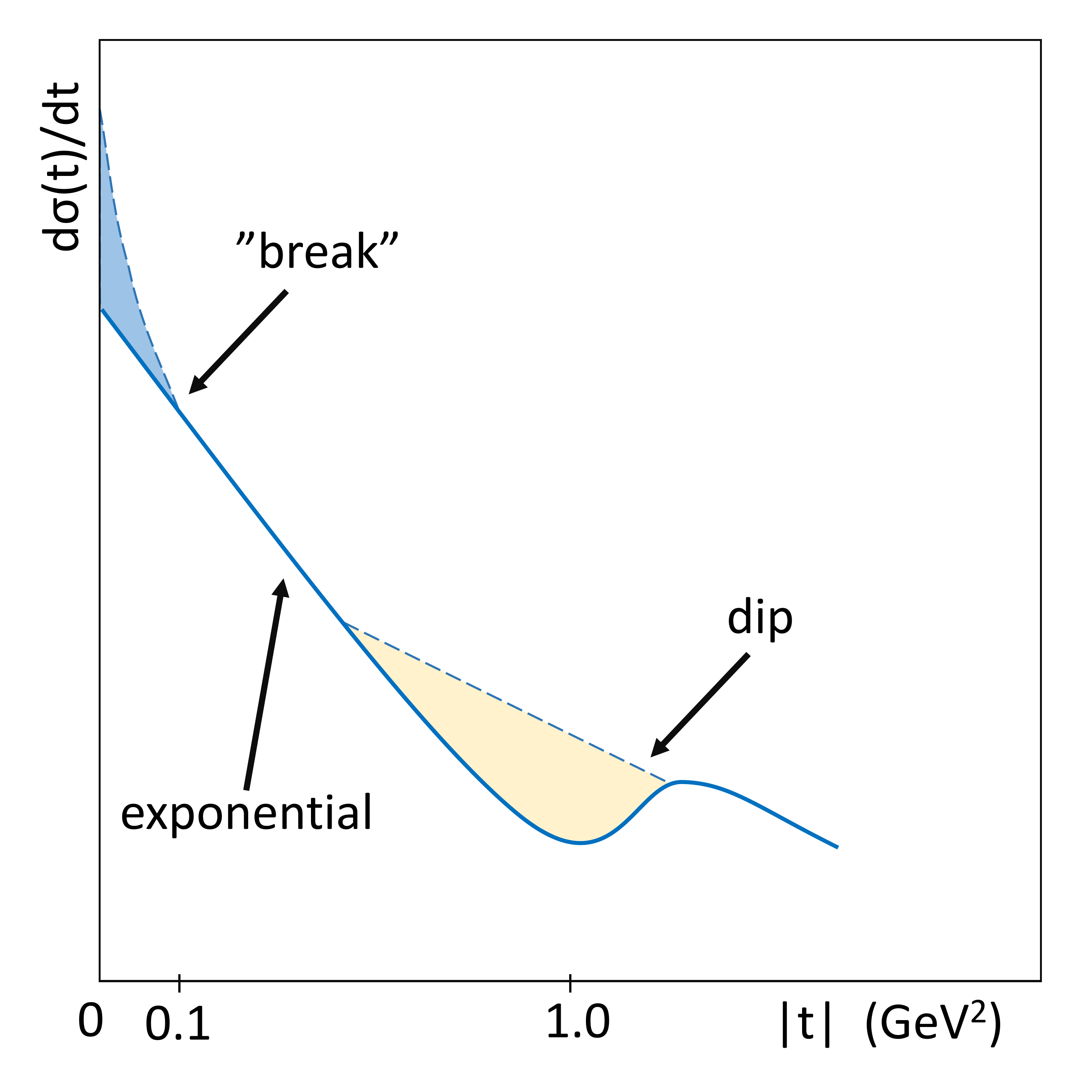}
		\includegraphics[width=0.35\linewidth]{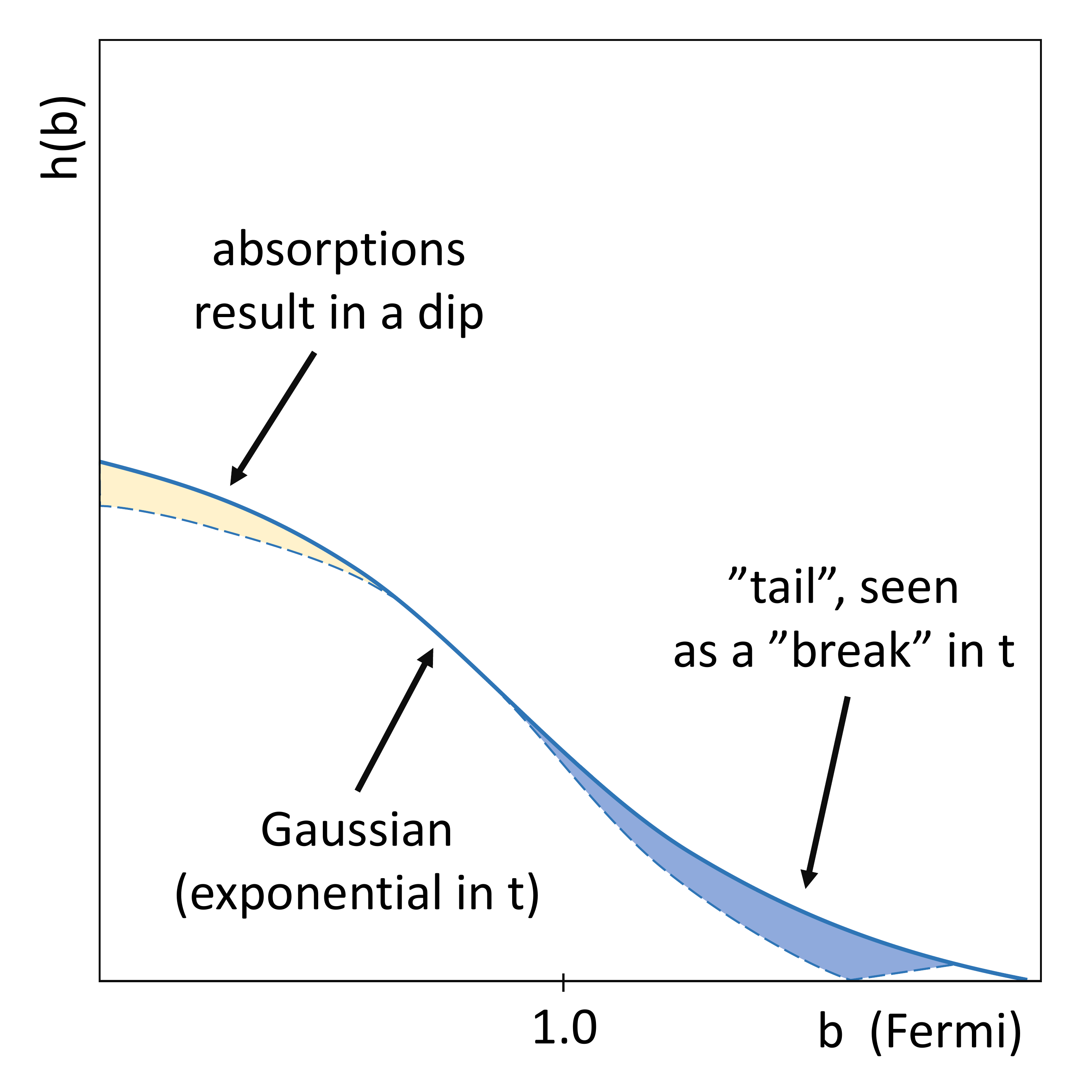}\\
		(a)\hspace{0.45\linewidth}(b)
	}
	\caption{Schematic (qualitative) view of the "break", followed by the diffraction minimum ("dip"), shown both as function in $t$ and its Fourier transform (impact parameter representation), in $b$.}
	\label{Fig:1}
\end{figure}

\begin{figure}[h] 
	\centering
	\includegraphics[width=1\textwidth]{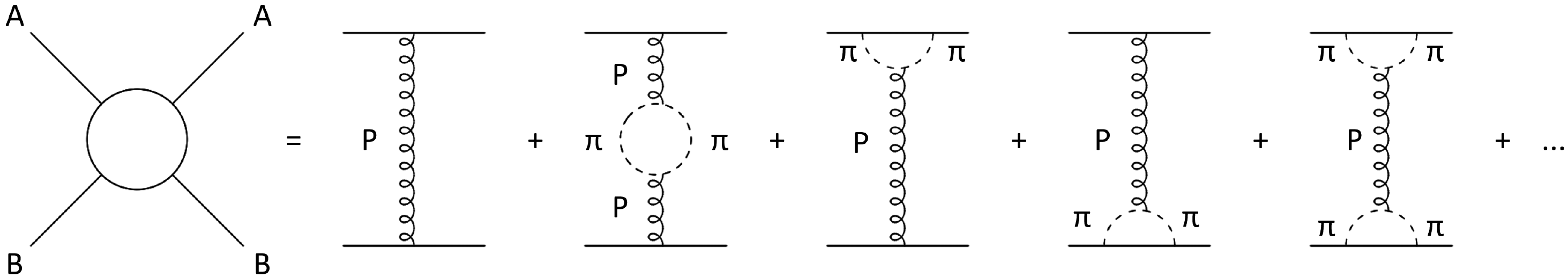}
	\caption{Diagram of elastic scattering with $t$-channel exchange containing a branch point at $t=4m_{\pi}^2$.} 
	\label{Fig:Diagram}
\end{figure}

The diffraction cone of high-energy elastic hadron scattering deviates from a purely exponential dependence on $t$ with two structures clearly visible in proton-proton scattering: the so-called "break" (in fact, a smooth curve with a concave curvature) near $t=-0.1$ GeV$^2$, whose position is nearly independent of energy and the prominent "dip" -- diffraction minimum, moving slowly (logarithmically) with $s$ towards smaller values of $|t|$. 
Physics of these two phenomena are quite different. As illustrated in Fig. \ref{Fig:1}, the "break" appears due to the "pion cloud", which controls  the ``static size" of nucleon. This effect, first observed in 1972 at the ISR, was interpreted \cite{RPM,JSZ2,JSZT, C-I1, C-I2} as the manifestation of $t$-channel unitarity, generating a two-pion loop in the cross channel, Fig. \ref{Fig:Diagram}, and was  referred to  by Bronzan \cite{Bronzan} as the ``fine structure" of the pomeron.
The dip (diffraction minimum), on the other hand, is generally accepted as a consequence of $s$-channel unitarity or absorption corrections to the scattering amplitude. 

\subsection{Low-$|t|$ "break" and proton shape}\label{ssec:break}
The deviation from the linear exponential behavior was confirmed by recent measurements by the TOTEM Collaboration at the CERN LHC, first at $8$ TeV (with significance greater then 7$\sigma$) \cite{totem8.3} and subsequently at $13$ TeV \cite{TOTEM_rho}. 
At the ISR the "break" was illustrated by plotting the local slope $B(t)$ for several $t$-bins at fixed values of $s$. 

At the LHC, the effect is of the same order of magnitude and is located near the same value of $t$. Different from the ISR \cite{Bar}, TOTEM quantifies the deviation from the exponential by normalizing the measured cross section to a linear exponential form, (see Eq.~(\ref{Eq:norm}) below). For the sake of completeness we will exhibit this ``break effect"  both in the normalized form and for $B(t)$.    

The new LHC data from TOTEM at $8$ TeV confirm the conclusions \cite{LNC, C-I1, C-I2} about the nature of the break and call for a more detailed analysis and better understanding of this phenomenon. The new data triggered intense theoretical work in this direction \cite{JL, Brazilb, RPM}, but many issues still remain open. 
While the departure from a linear exponential was studied in details both at the ISR and LHC energies, an interpolation between the two is desirable to clarify the uniqueness of the phenomenon. This is a challenge for the theory, and it can be done within Regge-pole models. Below we do so by adopting a simple Regge-pole model, with two leading {pomeron and odderon} and two secondary reggeons, $f$ and $\omega$ exchanges.    

Having identified \cite{LNC, C-I1, C-I2} the observed "break" with a two-pion exchange effect, 
we investigate further two aspects of the phenomenon, namely: 1) to what extent is the "break" observed recently at the LHC a "recurrence"  of that seen at the ISR (universality)? 2) what is the relative weight of the Regge residue (vertex) compared to the trajectory (propagator) in producing the "break"? We answer these questions by means of a detailed fit to the elastic proton-proton scattering data in the relevant kinematic range $0.05<-t<0.3$ GeV$^2$ ranging between the ISR energies, and those available at the LHC. 

As shown by Barut and Zwanziger \cite{Barut}, $t$-channel unitarity constrains the Regge trajectories near the threshold, $t\rightarrow t_0$ by
\begin{equation} \label{Eq:Barut}
Im\, \alpha(t)\sim (t-t_0)^{\Re e\, \alpha(t_0)+1/2},
\end{equation} 
where $t_0$ is the lightest threshold, $4m_{\pi}^2$ in the case of the vacuum quantum numbers (pomeron or $f$ meson). Since the asymptotic behavior of the trajectories is constrained by dual models with Mandelstam analyticity by square-root (modulus $\ln t$):
$\mid\frac{\alpha(t)}{\sqrt{t}\ln t}\mid_{t\rightarrow \infty}\leq {\rm const}$, 
(see Ref.~\cite{LNC} and references therein), for practical reasons it is convenient to approximate, for the region of $t$ in question,  the trajectory as a sum of square roots. Higher thresholds, indispensable in the trajectory, may be approximated by their power expansion, {\it i.e.} by a linear term, matching the threshold behavior with the asymptotic. 

At the ISR, the proton-proton differential cross section was measured at $\sqrt {s}=23.5, 30.7, 44.7, 52.8$ and $62.5$ GeV \cite{ISR}. In all the above energies the differential cross section changes its slope near $-t=0.1$ GeV$^2$. By using a simple Regge-pole model we have mapped the "break" fitted at the ISR onto the LHC TOTEM 8 and 13 TeV data. The simple Regge-pole model is constructed by two leading (pomeron and odderon) and two secondary, $f$ and $\omega$ contributions. 

\begin{figure}[h] 
	\centering
	\subfloat[52.8 GeV\label{fig:B2}]{%
		\includegraphics[scale=0.15]{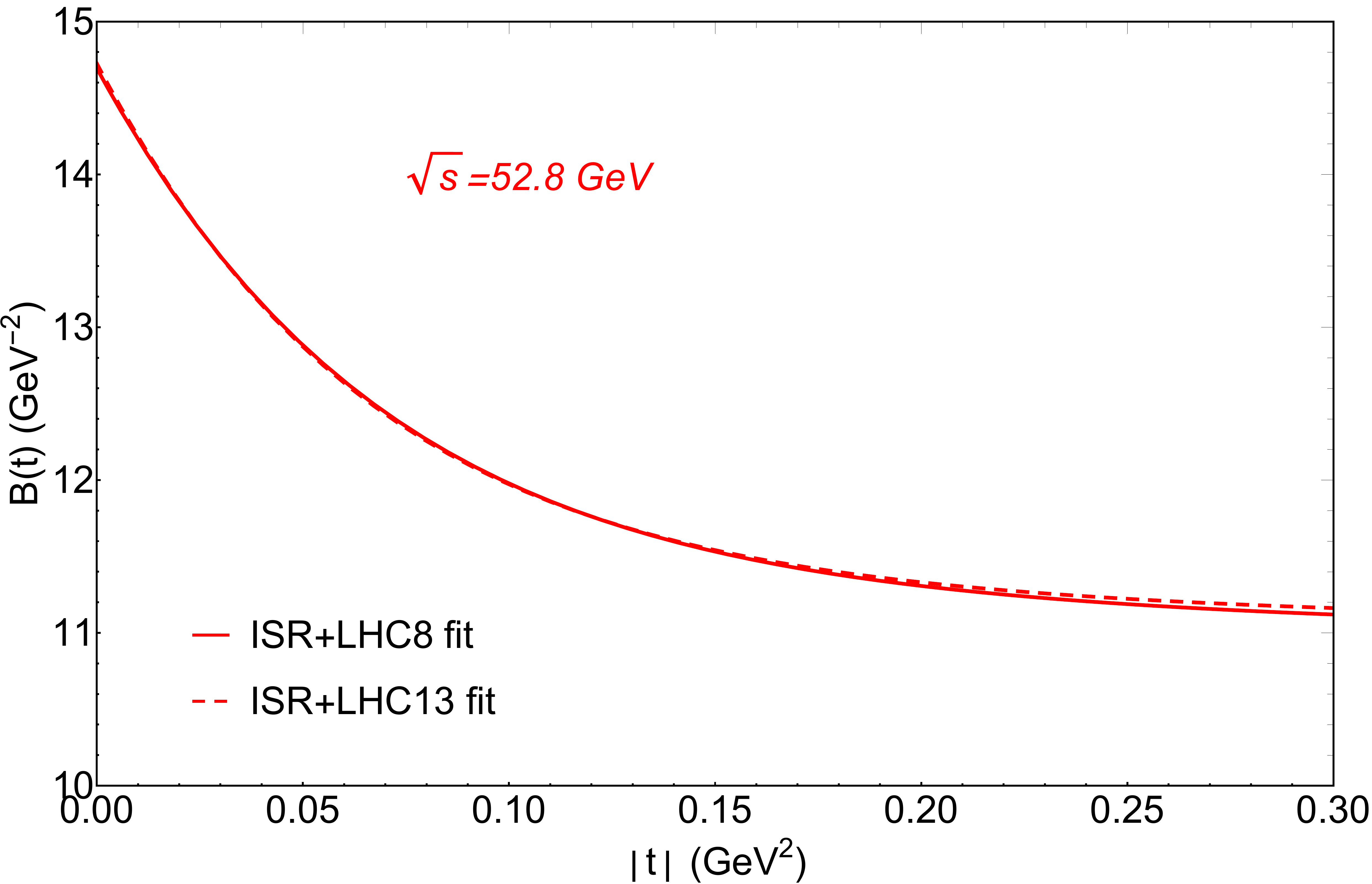}%
	}\hfil
	\subfloat[8 TeV\label{fig:B4}]{%
		\includegraphics[scale=0.15]{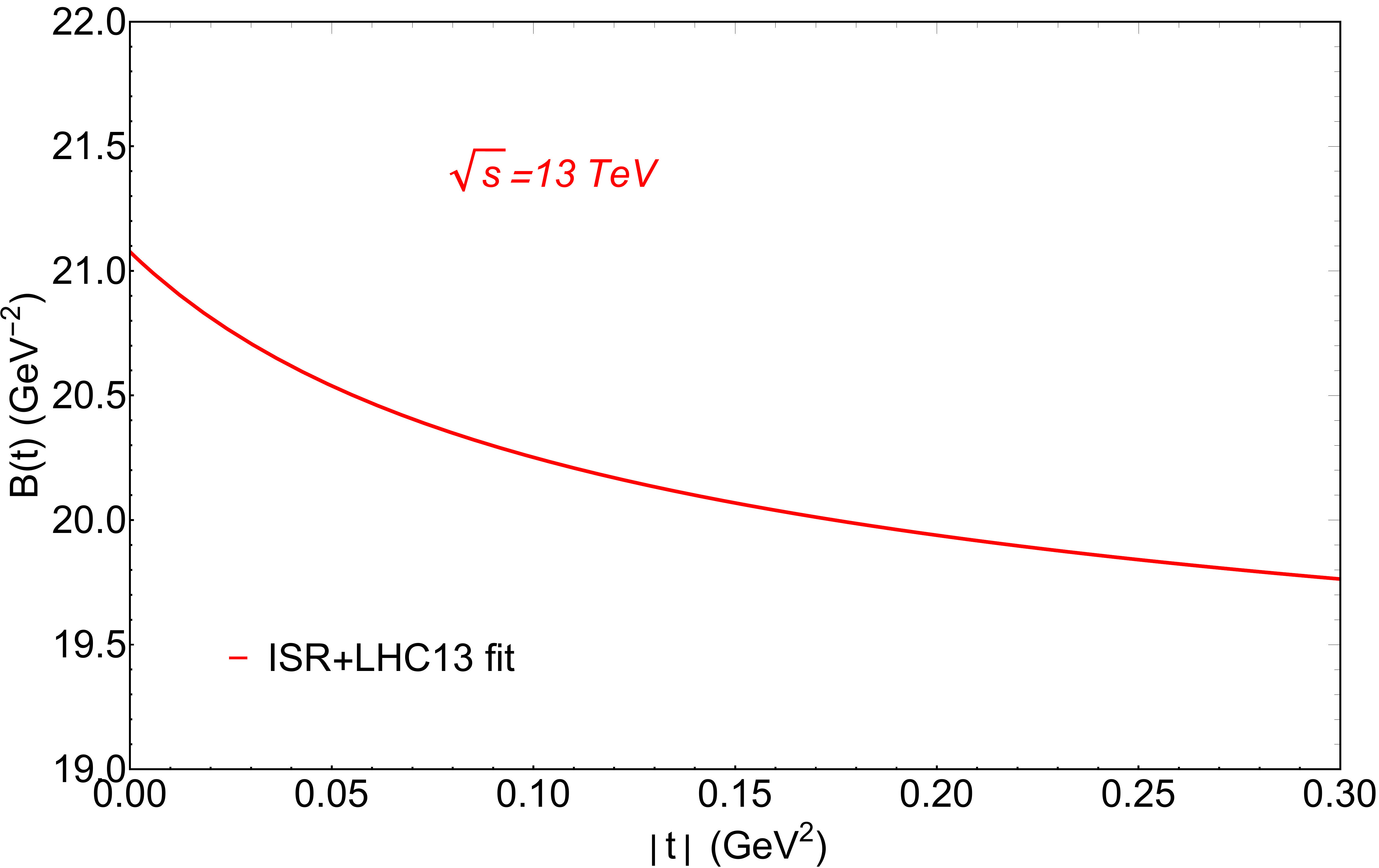}%
	}
	
	\caption{Local slopes.}
	\label{Fig:localslope}
\end{figure}

\begin{figure}[h] 
	\centering
	\subfloat[52.8 GeV\label{fig:R2}]{%
		\includegraphics[scale=0.15]{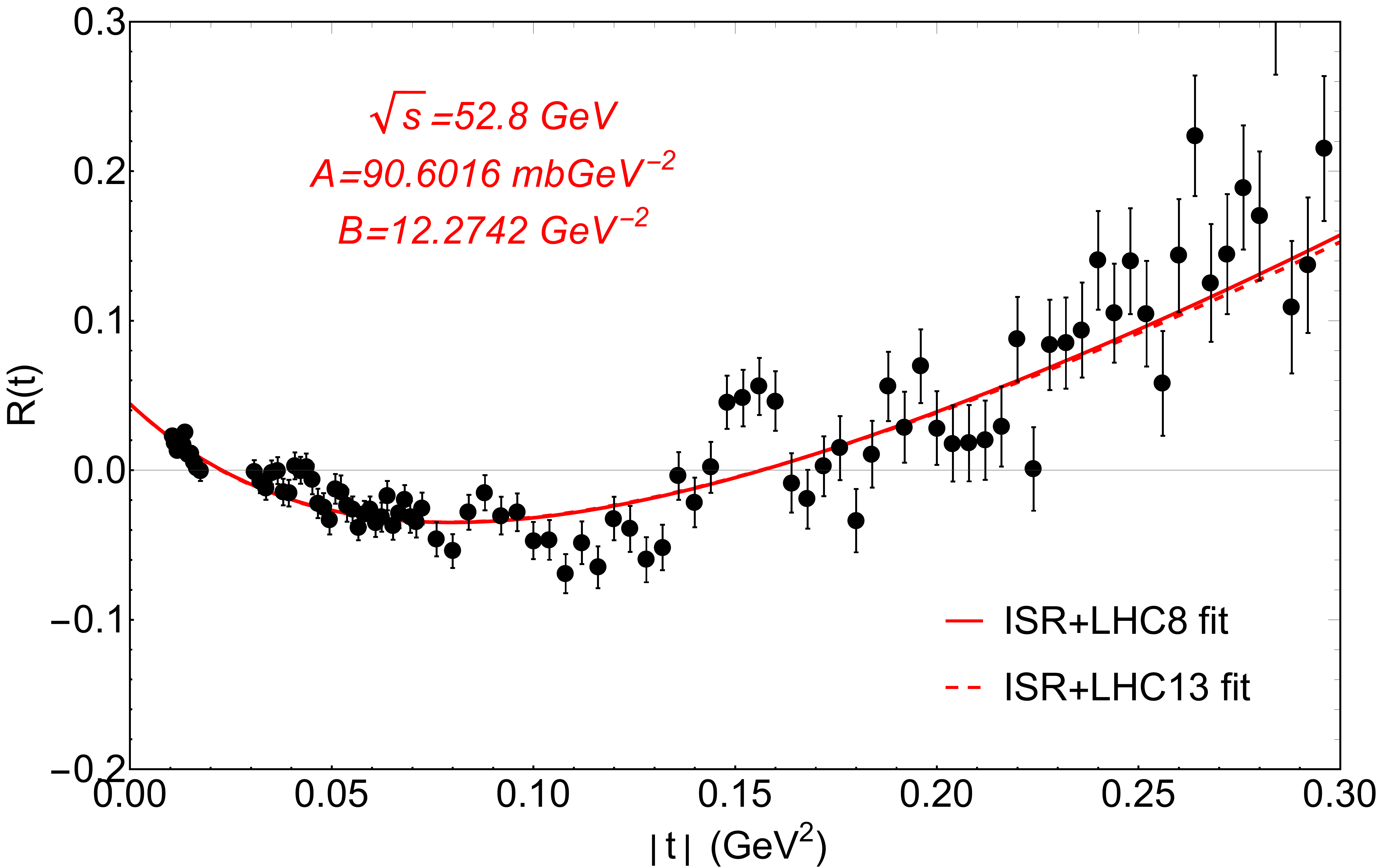}%
	}\hfil
	\subfloat[8 TeV\label{fig:R4}]{%
		\includegraphics[scale=0.15]{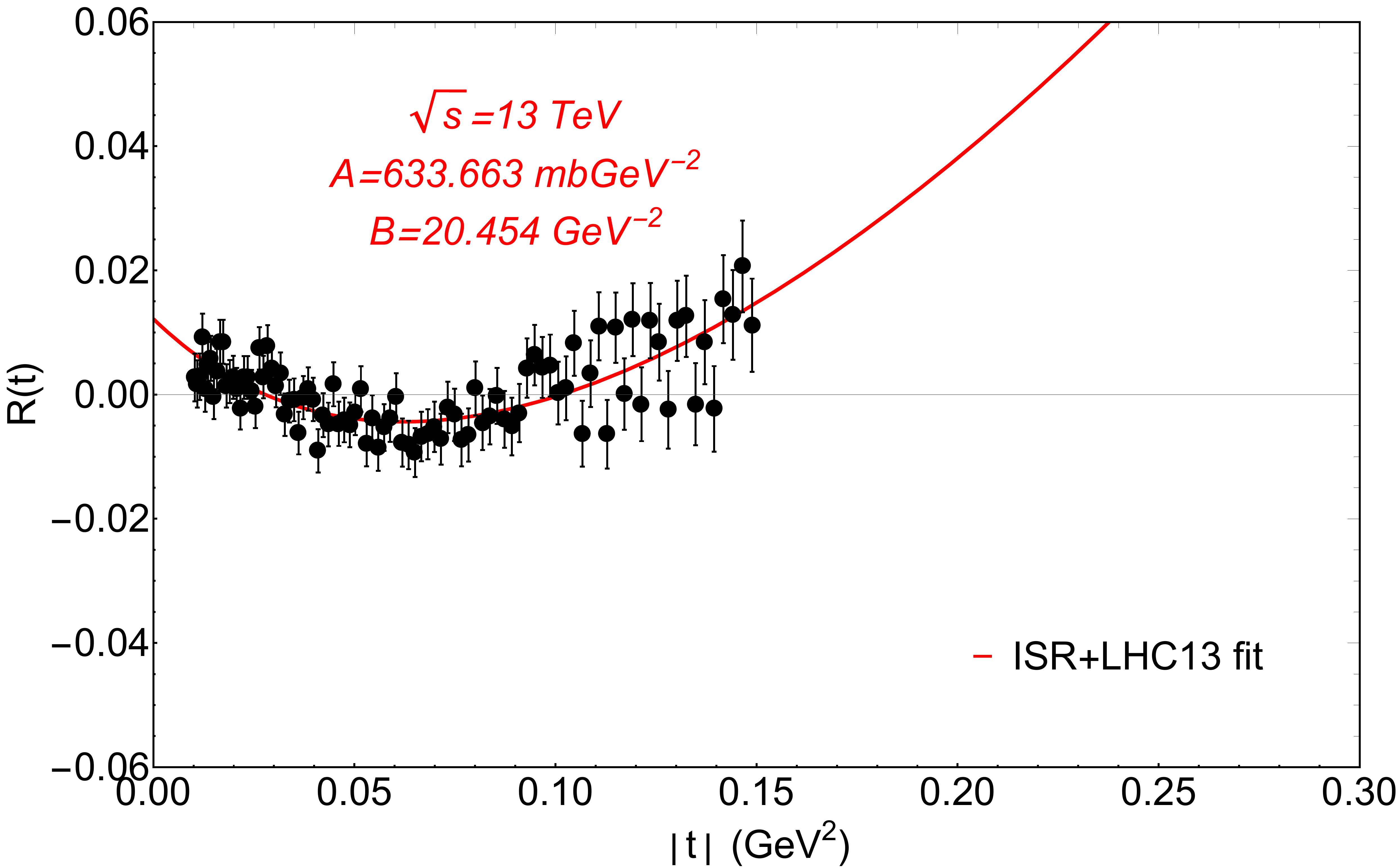}%
	}
	
	\caption{$R(t)$ ratios.}
	\label{Fig:Rratio}
\end{figure}

Detailed results of fits and the parameters are presented in Ref~\cite{JSZT}. To demonstrate the important features more clearly, we show the results of the mapping in higher resolution in Fig.~\ref{Fig:localslope} and Fig.~\ref{Fig:Rratio}. In Fig.~\ref{Fig:localslope}, we  exhibit  the shape of local slopes,  defined  by
\begin{equation}
B(s,t)=\frac{d}{dt} \ln (d\sigma/dt ) \,.
\end{equation}
To  demonstrate better the quality of our fit and to anticipate comparison with the LHC data, we  present  in Fig.~\ref{Fig:Rratio} also the ISR data  in normalized form as used by TOTEM \cite{TOTEM_rho}: 
\begin{equation} \label{Eq:norm}
R=\frac{d\sigma/dt}{d\sigma/dt_{ref}}-1,
\end{equation}
where $d\sigma/dt_{ref}=Ae^{Bt}$, with $A$ and $B$ are constants determined from a fit to the experimental data.   

Both  Fig.~\ref{Fig:localslope} and Fig.~\ref{Fig:Rratio} re-confirm the earlier finding that the ``break" can be attributed the presence of two-pion branch cuts in the Regge parametrization. 

\begin{figure}[H]
	\center{
		\centering
		\subfloat[]{%
			\includegraphics[width=0.5\linewidth]{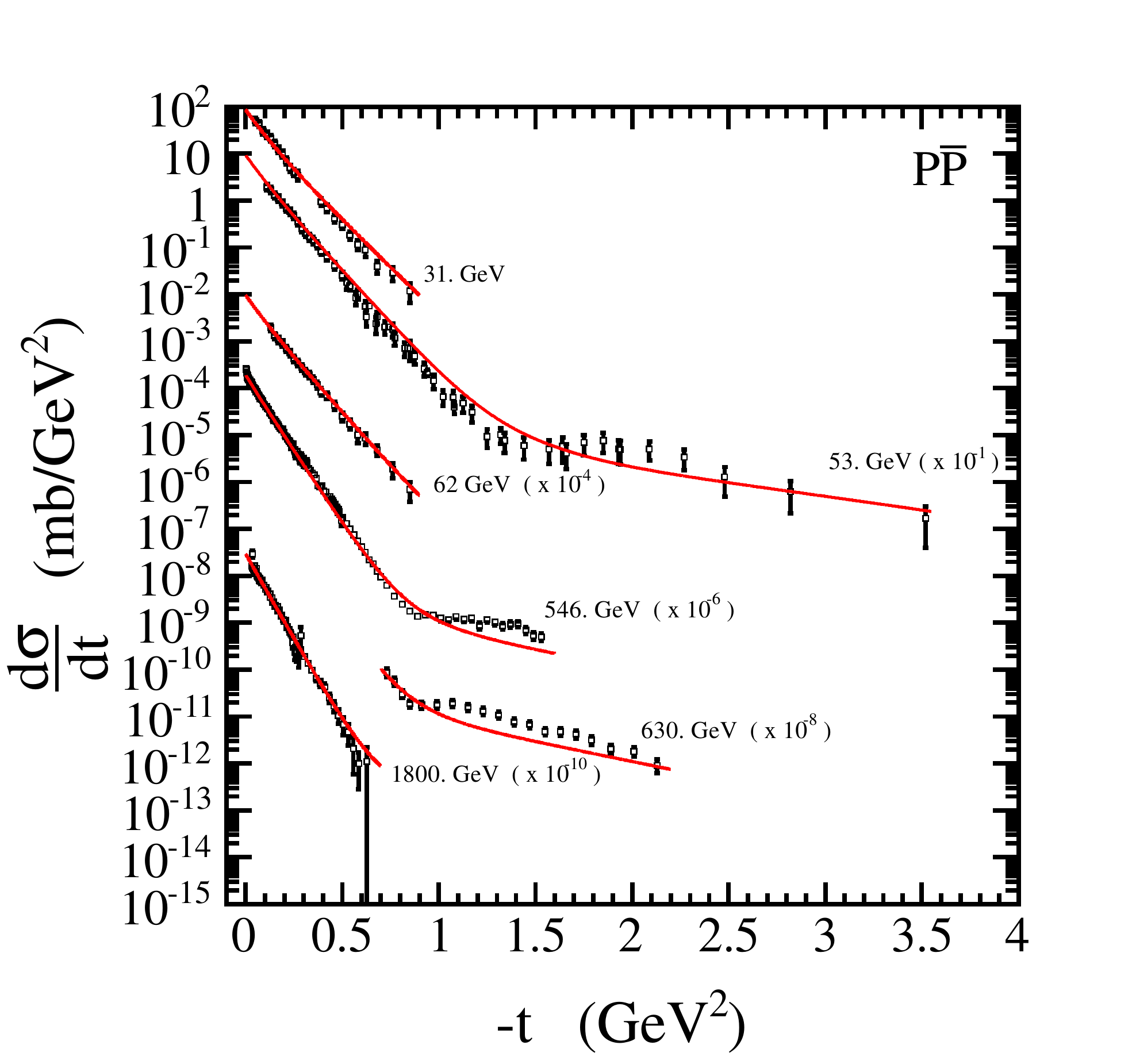}%
		}\hfil
		\subfloat[]{%
			\includegraphics[width=0.5\linewidth]{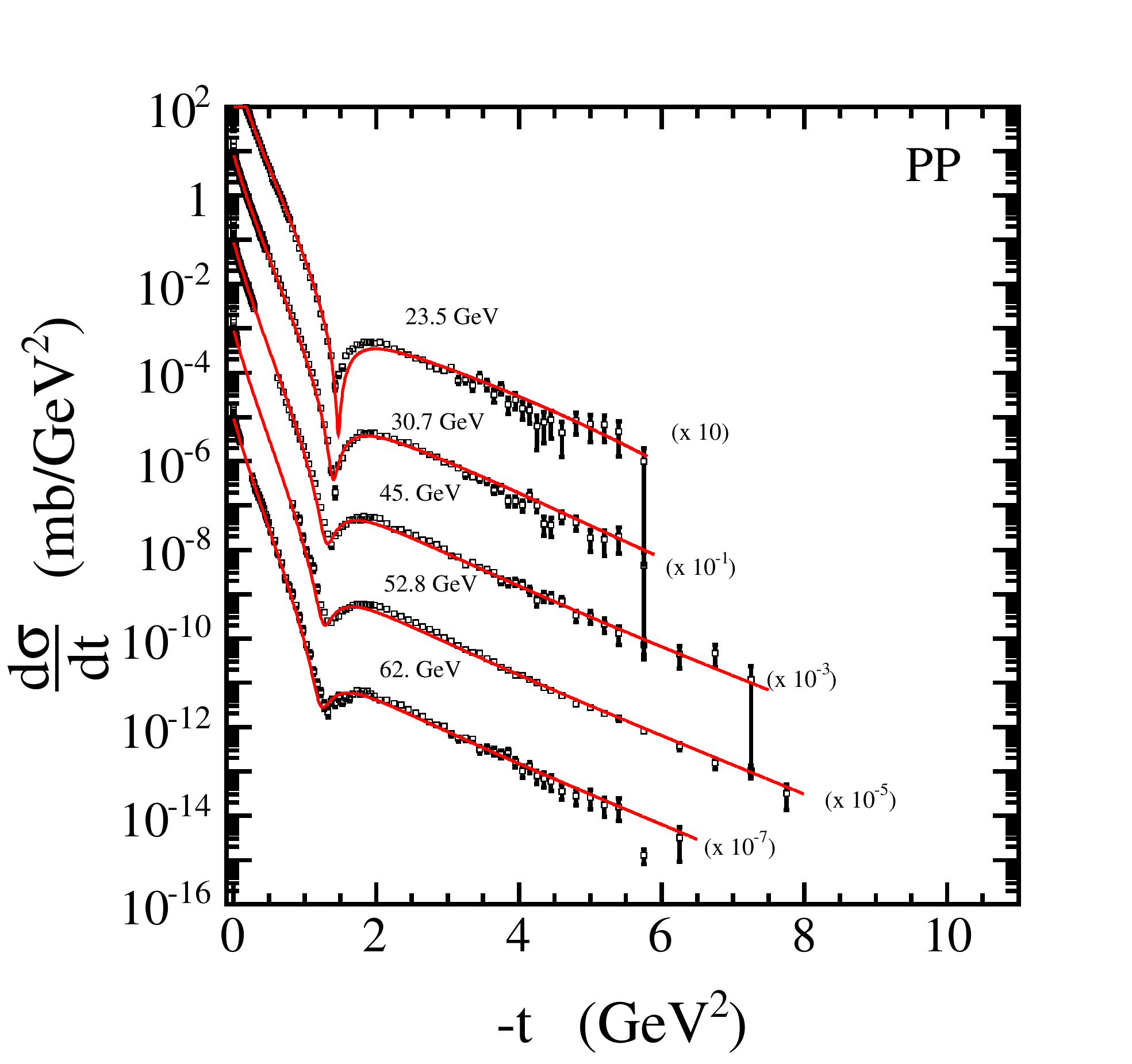}%
		}
	}
	\caption{
		(a) $\bar{p}p$ differential cross sections calculated
		from the model, Eqs.~(\ref{Eq:Amplitude}-\ref{Odd}), and fitted to the data, and fitted to the data in the range $-t$ = 0.1 --- 8~GeV$^2$.
		(b) $pp$ differential cross sections calculated
		from model and fitted to the data.
	}
	\label{fig:pp_all}
\end{figure}

\begin{figure}[H]
	\center{
    \includegraphics[angle=0,width=0.5\textwidth]{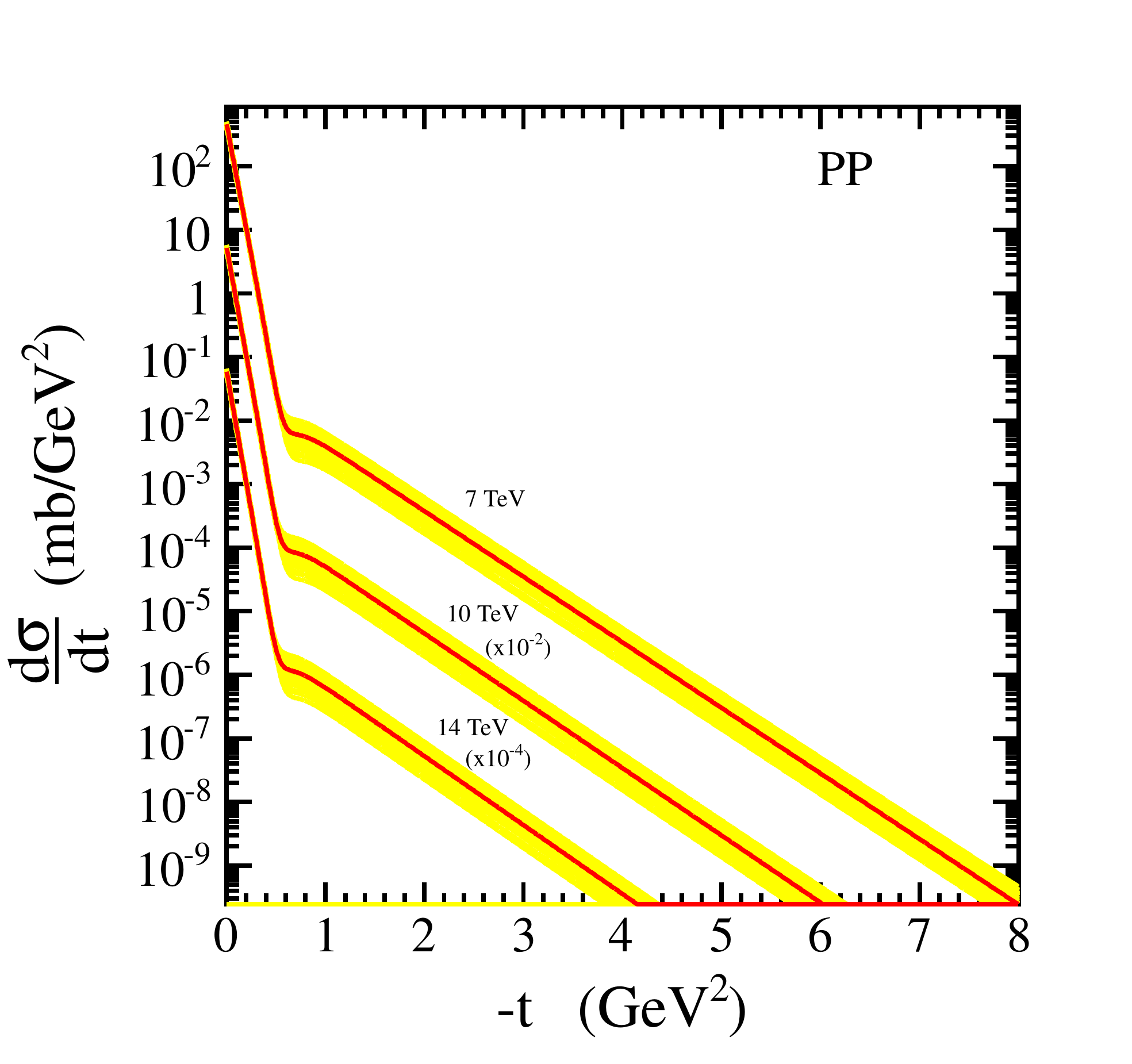}
    }
	\caption{Predictions for the $pp$ differential cross section calculated
		from the model for three different LHC energies.}
	\label{fig:pp_pre}
\end{figure}

Figures~\ref{fig:pp_all} and \ref{fig:pp_pre} show the fitted $\bar{p}p$ and $pp$ differential cross sections and predictions for three different center of mass energies using the model Eqs.~(\ref{Eq:Amplitude}-\ref{Odd}) (see more detail and parameters in Ref.~\cite{JLL}). The yellow area exhibits the statistical uncertainty on the calculations, described earlier. 

\section{Single and double diffraction dissociation}\label{Sec:DD}

Measurements of single (\textbf{SD}), double(\textbf{DD}) and central(\textbf{CD}) diffraction dissociation is among the priorities of the LHC research program.

In the past, intensive studies of high-energy diffraction dissociation were performed at the Fermilab,
on fixed deuteron target, and at the ISR, see Ref.~\cite{Goulianos} for an overview and relevant references. Fig.~\ref{fig:2SD.Fermilab}
shows representative curves of low-mass SD as measured at the Fermilab. One can see the rich resonance structure there,
typical for low missing masses, often ignored by extrapolating whole region by a simple $1/M^2$ dependence.
When extrapolating (in energy), one should however bear in mind that, in the ISR region, secondary reggeon contributions are still important (their relative contribution depends on momenta transfer considered), amounting to nearly $50\%$ in the forward direction. At the LHC, however, their contribution in the nearly forward direction in negligible, i.e. less than the relevant error bars in the measured total cross section \cite{JLL}.

\begin{figure}[!ht]
	\centering
	\includegraphics[width=0.8\textwidth ,bb= 0 0 730 180]{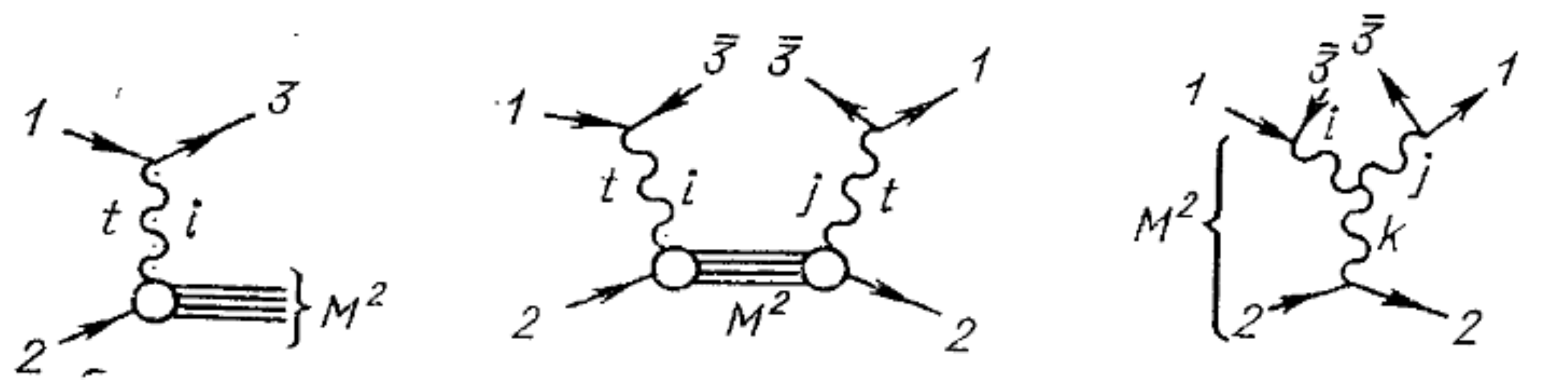}
	\caption{From SD to the triple Regge limit.}
	\label{fig:TripleReggeLimit}
\end{figure}

\begin{figure}[!ht]
	\centering{\includegraphics[width=0.4\linewidth]{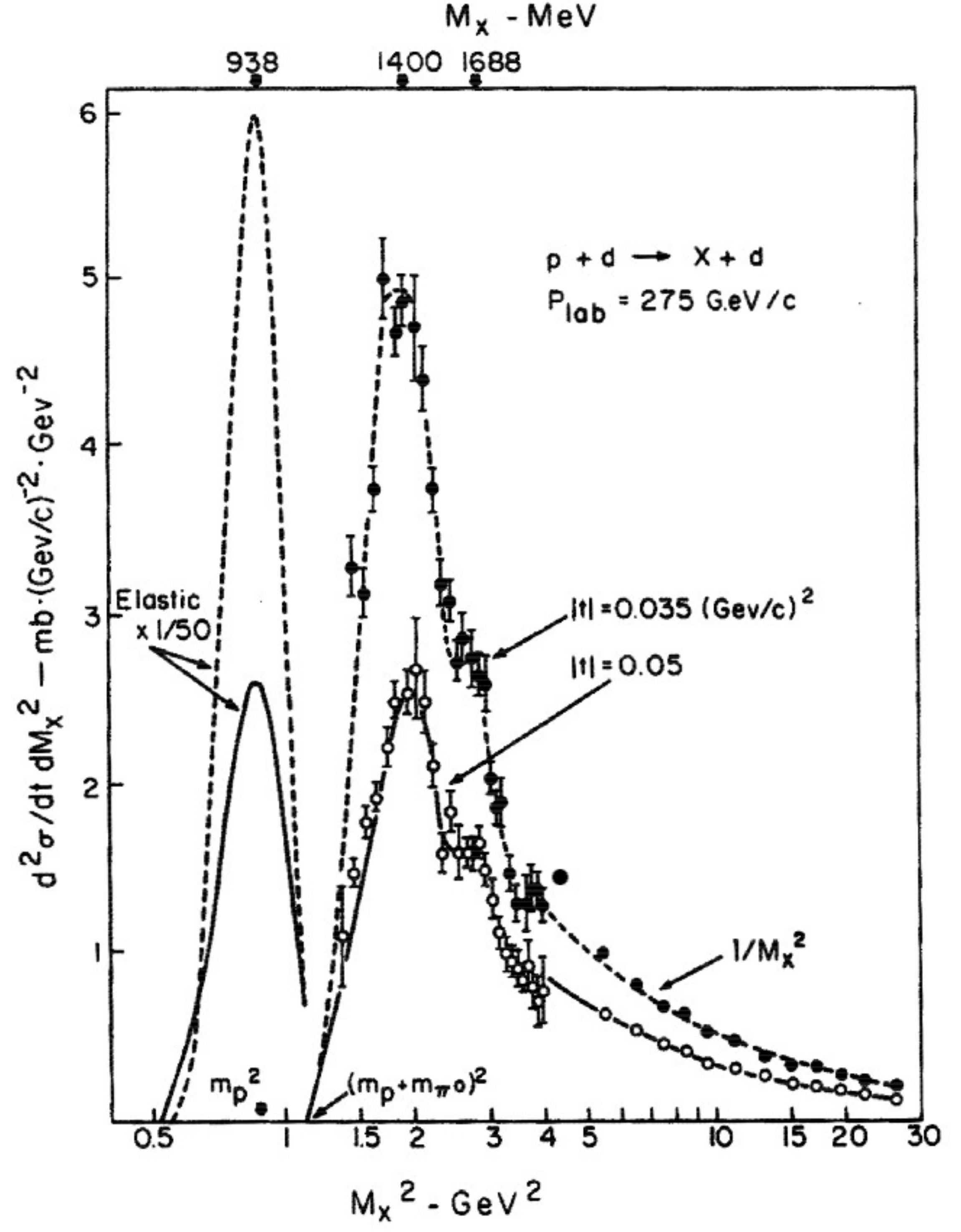}
		\includegraphics[width=0.55\linewidth]{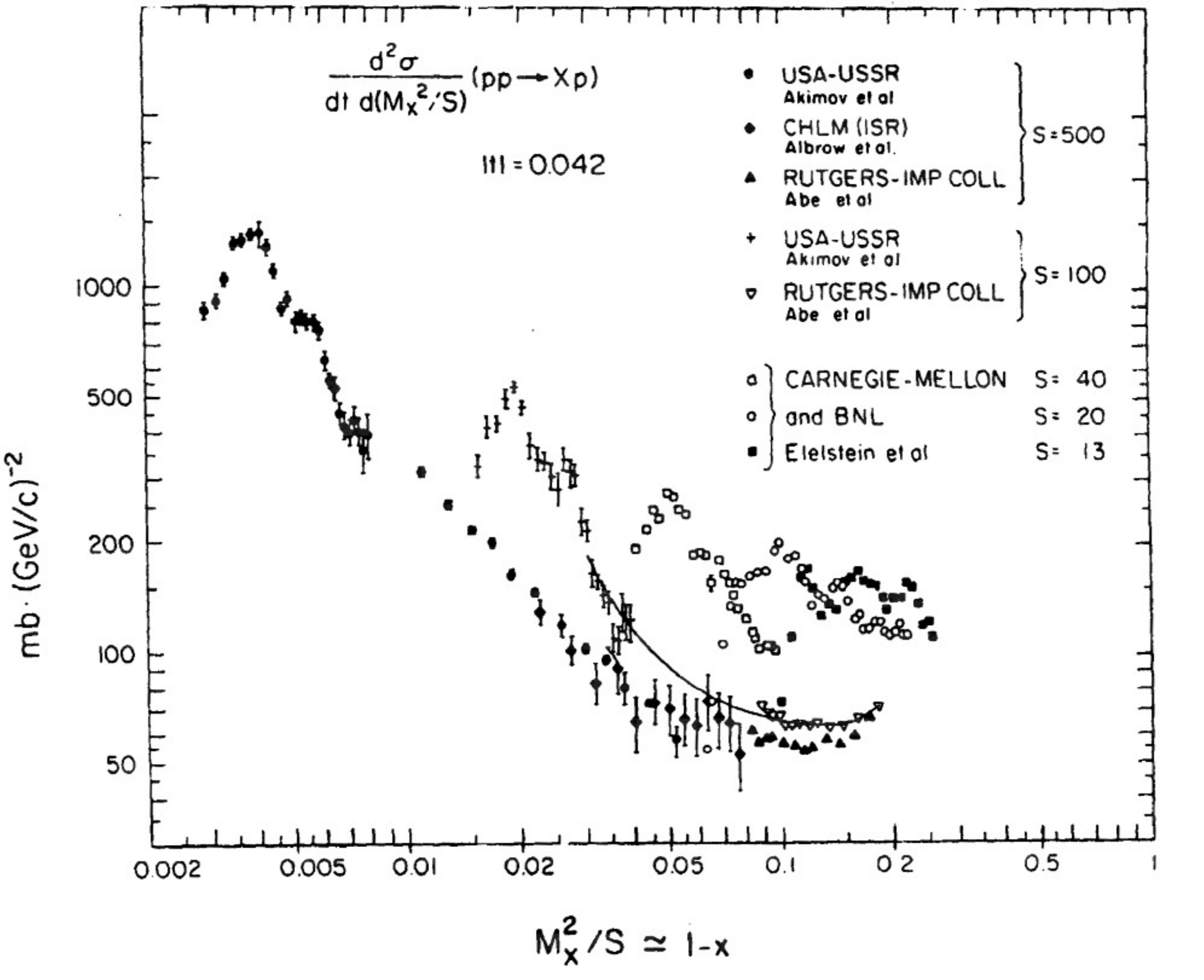}\\
		(a)\hspace{0.45\linewidth}(b)}
	\caption{Compilation of low-mass SD data form Fermilab experiments, see Ref.~\cite{Goulianos}.}
	\label{fig:2SD.Fermilab}
\end{figure}

In most of the papers on the subject SD is calculated from the triple Regge limit of an
inclusive reaction, as shown in Fig.~\ref{fig:TripleReggeLimit}.

In that limit, the double diffraction cross section can be written as \cite{Collins, Goulianos}
$$
\frac{d^2\sigma}{dtdM^2_x}=
\frac{G^{PP, P(t)}_{1 3, 2}}{16\pi^2s_0^2}\left(\frac{s}{s_0}\right)^{2\alpha_P(t)-2}\left(\frac{M^2}{s_0}\right)^{\alpha_P(0)-\alpha_P(t)}.
$$

This approach has two shortcomings. The first one is that it ignores the small-$M^2$ resonance region. The second one is connected with the fact that, whatever the pomeron, the (partial) SD cross section overshoots the total one, thus obviously conflicting with unitarity, see {\it e.g.} Sec. 10.8 in Ref. \cite{Collins} for a clear exposition of this paradox).   

Various ways of resolving this deficiency
are known from the literature, including the vanishing (decoupling) of the triple pomeron coupling (see the same reference), but none of them can be considered completely satisfactory. Any decoupling of the triple pomeron vertex looks unnatural, however, finite values are arbitrary, thus
unsatisfactory.  

To remedy the rapid increase of the "partial" cross section of diffraction dissociation K. Goulianos "renormalizes" the above cross section \cite{Goulianos_renorm} by multiplying it by a step function moderating its rise and thus resolving the conflict. 

We instead follow the idea put forward in paper Ref.~\cite{JLa1,JLa2} and developed further according to which the reggeon (here, the pomeron) is similar to the photon and that the reggeon-nucleon interaction is similar to deep-inelastic photon-nucleon scattering (DIS), with the replacement $-Q^2=q^2\rightarrow t$ and $s=W^2\rightarrow M^2_x$. There is an obvious difference between the two: while the $C$ parity of the photon is negative, it is positive for the pomeron. We believe that while the dynamics is essentially invariant under the change of $C$, the difference between the two being accounted for by the proper choice of the parameters. Furthermore, while Jaroszewicz and Landshoff \cite{JLa1} (see also Ref.~\cite{JLa2}), in their pomeron-nucleon DIS structure function (SF) (or $Pp$ total cross section) use the Regge asymptotic limit, we include also the low missing mass, resonance behavior. As is known, gauge invariance requires the DIS SF to vanish as $Q^2$ (here, $t)\ \rightarrow 0$. This property
is inherent of the SF, see  Refs.~\cite{JLa1, JLa2}, and it has important consequences for the behavior of the resulting cross sections at low $t$,  not shared by models based on the triple Regge limit, see Refs.~\cite{Collins, Goulianos}.

It is evident that Regge factorization is essential in both approaches (triple Regge and the present one). It is feasible when Regge singularities are isolated poles. While the pre-LHC data require the inclusion of secondary Reggeons, at the LHC we are in the fortunate situation of a single pomeron exchange (pomeron dominance) in the $t$ channel in single and double diffraction (not necessarily so in central diffraction, to be treated elsewhere). Secondary Regge pole exchanges will appear however, in our  dual-Regge treatment of $Pp$ scattering (see below), not to be confused with the the $t$ channel of $pp$.
This new situation makes diffraction at the LHC unique in the sense that for the first time Regge-factorization is directly applicable. We make full use of it.

As shown in Figs.\ref{fig:2SD.Fermilab}, there is a rich resonance structure in the small $M^2$ region. In most of the papers on the subject, this resonance structure is ignored and replaced by a smooth function $\sim M^{-2}$. Moreover, this simple power-like behavior is extended to the largest available missing masses. 
By duality, the averaged contribution from resonances sums up to produce high missing mass Regge behavior $~(M^2)^{-n},$ where $n$ is related to the intercept of the exchanged trajectory and may be close (but not necessarily equal) to the above-mentioned empirical value $\sim 1$. Of course, this number depends on the type trajectory exchanged; this interesting point deserves further study. 

\subsection{Duality and low missing-mass resonances}
With the advent of the LHC, diffraction, elastic and inelastic scattering entered a new area, where it can be seen uncontaminated by non-diffraction
events. In terms of the Regge-pole theory this means, that the scattering amplitude is completely determined by a pomeron exchange, and in a
simple-pole approximation, Regge factorization holds and it is of practical use! Remind that the pomeron is not necessarily a simple pole:
perturbative QCD suggests that the pomeron is made of an infinite number of poles (useless in practice), and the unitarity condition requires corrections
to the simple pole, whose calculation is far from unique. Instead a simple pomeron pole approximation \cite{DL} is efficient in describing
a variety of diffraction phenomena.

The elastic scattering amplitude in the simple model of Donnachie and Landshoff \cite{DL} is
\begin{equation}\label{DL}
A(s,t)=\xi(t)\beta(t)^2(s/s_0)^{\alpha_P(t)-1}+A_R(s,t),
\end{equation}
where $\xi(t)$ is the signature factor, and $\alpha_P(t)$ is the (linear) pomeron trajectory. The signature factor can be
written as $\xi(t)=e^{-i\pi\alpha_P(t)/2}$, however it is irrelevant here, since below we shall use only cross sections (squared modules of the amplitude), where it reduces to unity. The residue is chosen to be a simple exponential, $\beta(t)=e^{b_Pt}$.
"Minus one" in the propagator term $(s/s_0)^{\alpha_P(t)-1}$ of (\ref{DL}) correspond for normalization $\sigma_{T}(s)=Im A(s,t=0)$. The scale parameter $s_0$ is not fixed by the Regge-pole theory: it can be fitted do the data or fixed to a "plausible" value of a hadronic mass, or to the inverse "string tension" (inverse of the pomeron slope), $s_0=1/\alpha'_P$. The second term in Eq.~(\ref{DL}), corresponding to sub-leading Reggeons, has the same functional form as the first one (that of the pomeron), just the values of the parameters differ. We ignore this term.

Fig.~\ref{Fig:Factor} shows the simplest configurations of Regge-pole diagrams for elastic, single- and double diffraction dissociation, as well as central diffraction dissociation (CD). In this Section we consider only SD and DD. CD will be treated in the next Section.

Having accepted the factorized form of the scattering amplitude, Sec. \ref{Regge},
the main object of our study is now the inelastic proton-pomeron vertex or transition amplitude.
As argued in Ref.~\cite{JLa1,JLa2}, it can be treated as the proton structure function (SF), probed by the pomeron, and proportional to the
pomeron-proton total cross section, $\sigma_T^{Pp}(M^2_x, t),$ with the norm $\sigma_T^{Pp}(M^2_x, t)=Im A(M^2,t),$
in analogy with the proton SF probed by a photon (in $ep$ scattering e.g. at HERA or JLab).
$$\nu W_2(M_x^2,t)=F_2(x,t)=\frac{4(-t)(1-x)^2}{\alpha(M_x^2-m_p^2)(1+4m_p^2x^2/(-t))^{3/2}}Im A(M^2,t),$$
where $\alpha$ is the fine structure constant, $\nu=\frac{M_x^2-m_p^2-t}{2m_p}$, and $x=\frac{-t}{2m_p\nu}$ is the Bjorken variable.

The only difference is that the pomeron's (positive) $C$ parity is opposite to that of the photon. This difference is evident in the values of the parameters but is unlikely to affect the functional form of the SF itself, for which we choose its high-$M_x^2$ (low Bjorken $x$) behavior. Notice that the the total energy in this subprocess, the analogy of $s=W^2$ in DIS, here is $M^2_x$ and $t$ here replaces $q^2=-Q^2$ of DIS.
Notice that gauge invariance requires that the SF vanishes towards $Q^2\rightarrow 0$ (here, $t$), resulting in the dramatic vanishing of the SD and DD differential cross section towards $t=0.$  How fast does the SF (and relevant cross sections) recover from $t=0$ a priori is not known.

\begin{figure}[!ht]
	\centering
	\includegraphics[width=0.24\linewidth,bb=0 0 480 480,clip]{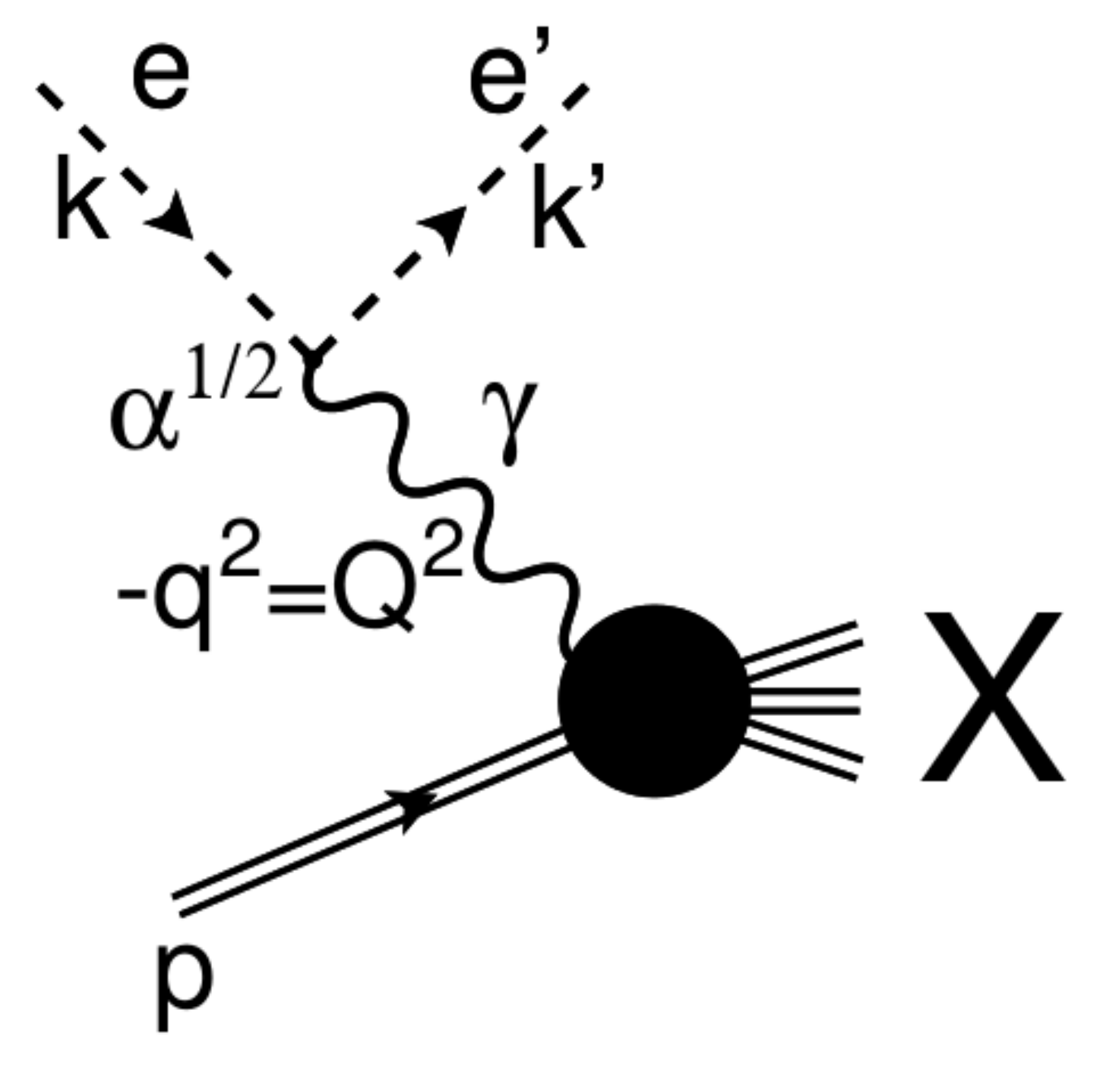}
	\caption{ Virtual photon + proton $\rightarrow M_x^2$ transition.}
	\label{fig:gamma_p}
\end{figure}

Furthermore, according to the ideas of two-component duality, the cross sections of any process, including that of $pP\rightarrow X,$ is a sum of a non-diffraction component, in which resonances sum up in high-energy (here: mass $M^2$ plays the role of energy $s$) Regge exchanges, and the smooth background (below the resonances), dual to the pomeron exchange. The dual properties of diffraction dissociation can be quantified also by finite mass sum rules, see Ref.~\cite{Goulianos}. In short: the high-mass behavior of the $pP\rightarrow X$ cross section is a sum of a decreasing term going like $\sim \frac{1}{M^m}$,$m\approx 2$ and a "pomeron exchange" increasing slowly with mass. All this has little affect on the low-mass behavior at the LHC, however normalization implies calculation of cross sections integrated over all physical values of $M^2$, i.e. until $M^2<0.05s$.

The background in the reactions (SD, DD) under consideration there are two sources of the background. One is related to the $t$ channel exchange in Fig \ref{fig:2SD.Fermilab}(b) and it can be accounted for by rescaling the parameter $s_0$ in the denominator of the pomeron propagator. In any case, at high energies, those of the LHC, this background is included automatically in the pomeron. The second component of the background comes from the subprocesses $pP\rightarrow X$. Its high-mass behavior is not known experimentally and it can be only conjecture on the bases of the known energy dependence of the typical meson-baryon processes appended by the ideas of duality. The conclusion is that the $Pp$ total cross section at high energies (here: missing masses $M$) has two components:
a decreasing one, dual to direct-channel resonances and going as $\sigma_{tot}^{Pp}\sim \sum_R(s')^{\alpha_R(0)-1}= \sum_R\left(M^2\right)^{\alpha_R(0)-1},$ where $R$ are non-leading Reggeons, and a slowly rising pomeron term producing $\sim M^{2\cdot0.08}$.

\subsection{Resonances in the $Pp$ system; the $N^*$ trajectory}\label{Sec:trajectory}
The Pp total cross section at low missing masses is dominated by nucleon resonances. In the dual-Regge approach the relevant cross section is a "Breit-Wigner" sum Eq.~(\ref{total}), in which the direct-channel trajectory is that of $N^*$. An explicit model of the nonlinear $N^*$ trajectory was elaborated in paper Ref.~\cite{Paccanoni}.

\begin{equation} \label{total}
\sigma^{Pp}_T(M_x^2,t)=Im\, A(M_x^2,t)=Im\left(\sum_{n=0,1,...}\frac{a f(t)^{2(n+1)}}{2n+0.5-\alpha_{N^*}(M_x^2)}\right).
\end{equation}
%

The pomeron-proton channel, $Pp\rightarrow M_X^2$  couples to the
proton trajectory, with the $I(J^P)$ resonances: $1/2(5/2^+),\
F_{15},\ m=1680$~MeV, $\Gamma=130$~MeV; $1/2(9/2^+),\ H_{19},\
m=2200$~MeV, $\Gamma=400$~MeV; and $1/2(13/2^+),\ K_{1,13},\
m=2700$~MeV, $\Gamma=350$~MeV. The status of the first two is
firmly established, while the third one,
$N^*(2700),$ is less certain, with its width varying between
$350\pm 50$ and $900\pm 150$~MeV . Still, with the
stable proton included, we have a fairly rich trajectory,
$\alpha(M^2)$.

Despite the seemingly linear form of the trajectory, it is not
that: the trajectory must contain an imaginary part corresponding
to the finite widths of the resonances on it. The non-trivial
problem of combining the nearly linear and real function with its
imaginary part was solved in Ref.~\cite{Paccanoni} by means of
dispersion relations.

We use  the explicit form of the trajectory derived in Ref.~\cite{Paccanoni}, ensuring  correct behavior of both its real and
imaginary parts. The imaginary part of the trajectory can be
written in the following way:
\begin{equation}
Im\, \alpha(s)=s^{\delta} \sum_n c_n
\left(\frac{s-s_n}{s}\right)^{\lambda_n} \cdot \theta(s-s_n)\,,
\label{b2}
\end{equation}
where $\lambda_n=Re\ \alpha(s_n)$. Eq.~~(\ref{b2}) has the
correct threshold behavior, while analyticity requires that
$\delta <1$. The boundedness of $\alpha(s)$ for $s \to \infty$
follows from the condition that the amplitude, in the Regge form,
should have no essential singularity at infinity in the cut plane.

\subsection{Compilation of the basic formulae}\label{Sub:Formulae}
This subsection contains a compilation of the main formulae used in the calculations and fits to the data.

The elastic cross section is:
\begin{equation}\label{elastic}\frac{d\sigma_{el}}{dt}=A_{el}{F_p}^4(t)\left(\frac{s}{s_0}\right)^{2(\alpha(t)-1)}.\end{equation}
The single diffraction (SD) dissociation cross section is:
\begin{equation}\label{SD}2\cdot\frac{d^2\sigma_{SD}}{dtdM_x^2}={F_p}^2(t){F_{inel}}^2(t,M_x^2) \left(\frac{s}{M_x^2}\right)^{2(\alpha(t)-1)}.\end{equation}
Double diffraction (DD) dissociation cross section:
\begin{equation}\label{DD}\frac{d^3\sigma_{DD}}{dtdM_1^2dM_2^2}=N_{DD}{F_{inel}}^2(t,M_1^2){F_{inel}}^2(t,M_2^2)\left(\frac{ss_0}{M_1^2M_2^2}\right)^{2(\alpha(t)-1)}.\end{equation}
with the norm $N_{DD}=\frac{1}{4A_{el}},$
and inelastic vertex:
\begin{equation}\label{eq:inelVertex}
{F_{inel}}^2(t,M_x^2)=A_{res}\frac{1}{M_x^4}\sigma_T^{Pp}(M_i^2,t)+C_{bg}\sigma_{Bg},
\end{equation}
where the pomeron-proton total cross section is the sum of $N^*$ resonances and the Roper resonance:
\begin{eqnarray}\label{eq:sigma_tot}
\sigma_T^{Pp}(M_x^2,t)=
R\frac{[f_{res}(t)]^{2} \cdot M_{Roper}\left(\frac{\Gamma_{Roper}}{2}\right)}
{{\left(M_x^2-M_{Roper}^2\right)}^2+{\left(\frac{\Gamma_{Roper}}{2}\right)}^2}\\ \nonumber
+[f_{res}(t)]^{4}\sum_{n=1,3} \frac{Im\,\alpha(M^2_x)}{(2n+0.5-Re\, \alpha(M_x^2))^2+( Im\,\alpha(M^2_x))^2},
\end{eqnarray}
$A_{res}$, $C_{bg}$ and $R$ being adjustable parameters. A reasonable approximation for the background, corresponding to non-resonance contributions is:
\begin{equation}\label{eq:sigma_tot}
\sigma_{Bg}=\frac{f_{bg}(t)}{2M_x^2-(m_p+m_{\pi})^2}.
\end{equation}


The pomeron trajectory is \cite{JLL}:
$$\alpha(t)=1.075+0.34t,$$
and the  $t-$dependent elastic and inelastic form factors are:
$$F_p(t)=e^{b_{el}t}, \qquad f_{res}(t)=e^{b_{res}t}, \qquad f_{bg}(t)=e^{b_{bg}t}.$$
The slope of the cone is defined as:
\begin{equation}\label{eq:B}
B=\frac{d}{dt}\ln\frac{d\sigma}{dt},
\end{equation}
where $\frac{d\sigma}{dt}$ stands for $\frac{d\sigma_{el}}{dt},$  $\frac{d\sigma_{SD}}{dt},$ or $\frac{d\sigma_{DD}}{dt}$,  defined by Eq.~(\ref{elastic}), (\ref{SD}) and  (\ref{DD}), respectively.

The local slope $B_M$ at fixed $M_x^2$ is defined in the same way:
\begin{equation}\label{eq:B_M}
B_{M}=\frac{d}{dt}\ln\frac{d^2\sigma}{dtdM_x^2}
\end{equation}

The integrated cross sections are calculated as:
\begin{equation}\label{eq:dcsdt_SD}
\frac{d\sigma_{SD}}{dt}=\int_{M^2_1}^{M^2_2}\frac{d^2\sigma_{SD}}{dtdM_x^2}dM_x^2
\end{equation}
for the case of SD and:
\begin{equation}\label{eq:dcsdt_DD}
\frac{d\sigma_{DD}}{dt}= \int\int_{f(M^2_{x_1},M^2_{x_2})}\frac{d^3\sigma_{SD}}{ dtdM_{x_1}^2 dM_{x_2}^2 }dM_{x_1}^2dM_{x_2}^2
\end{equation}
for the case of DD.\\

The integrated cross sections are:
\begin{equation}
\sigma_{SD}=\int_{0}^{1}dt\int_{M^2_{th}}^{0.05s}dM_{x}^2 \frac{d^2\sigma_{SD}}{dtdM_{x}^2},
\end{equation}
\begin{equation}
\sigma_{DD}=\int_{0}^{1}dt\int\int_{\Delta\eta>3}dM_{x_1}^2dM_{x_2}^2 \frac{d^3\sigma_{DD}}{dtdM_{x_1}^2dM_{x_2}^2},
\end{equation}
where integration in $M_1^2$ and $M_2^2$ comprises the range $\Delta\eta>3$, $\Delta\eta=\ln\left(\frac{ss_0}{M_1^2M_2^2}\right)$, and
\begin{equation}
\frac{d^2\sigma_{DD}}{dM_{x_1}^2dM_{x_2}^2}= \int_{0}^{1}\frac{d^3\sigma_{DD}}{dtdM_{x_1}^2dM_{x_2}^2}dt.
\end{equation}

\subsection{Summary of the results on SD and DD and (temporary) conclusions} \label{sec:Results}

Single diffraction dissociation (SD) is an important pillar in our fitting procedure. 

At low $t$ (below $0.5$~GeV$^2$), the $t$-dependence of SD cross section are well described by an exponential fit, however beyond this region the cross sections start flattening due to transition effects towards hard physics.

Low-energy data $\sqrt{s}<100$~GeV  require the inclusion of non-leading Reggeons, so they are outside our single pomeron exchange in the $t$ channel.

Double diffraction dissociation (DD) cross sections follow, up to some fine-tuning of the parameters, from our fits to SD and factorization relation.


\begin{figure}[!ht]
	\centering
	\includegraphics[width=1\linewidth,bb=0 0 1400 900,clip]{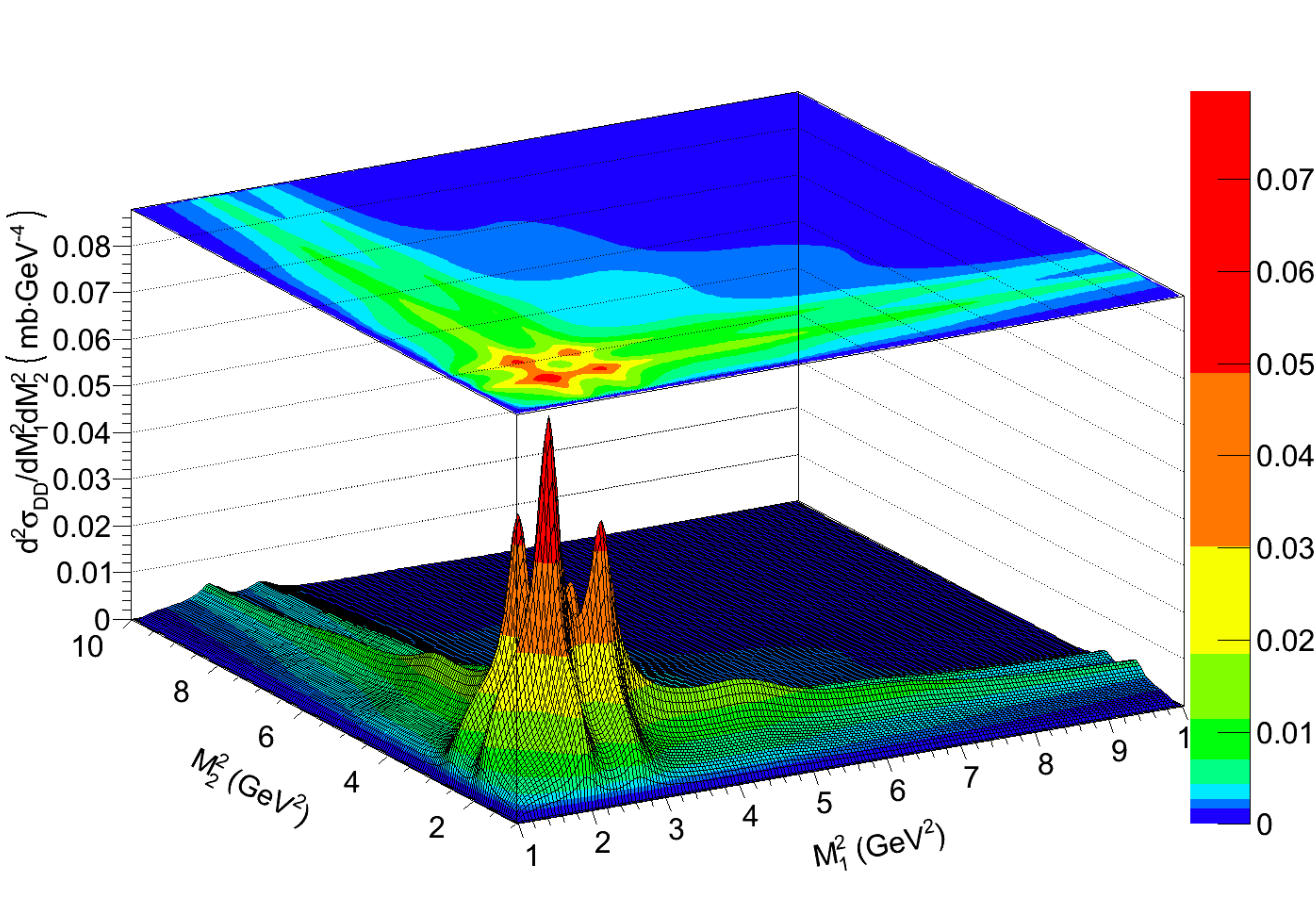}
	\caption{Double differential DD cross section as a function of $M_1^2$ and $M_2^2$ integrated over $t$; see Eq.~(\ref{DD}).}\label{int.d2cs|dcs.DD}
\end{figure}

Below is a brief summary on SD a DD at the LHC:

\begin{itemize}
	\item At the LHC, in the diffraction cone region ($t<1$~GeV$^2$) proton-proton scattering is dominated (over $95 \%$) by pomeron exchange (quantified in Ref.~\cite{JLL}). This enables full use of factorized Regge-pole models. Contributions from non-leading (secondary) trajectories can (and should be) included in the extension of the model to low energies, e.g. below those of the SPS.
	
	\item Unlike to the most of the approaches which use the triple Regge limit for construction of inclusive diffraction, approaches based on the assumed similarity between the pomeron-proton and virtual photon-proton (Fig.\ref{fig:gamma_p}) scattering. The proton structure function (SF) probed by the pomeron is the central object of our studies. This SF, similar to the DIS SF, is exhibits direct-channel (i.e. missing mass, $M$) resonances transformed in resonances in single- double- and central diffraction dissociation. The high-$M$ behavior of the SF (or pomeron-proton cross section) is Regge-behaved and contains two components: one decreasing roughly like $M^{-m},\ \ m\approx 2$ due to the exchange of a secondary reggeon 
	(not to be confused with the pomeron exchange in the $t$ channel!). The latter dominates the large-$M$ part of the cross sections. 
	On the other hand, the large-$M$ region is the border
	of diffraction, $\xi>0.05$ ($\xi=M^2/s$).
	
	\item The present approach is inclusive, ignoring e.g. the angular distribution of the produced particles from decaying resonances.
	All resonances, except Roper, lie on the $N^*$ trajectory. Any complete study of the final states should included also spin degrees of freedom, ignored in the present model.
	
	\item For simplicity, here we use linear Regge trajectories and exponential residue functions, thus limiting the applicability of our model to low and intermediate values of $|t|$. Its extension to larger $|t|$ is straightforward and promising. It may reveal new phenomena, such as the the possible dip-bump structure is SD and DD as well as the transition to hard scattering at large momenta transfers, although it should be remembered that diffraction (coherence) is limited (independently) both in $t$ and $\xi$.
	
\item Last but not least: Fig. \ref{int.d2cs|dcs.DD} summarizes the rich potential of present approach to diffraction dissociation. The model reproduces and predicts both the resonance (here, in the missing masses) and smooth asymptotic behavior of the differential cross sections. Integrated cross section, in any variable can be calculated from the formulae presented in \ref{Sub:Formulae}.   	
\end{itemize}

\section{Central exclusive diffraction (CED)} \label{Sec:CED}

Central production in proton-proton collisions has been studied in the energy range from the ISR at CERN up to the presently highest LHC energies \cite{Albrow1}. Ongoing data analysis include data taken by the COMPASS collaboration at the SPS \cite{COMPASS}, the CDF collaboration at the TEVATRON \cite{CDF}, the STAR collaboration at RHIC \cite{STAR},  and the ALICE, ATLAS, CMS and LHCb collaborations at the LHC \cite{ALICE,LHCb,ATLAS,CMS}. The analysis of events recorded by the large and complex detector systems requires the simulation of such events to study the experimental acceptance and efficiency.  Much larger data samples are expected in the next few years both at RHIC and at the LHC allowing the study of differential distributions with much improved statistics. The purpose of the ongoing work presented here is the formulation of a Regge pole model for simulating such differential distributions.


The study of central production in hadron-hadron collisions is interesting for a variety of 
reasons. Such events are characterized by a hadronic system formed at mid-rapidity,
and by the two very forward scattered protons, or remnants thereof. The rapidity gap
between the mid-rapidity system and the forward scattered proton is a distinctive 
feature of such events. Central production events can hence be tagged by measuring 
the forward scattered protons and/or by identifying the existence of rapidity gaps. 
Central production is dominated at high energies by pomeron-pomeron exchange. 
The hadronization of this gluon-dominated environment is expected to produce with 
increased probability gluon-rich states, glueballs and hybrids.  Of particular interest 
are states of exotic nature, such as tetra-quark ($q\bar{q}$ + $\bar{q}q$) 
configurations, or gluonic hybrids ($q\bar{q} + gluon$).

The production of central events can take place with the protons remaining
in the ground state, or with diffractive excitation of one or both of the 
outgoing protons.  	
\begin{figure}[t]
	\begin{center}
		\includegraphics[width=.316\textwidth]{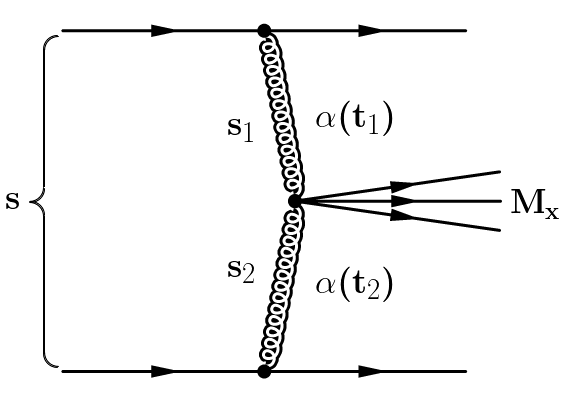}
		\includegraphics[width=.286\textwidth]{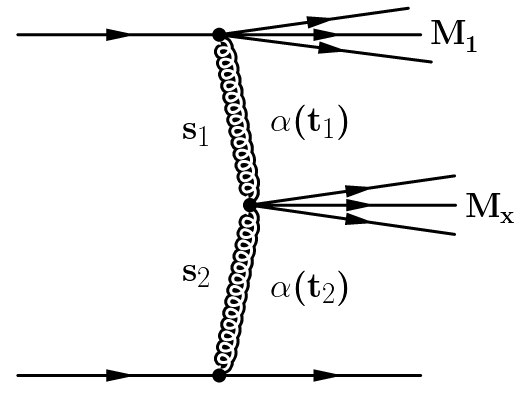}
		\includegraphics[width=.280\textwidth]{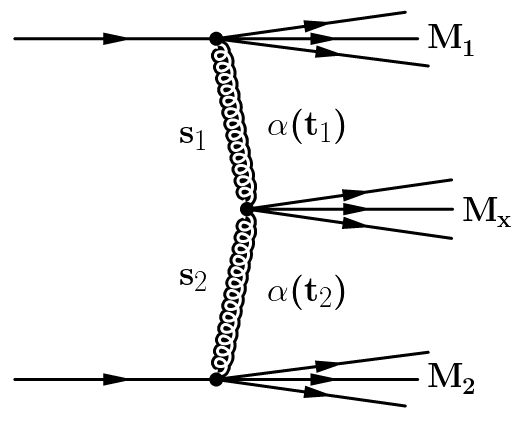}
		\caption{Central production event topologies.}
		\label{fig1}
	\end{center}
\end{figure}

The topologies of central production are shown in Fig. \ref{fig1}.
This figure shows central production with the two protons in the
ground state on the left, and with one and both protons getting diffractively
excited in the middle and on the right, respectively. These reactions 
take place  by the exchange of Regge trajectories $\alpha(t_1)$ and $\alpha(t_2)$ 
in the central region where a system of mass M$_{x}$ is produced. The total 
energy $s$ of the reaction is shared by the subenergies $s_1$ and $s_2$ 
associated to the trajectories $\alpha(t_1)$ and $\alpha(t_2)$, respectively. 
The LHC energies of $\sqrt{s}$ = 8 and 13 TeV are large enough to provide 
pomeron dominance. Reggeon exchanges can hence be neglected which was not 
the case at the energies of previous accelerators.

The main interest in the study presented here is the central part of the diagrams
shown in Fig. \ref{fig1}, i.e. pomeron-pomeron ($PP$) scattering producing 
mesonic states of mass M$_{x}$. We isolate the pomeron-pomeron-meson vertex 
and calculate the $PP$ total cross
section as a function of the centrally produced system of mass M$_{x}$.
The emphasis here is the behavior in the low mass resonance region 
where perturbative QCD approaches are not applicable. 
Instead, similar to \cite{Jenk1}, we use the pole decomposition of a dual 
amplitude with relevant direct-channel trajectories $\alpha(M^2)$ for 
fixed values of pomeron virtualities, $t_1=t_2=const.$ 
Due to Regge factorization, the calculated pomeron-pomeron cross section 
is part of the measurable proton-proton cross section \cite{Jenk2}.

\subsection{Dual resonance model of pomeron-pomeron scattering}
Most of the existing studies on diffraction dissociation, single, double and 
central, are done within the framework of the triple reggeon approach. This 
formalism is useful beyond the resonance region, but is not valid for the low 
mass region which is dominated by resonances. A formalism to account for 
production of resonances was formulated in Ref.~\cite{Fiore1}. This formalism is based 
on the idea of duality with a limited number of resonances represented by 
nonlinear Regge trajectories.

\begin{figure}[htb]
	\includegraphics[width=.19\textwidth]{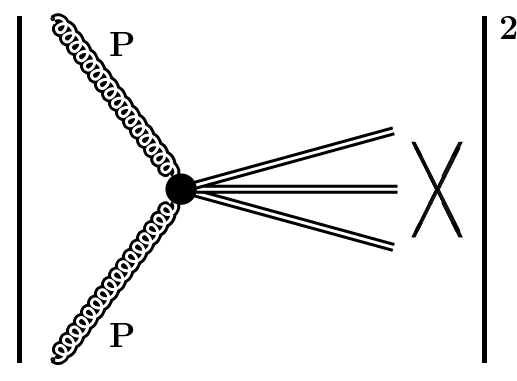}
	\includegraphics[width=.038\textwidth]{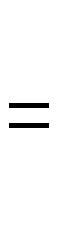}
	\hspace{-0.2cm}
	\includegraphics[width=.154\textwidth]{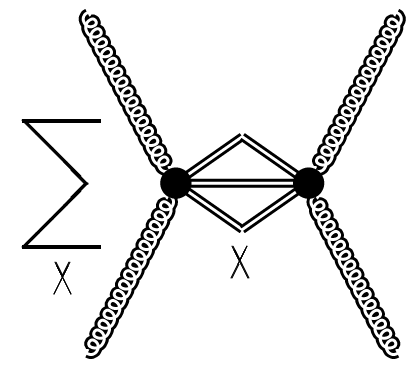}
	\hspace{0.2cm}
	\includegraphics[width=.037\textwidth]{pp0_0}
	\begin{overpic}[width=.12\textwidth]{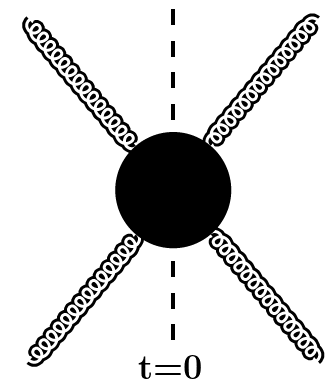}
		\put(-30.,14.){\footnotesize{\bf Unitarity}}
	\end{overpic}
	\hspace{-0.2cm}
	\includegraphics[width=.037\textwidth]{pp0_0}
	\includegraphics[width=.132\textwidth]{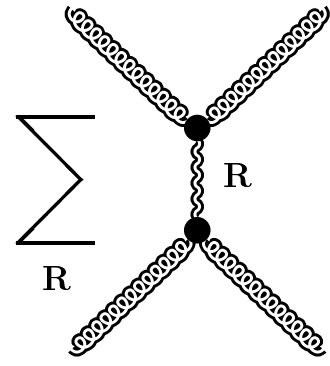}
	\hspace{0.2cm}
	\includegraphics[width=.038\textwidth]{pp0_0}
	\hspace{0.2cm}
	\begin{overpic}[width=.14\textwidth]{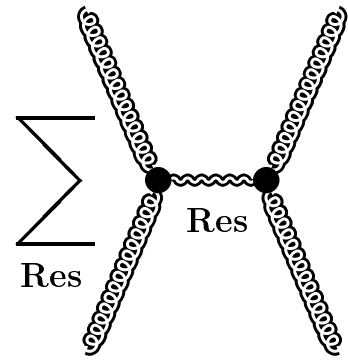}
		\put(-44.,15.){\footnotesize{\bf Veneziano }}
		\put(-32.,5.){\footnotesize{\bf duality }}
	\end{overpic}
	\caption{Connection, through unitarity (generalized optical theorem)
		and Veneziano-duality, between the pomeron-pomeron cross section 
		and the sum of direct-channel resonances: by the optical theorem (unitarity) the squared modules of a production amplitude summed over all intermediate states is the imaginary part of the forward elastic amplitude, that, by Veneziano-duality can be written equally as an infinite sum of direct-channel resonances or Regge exchanges.}
	\label{fig2}
\end{figure}

The motivation of this approach consists of using dual amplitudes with Mandelstam
analyticity (DAMA), and is shown in Fig. \ref{fig2}. 
For $s\rightarrow\infty$ and fixed $t$ it is Regge-behaved.
Contrary to the Veneziano model, DAMA not only allows
for, but rather requires the use of nonlinear complex trajectories
which provide the resonance widths via the imaginary part of the
trajectory. A finite number of resonances is produced for limited real part of 
the trajectory. 

For our study of central production, the direct-channel pole decomposition 
of the dual amplitude $A(M_{X}^{2},t)$ is relevant. This amplitude receives 
contributions from different trajectories $\alpha_{i}(M_X^2)$, with $\alpha_{i}(M_X^2)$ a 
nonlinear, complex Regge trajectory in the pomeron-pomeron system,
\begin{eqnarray}
A(M_X^2,t)=a\sum_{i=f,P}\sum_{J}\frac{[f_{i}(t)]^{J+2}}{J-\alpha_i(M_X^2)}.
\label{eq2}
\end{eqnarray}

The pole decomposition of the dual amplitude  $A(M_{X}^{2},t)$ is shown 
in Eq. (\ref{eq2}), with $t$ the squared momentum transfer in the 
$PP\rightarrow PP$ reaction. The index $i$ sums over the trajectories which 
contribute to the amplitude. Within each trajectory, the second sum extends 
over the bound states of spin $J$. The prefactor $a$ in Eq. (\ref{eq2}) is 
of numerical value a = 1 GeV$^{-2}$ = 0.389 mb.

The imaginary part of the amplitude $A(M_X^2,t)$ given in Eq. (\ref{eq2})
is defined by

\begin{equation} \label{ImA}
Im\, A(M_{X}^2,t)=a\sum_{i=f,P}\sum_{J}\frac{[f_{i}(t)]^{J+2} 
	Im\,\alpha_{i}(M_{X}^2)}{(J-Re\,\alpha_{i}(M_{X}^2))^2+ 
	(Im\,\alpha_{i}(M_{X}^2))^2}.
\end{equation}

For the $PP$ total cross section we use the norm

\begin{eqnarray}
\sigma_{t}^{PP} (M_{X}^2)= {Im\; A}(M_{X}^2, t=0).
\label{eq:ppcross}
\end{eqnarray}

The pomeron-pomeron channel, $PP\rightarrow M_X^2$, couples to the pomeron
and $f$ channels due to quantum number conservation.
For calculating the $PP$ cross section, we therefore take into account
the trajectories associated to the f$_0$(980) and to the f$_2$(1270) resonance, 
and the pomeron trajectory.

\subsection{Non-linear, complex meson Regge trajectories}
Analytic models of Regge trajectories relate the imaginary 
part of the trajectory with their nearly linear real part by means of dispersion relations, as suggested in  Ref.~\cite{Fiore2}, 

\begin{eqnarray}
Re\:\alpha(s) = \alpha(0) + \frac{s}{\pi} 
PV \int_0^{\infty} ds^{'} \frac{Im\:\alpha(s^{'})}{s^{'}(s^{'}-s)}. 
\label{eq:disp}
\end{eqnarray}

In Eq. \ref{eq:disp}, the dispersion relation connecting the real and imaginary 
part is shown. The imaginary part of the trajectory is related to the decay width by

\vspace{-0.6cm}
\begin{eqnarray}
\Gamma(M_{R}) = \frac{Im\: \alpha(M_{R}^{2})}{\alpha^{'}\:M_{R}}.
\label{eq:width}
\end{eqnarray}

Apart from the pomeron trajectory, the direct-channel $f$ trajectory is essential 
in the PP system.  Guided by conservation of quantum numbers, we include two 
$f$ trajectories, labeled $f_1$ and $f_2$, with mesons lying on these trajectories 
as specified in Table \ref{table1}.

\begin{table}[h]
	\begin{center}
		\begin{tabular}{| c | c c | c || c | c | c ||}
			\hline
			& I$^{G} $& J$^{PC}$ & traj. & M (GeV) & M$^{2}$ (GeV$^{2}$) &  $\Gamma$ (GeV) \\ 
			\cline{1-7}
			f$_{0}$ (500)& 0$^{+}$ &0$^{++}$ & $f_{0}$ &$\approx$ 0.475$\pm$0.125  & $\approx$ 0.23 &0.550$\pm 0.150$\\ 			
			f$_{0}$(980) & 0$^{+}$ &0$^{++}$ & $f_{1}$ &0.990$\pm$0.020 &0.980$\pm$0.040 &0.070$\pm$ 0.030\\ 
			f$_{1}$(1420) & 0$^{+}$ &1$^{++}$ & $f_{1}$ &1.426$\pm$0.001 &2.035$\pm$0.003 &0.055$\pm$ 0.003\\ 
			f$_{2}$(1810) & 0$^{+}$ &2$^{++}$ & $f_{1}$ &1.815$\pm$0.012 &3.294$\pm$0.044 &0.197$\pm$ 0.022\\ 
			f$_{4}$(2300) & 0$^{+}$ &4$^{++}$ & $f_{1}$ &2.320$\pm$0.060 &5.382$\pm$0.278 &0.250$\pm$ 0.080\\ 
			f$_{2}$(1270) & 0$^{+}$ &2$^{++}$ & $f_{2}$ &1.275$\pm$0.001 &1.6256$\pm$0.003 &0.185$\pm$ 0.003\\ 
			f$_{4}$(2050) & 0$^{+}$ &4$^{++}$ & $f_{2}$ &2.018$\pm$0.011 &4.0723$\pm$0.044 &0.237$\pm$ 0.018\\ 
			f$_{6}$(2510) & 0$^{+}$ &6$^{++}$ & $f_{2}$ &2.469$\pm$0.029 &6.096$\pm$0.143 &0.283$\pm$ 0.040\\ 
			\hline
		\end{tabular}   
		\caption{Parameters of resonances belonging to the $f_1$ and $f_2$ trajectories.} 
		\label{table1}
	\end{center}
\end{table}

The experimental data on central exclusive pion-pair production measured at the 
energies of the ISR, RHIC, TEVATRON and the LHC collider all show a broad 
continuum for pair masses m$_{\pi^+\pi^-} <$ 1 GeV/c$^{2}$. 
The population of this  mass region is attributed to the $f_{0}$(500).
This resonance $f_{0}$(500) is of prime importance for the understanding 
of the attractive part of the nucleon-nucleon interaction, as well as for
the mechanism of spontaneous breaking of chiral symmetry. Therefore, in spite of the above uncertainties, we have included it in the above tabel and in our analyses. 

The real and imaginary part of the $f_{1}$ and $f_{2}$ trajectories can 
be derived from the parameters of the f-resonances listed in Table \ref{table1},
and has explicitly been derived in Ref.~\cite{Fiore3}.

While ordinary meson trajectories may be fitted both in the resonance or 
scattering region, corresponding to positive or negative values of the 
argument, the parameters of the pomeron trajectory can only be determined in 
the scattering region $M^2<0$. A fit to high-energy $pp$ and 
$p\bar{p}$ of the nonlinear pomeron trajectory  is discussed in Ref.~\cite{Jenk2}
\begin{eqnarray}
	\alpha_P(M^2) = 1. + \varepsilon + \alpha^{'} M^2 - c\sqrt{s_{0}-M^2},
	\label{eq:pom1}
\end{eqnarray}
with $\varepsilon$\,=\,0.08, $\alpha^{'}$\,=\,0.25 GeV$^{-2}$,
the two-pion threshold $s_0$\,=\,4m$_{\pi}^{2}$, and $c\approx \alpha '/10\approx 0.025$ GeV$^{-1}$.

The above values of the parameters are those advocated by Donnachie and Landshoff (DL) \cite{DL} for the pomeron trajectory, based on efficient fits with to various data. The only difference between our Eq. \ref{eq:pom1} and DL's pomeron is that we allow for a small imaginary part to produce resonances in the direct channel. Its weight is determined empirically.  

For consistency with the mesonic trajectories, the
linear term in Eq. (\ref{eq:pom1}) may be replaced by a heavy threshold mimicking
linear behavior in the mass region of interest (M $<$ 5 GeV),
\begin{eqnarray}
	\alpha_{P}(M^2)=\alpha_0+\alpha_1(2m_{\pi}-\sqrt{4m_{\pi}^2-M^2})+\alpha_2(\sqrt
	{M^2_H}-\sqrt{M^2_H-M^2}),
	\label{eq:pom2}
\end{eqnarray}
with $M_H$ an effective heavy threshold $M=4.5$ GeV. This value may be determined only empirically. The
coefficients $\alpha_{0}, \alpha_{1}$ and $\alpha_{2}$ 
were chosen in Ref. \cite{Fiore3} such that the pomeron trajectory of Eq. (\ref{eq:pom2}) 
has a low energy behaviour as defined by Eq. (\ref{eq:pom1}).

Later on \cite{Fiore4} it was realized that the above trajectories result in (non-physical) decreasing-widths (narrowing) glueball resonances. A remedy of this problem was found in Ref. \cite{Fiore4} by modifying the 
the above pomeron trajectory as follows: 
\begin{eqnarray}
	\alpha_P(M^2) = \frac{1. + \tilde\varepsilon + \tilde\alpha^{'} M^2}{\tilde c\sqrt{s_{0}-M^2}},
	\label{eq:pom3}
\end{eqnarray} 
producing glueballs with increasing widths. The values of parameters $\tilde\varepsilon$, $\tilde\alpha^{'}$ and $\tilde c$ can be determined from fits, but they are close to the values of parameters of Eq.~\ref{eq:pom1} {\it i.e.} $\varepsilon$, $\alpha^{'}$ and $c$.  

\subsection{Pomeron-pomeron total cross section}

The pomeron-pomeron cross section is calculated from the 
imaginary part of the amplitude by use of the optical theorem
\begin{eqnarray}
\label{eq:imampl}
	\sigma_{t}^{PP} (M^2) \;\; = \;\; {Im\; A}(M^2, t=0) \;\; = \\ \nonumber
	a\sum_{i=f,P}\sum_{J}\frac{[f_{i}(0)]^{J+2}\; Im\;\alpha_{i}(M^{2})}
	{(J-Re\;\alpha_{i}(M^{2}))^{2}+(Im\;\alpha_{i}(M^{2}))^{2}}.
\end{eqnarray}
In Eq. (\ref{eq:imampl}), the index $i$ sums over the trajectories which
contribute to the cross section, in our case the $f_{1}$, $f_{2}$
and the pomeron trajectory discussed above.
Within each trajectory, the summation extends over the bound states
of spin $J$ as expressed by the second summation sign.
The value $f_{i}(0) =f_{i}(t)\big |_{t=0}$ is not known a priori.
The analysis of relative strengths of the states of trajectory $i$ will, 
however, allow to extract a numerical value for $f_{i}$(0) from the 
experimental data.

\begin{figure}[H]
	\begin{minipage}[t]{.97\textwidth}
		\begin{center}
			\begin{overpic}[width=.90\textwidth]{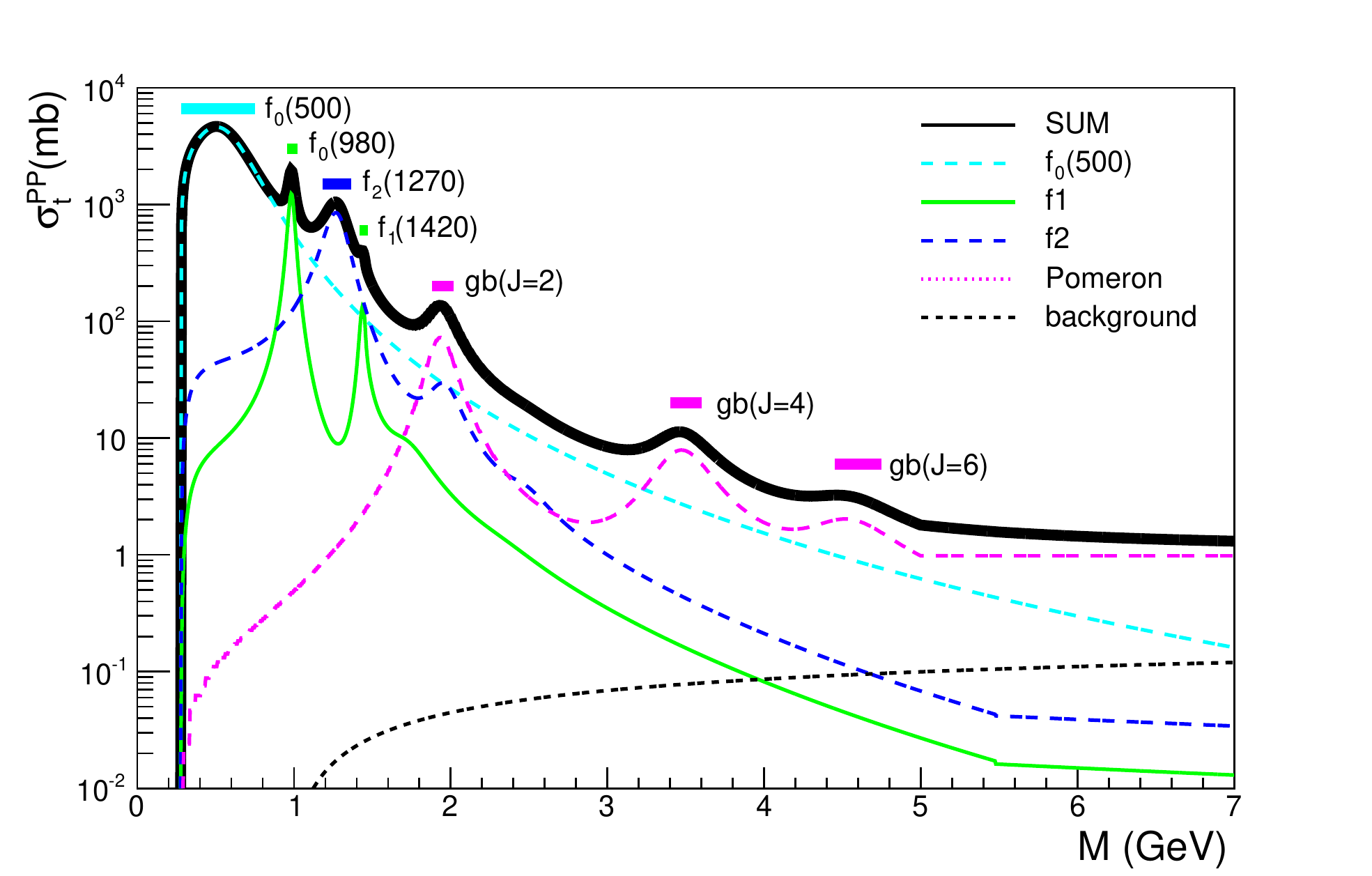}
			\end{overpic}
		\end{center}
	\end{minipage}
	\caption{Contributions of the f$_{0}$(500) resonance, the $f_{1}$, $f_{2}$ and the pomeron trajectory, Eq.(\ref{eq:pom3}) as well as the background to PP total cross section.}
	\label{fig3}
\end{figure}

The  pomeron-pomeron total cross section is calculated by summing over the 
contributions discussed above, and is shown in Fig. \ref{fig3}
by the solid black line. The prominent structures seen in the total cross
section are labeled by the resonances generating the peaks.
The model presented here does not specify the 
relative strength of the different contributions shown in 
Fig. \ref{fig3}. A Partial Wave Analysis of experimental data on 
central production events will be able to extract the quantum numbers 
of these resonances, and will hence allow to associate each resonance
to its trajectory. The relative strengths of the  contributing
trajectories need to be taken from the experimental data. 

\section{Diffractive deep-inelastic scattering; how many pomerons (“soft” and “hard”?)}	
\begin{figure}[ht]
 \vspace{-0.2cm}
 \begin{center}\includegraphics[trim = 0mm 8mm 10mm 14mm,clip,scale=0.6]
{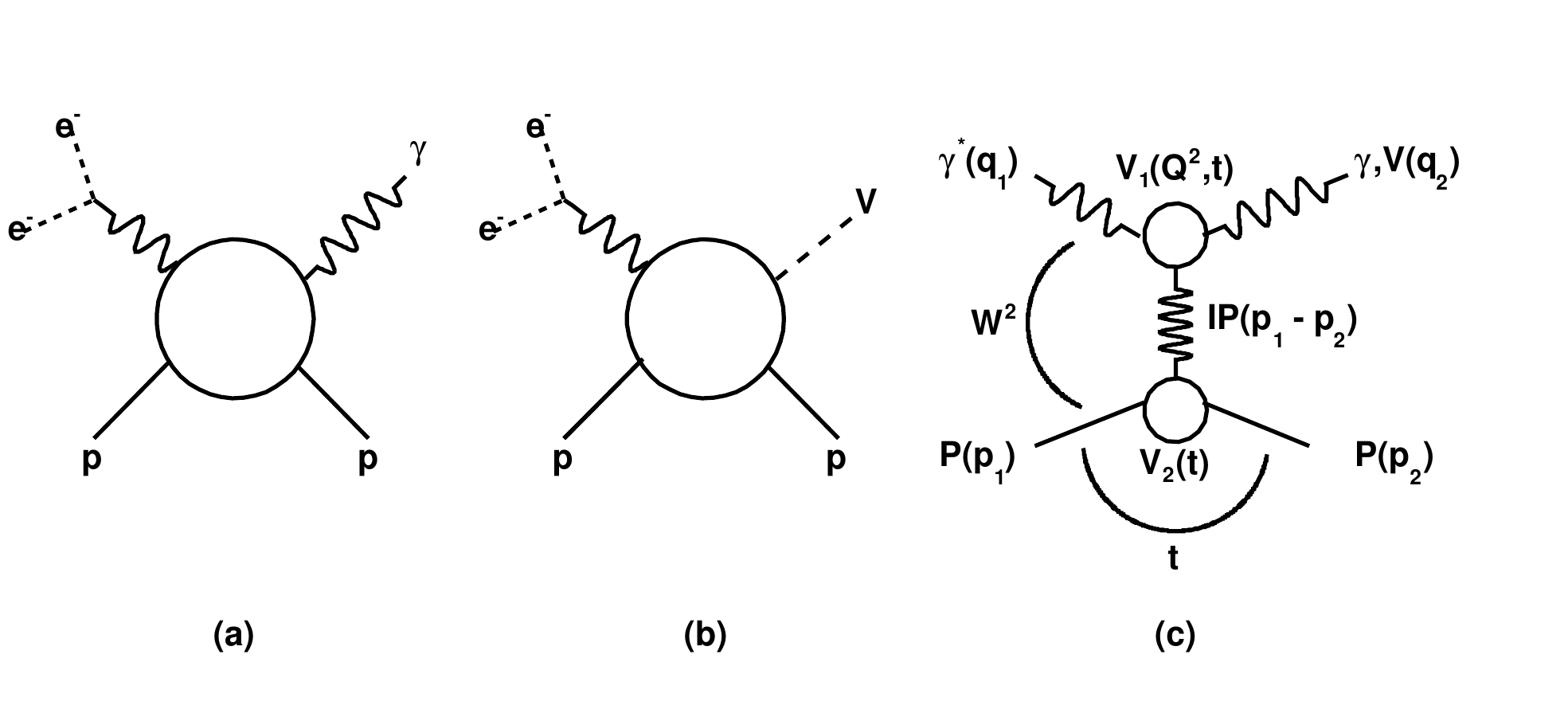}\end{center}
  \caption{ \label{fig:diagram} Diagrams of DVCS (a) and VMP (b); (c) DVCS 
(VMP) amplitude in a Regge-factorized form.}
\end{figure}

According to perturbative QCD calculations, the pomeron corresponds to the exchange of an infinite gluon ladder, producing an infinite set of moving Regge poles, the so-called BFKL pomeron \cite{BFKL1, BFKL2}, whose highest intercept $\alpha(0)$ is near $1.3\divisionsymbol1.4$. Phenomenologically,
``soft" (low virtuality $Q^2$) and ``hard'' (high virtuality $Q^2$) diffractive ({\it i.e.} small squared momentum transfer $t$)
processes with pomeron exchange are described by the exchange of two different objects in the $t$ channel, a ``soft'' and a ``hard'' pomeron (or their QCD gluon images), (see, for instance, Refs.~\cite{BP, DDLN}). This implies the existence of two (or even more) scattering amplitudes, differing by the values of the parameters of the pomeron trajectory, their intercept $\alpha(0)$ and slope $\alpha'(t=0)$, typically $(1.08\divisionsymbol 1.09)$ and $(0.25),$ respectively, for the ``soft'' pomeron, and
$(1.3\divisionsymbol 1.4),$ and $(0.1$ or even less$)$ for the ``hard" one, each attached to vertices of the relevant reaction or kinematical region.
A simple ``unification" is to make theses parameters $Q^2$-dependent. This breaks Regge
factorization, by which Regge trajectories should not depend on $Q^2$, see Fig. \ref{fig:diagram}.

In the present approach, initiated in Refs. \cite{Acta,Capua}, we postulate that

1. Regge factorization holds, {\it i.e.} the dependence on the virtuality of the external particle (virtual photon) enters only the relevant vertex, not the propagator;

2. there is only one pomeron in Nature and it is the same in all reactions. It may be 
complicated, e.g. having many, at least two, components.

The first postulate was applied, for example, in Refs.~\cite{Francesco, Capua, FazioPhysRev} to study the deeply virtual Compton scattering (DVCS) and the vector meson production (VMP). In Fig. \ref{fig:diagram}, where diagrams (a) and (b) represent the DVCS and the VMP, respectively, the $Q^2$ dependence enters only the upper vertex of the diagram (c).
The particular form of this dependence and its interplay with $t$ is not unique.

In Refs.~\cite{Capua,FazioPhysRev} the interplay between $t$ and $Q^2$ was realized by the introduction of a new variable, $z=t-\widetilde {Q^2}$, where $\widetilde {Q^2}$ is the familiar variable $\widetilde {Q^2} = Q^2 + M_V^2$, $M_V$ being the vector meson mass. The model (called also ``scaling model") is simple and fits the data on DVCS (in this case $M_V = 0$) and VMP, although the physical meaning of this new variable is not clear.

In a series of subsequent papers (see Refs.~\cite{Confer1,Confer2,Confer3,Confer4,Acta,FazioPhysRev}), $\widetilde {Q^2}$ was incorporated in a ``geometrical'' way reflecting the observed trend in the decrease of the forward slope as a function of  $\widetilde {Q^2}$. This geometrical approach, combined with the Regge-pole model was named ``Reggeometry''. A Reggeometric amplitude dominated by a single pomeron  shows \cite{Acta} reasonable agreement with the HERA data on VMP and DVCS, when fitted separately to each reaction, i.e. with a large number of parameters adjusted to each particular reaction.

As a further step, to reproduce the observed trend of hardening\footnote{In what follows we use the variable $\widetilde {Q^2}$ as a measure of
	``hardness''.}, as $\widetilde{Q^2}$ increases, and following Refs.~\cite{L,DL}, a two-term amplitude, characterized by a two-component - ``soft'' + ``hard'' - pomeron,  was suggested \cite{Acta}. We stress that the pomeron is unique, but we construct it as a sum of two terms. Then, the amplitude is defined as
\begin{equation}
A(\widetilde {Q^2},s,t)=A_s (\widetilde {Q^2},s,t)+A_h(\widetilde {Q^2},s,t)
\label{two-term-amp}
\end{equation}
($s=W^2$ is the square of the c.m.s. energy),
such that the relative weight of the two terms changes with $\widetilde {Q^2}$ in a right way, i.e. such that the ratio $r=A_h/A_s$ increases as the reaction becomes ``harder'' and v.v. It is interesting to note that this trend is not guaranteed ``automatically'': both the ``scaling'' model  \cite{Capua, FazioPhysRev} or the Reggeometric one \cite{Acta} show the opposite tendency, that may not be merely an accident and whose reason should be better understood. This ``wrong'' trend can and should be corrected, and in fact it was corrected \cite{L, DL} by means of additional $\widetilde {Q^2}$-dependent factors $H_i(\widetilde {Q^2}),\ i=s,h$ modifying the $\widetilde {Q^2}$
dependence of the amplitude,
in a such way as to provide increasing of the weight of the hard component with increasing $\widetilde {Q^2}$. To avoid conflict with unitarity, the rise with $\widetilde {Q^2}$ of the hard component is finite (or moderate), and it terminates at some saturation scale,
whose value is determined phenomenologically. In other words, the ``hard" component, invisible at small $\widetilde {Q^2}$, gradually takes over as $\widetilde {Q^2}$ increases. An explicit example of these functions will be given below. More can be found in Ref.~\cite{FFJS}.

Hadron-hadron elastic scattering is different from exclusive VMP and DVCS not only because the photon is different from a hadron (although they are related by vector meson dominance), but even more so by the transition between space- and time-like regions: while the virtual photon's ``mass square" $q^2$ is negative, that of the hadron is positive and that makes this attempt interesting!

\subsection{Reggeometric pomeron}
\label{sec:Single}
We start by reminding the properties and some representative results of the single-term Reggeometric \footnote{Wordplay (pun): Regge+geometry=Reggeometry.} model \cite{Acta}.

The invariant scattering amplitude is defined as
\begin{equation}\label{Amplitude1}
A(Q^2,s,t)=\widetilde He^{-\frac{i\pi\alpha(t)}{2}}\left(\frac{s}{s_0}\right)^{\alpha(t)} e^{2\left(\frac{a}{\widetilde{Q^2}}+\frac{b}{2m_N^2}\right)t},
\end{equation}
where
\begin{equation}
\alpha(t)=\alpha_0+\alpha't
\end{equation}
is the linear pomeron trajectory, $s_0$ is a scale for the square of the total energy $s$, $a$ and $b$ are two parameters to be determined with the fitting procedure and $m_N$ is the nucleon mass. The coefficient $\widetilde H$ is a function providing the right behavior of elastic cross sections in 
$\widetilde{Q^2}$:
\begin{equation}
\widetilde H\equiv \widetilde H(\widetilde{Q^2})=\frac{\widetilde{A_0}}{\left(1+\frac{\widetilde{Q^2}}{{Q_0^2}}\right)^{n_s}},
\label{eq:norm}
\end{equation}
where $\widetilde{A_0}$ is a normalization factor, $Q_0^2$ is a scale for the virtuality and $n_s$ is a real positive number.

In this model we use an effective pomeron, which can be ``soft'' or ``hard'', depending on the reaction and/or kinematical region defining its ``hardness''. In other words, the  values of the parameters $\alpha_0$ and $\alpha'$ must be fitted to each set of the data. Apart from $\alpha_0$ and $\alpha'$, the model contains five more sets of free parameters, different for each reaction.
The exponent in the exponential factor in Eq.~(\ref{Amplitude1}) reflects the geometrical nature of the model: $a/\widetilde {Q^2}$  and $b/2m_N^2$ correspond to the ``sizes" of upper and lower vertices in Fig. \ref{fig:diagram}c.

By using the {Eq.~\ref{eq:norm}) wkth the norm
	\begin{equation}\label{eq:dcsdt_from_Amplitude}
	\frac{d\sigma_{el}}{dt}=\frac{\pi}{s^2}|A(Q^2,s,t)|^2,
	\end{equation}
	the differential and integrated elastic cross sections become,
	\begin{equation}\label{eq:dcsdt_1}
	\frac{d\sigma_{el}}{dt}=\frac{A_0^2}{\left(1+\frac{\widetilde{Q^2}}{{Q_0^2}}\right)^{2n}}\left(\frac{s}{s_0}\right)^{2(\alpha(t)-1)}e^{4\left(\frac{a}{\widetilde{Q^2}}+\frac{b}{2m_N^2}\right)t}
	\end{equation}
	and
	\begin{equation}\label{eq:cs_1}
	\sigma_{el}=\frac{A_0^2} {\left(1+\frac{\widetilde{Q^2}}{{Q_0^2}}\right)^{2n}}
	\frac{\left(\frac{s}{s_0}\right)^{2(\alpha_0-1)}}
	{4\left(\frac{a}{\widetilde{Q^2}}+\frac{b}{2m_N^2}\right)+2\alpha'\ln\left(\frac{s}{s_0}\right)},
	\end{equation}
	where
	$$A_0=-\frac{\sqrt{\pi}}{s_0}\widetilde{A_0}.$$
	
	Eqs.~(\ref{eq:dcsdt_1}) and (\ref{eq:cs_1}) (for simplicity we set $s_0 = 1$ GeV$^2$) were fitted \cite{FFJS} to the HERA data obtained the by ZEUS and H1 Collaborations on exclusive diffractive VMP.
	
	A shortcoming of the single-term Reggeometric pomeron model, Eq.~\eqref{Amplitude1}, is that the fitted parameters in this model acquire particular values for each reaction.
	
		\subsection{Two-component Reggeometric pomeron}
	\label{sec:Two-components model}
In this section we try to approach a complicated and controversial subject, namely the existence of two (or more) different pomerons: one "soft" responsible for on-mass-shall hadronic reactions, and the other one(s) applicable to off-mass-shall deep inelastic scattering. There are similarities between the two (Regge behavior) and differences. The main difference is that the Regge pole model, being part of the analytic $S$ matrix theory, strictly speaking, is applicable to asymptotically free states on the mass shall only. Another difference is that the "hard" (or "Lipatov") pomeron is assumed to follow from the local quantum field theory (QCD) with confined quarks and gluons. We do not know how can these two extremes be reconciled. Below we try to combine these two approaches by using a specific model, a "handle" combining three independent variables: $s,\ t$ and $Q^2.$

	We introduce a universal, ``soft'' and ``hard'', pomeron model.
	Using the Reggeometric ansatz of Eq.~\eqref{Amplitude1}, we write the universal, unique amplitude as a sum of two parts,
	corresponding to the "soft" and "hard" components of pomeron:
	\begin{eqnarray}\label{eq:Amplitude_hs_prime}
	A(Q^2,s,t)=
	\widetilde{H_s}\,e^{-i\frac{\pi}{2}\alpha_s(t)}\left(\frac{s}{s_{0s}}\right)^{\alpha_s(t)} e^{2\left(\frac{a_{s}}{\widetilde{Q^2}}+\frac{b_{s}}{2m_N^2}\right)t}\\ \nonumber
	+\widetilde{H_h}\,e^{-i\frac{\pi}{2}\alpha_h(t)}\left(\frac{s}{s_{0h}}\right)^{\alpha_h(t)} e^{2\left(\frac{a_{h}}{\widetilde{Q^2}}+\frac{b_{h}}{2m_N^2}\right)t}.
	\end{eqnarray}
	Here  $s_{0s}$ and $s_{0h}$
	are squared energy scales, and $a_i$ and $b_i$, with $i = s,h$, are parameters to be determined with the fitting procedure. The two coefficients $\widetilde{H_s}$ and $\widetilde{H_h}$ are functions similar to those defined in Ref.~\cite{DL}:
	\begin{equation}\label{eq:HsHh_tilde}
	\widetilde{H_s}\equiv\widetilde{H_s}(\widetilde{Q^2})=\frac{\widetilde{A_s}}{{\Bigl(1+\frac{\widetilde{Q^2}}{{Q_s^2}}\Bigr)}^{n_s}}, ~~~~~~\quad
	\widetilde{H_h}\equiv\widetilde{H_h}(\widetilde{Q^2})=\frac{\widetilde{A_h} \Bigl(\frac{\widetilde{Q^2}}{Q_h^2}\Bigr)}{{\Bigl(1+\frac{\widetilde{Q^2}}{{Q_h^2}}\Bigr)}^{n_h+1}},
	\end{equation}
	where $\widetilde{A_s}$ and $\widetilde{A_h}$ are normalization factors, $Q_s^2$ and $Q_h^2$ are scales for the virtuality, $n_s$ and $n_h$ are real positive numbers.
	Each component of Eq.~\eqref{eq:Amplitude_hs_prime} has its own, ``soft" or ``hard", Regge (here pomeron) trajectory:
	$$\alpha_s(t)=\alpha_{0s}+\alpha_s't, ~~~~~~~~~~~\quad  \alpha_h(t)=\alpha_{0h}+\alpha_h't.$$
	As an input we use the parameters suggested by Donnachie and Landshoff \cite{DL_tr}, so that
	$$\alpha_s(t) = 1.08 + 0.25t,~~~~~~~~~~\quad \alpha_h(t) = 1.40 + 0.1t.$$
	
	The ``pomeron"  amplitude (\ref{eq:Amplitude_hs_prime}) is unique, valid for all diffractive
	reactions, its ``softness'' or ``hardness'' depending on the relative $\widetilde{Q^2}$-dependent weight of the two components, governed by the relevant factors $\widetilde H_s(
	\widetilde Q^2)$ and $\widetilde {H_h}(\widetilde {Q^2)}$.
	
	Fitting Eq.~\eqref{eq:Amplitude_hs_prime} to the data, we have found that the parameters assume rather large errors and, in particular, the parameters $a_{s,h}$ are close to $0$. Thus, in order  to reduce the number of free parameters, we simplified the model, by fixing $a_{s,h}=0$ and substituting the exponent $2\left(\frac{a_{s,h}}{\widetilde{Q^2}}+\frac{b_{s,h}}{2m_N^2}\right)$ with $b_{s,h}$ in Eq.~\eqref{eq:Amplitude_hs_prime}. The proper variation with $\widetilde {Q^2}$ will be provided by the factors $\widetilde{H_s}(\widetilde{Q^2})$ and $\widetilde{H_h}(\widetilde{Q^2})$.
	
	Consequently, the scattering amplitude assumes the form
	\begin{equation}\label{eq:Amplitude_hs}
	A(s,t,Q^2,M_V^2)=
	\widetilde{H_s}\,e^{-i\frac{\pi}{2}\alpha_s(t)}\left(\frac{s}{s_{0s}}\right)^{\alpha_s(t)} e^{b_st}
	+\widetilde{H_h}\,e^{-i\frac{\pi}{2}\alpha_h(t)}\left(\frac{s}{s_{0h}}\right)^{\alpha_h(t)} e^{b_ht}.
	\end{equation}
	
	The ``Reggeometric'' combination $2\left(\frac{a_{s,h}}{\widetilde{Q^2}}+\frac{b_{s,h}}{2m_N^2}\right)$ was important for the description of the slope $B(Q^2)$ within the single-term pomeron model (see previous Section), but in the case of two terms the $Q^2$-dependence of $B$ can be reproduced without this extra combination, since each term in the amplitude \eqref{eq:Amplitude_hs} has its own $Q^2$-dependent factor $\widetilde{H_{\ }}_{\!\!s,h}(Q^2)$.
	
	By using the amplitude (\ref{eq:Amplitude_hs}) and Eq.~\eqref{eq:dcsdt_from_Amplitude}, we calculate the differential and elastic cross sections, by setting for simplicity $s_{0s} = s_{0h} = s_0$ to obtain
	\begin{equation}\label{eq:dcsdt(h+s)}
	\frac{d\sigma_{el}}{{dt}}=H_s^2e^{2\{L(\alpha_s(t)-1)+{b_s}t\}}+H_h^2e^{2\{L(\alpha_h(t)-1)+{b_h}t\}}
	\end{equation}
	$$+2H_sH_he^{\{L(\alpha_s(t)-1)+L(\alpha_h(t)-1)+({b_s}+{b_h})t\}}\cos\Bigl(\frac{\pi}{2}(\alpha_s(t)-\alpha_h(t))\Bigr),$$
	\begin{equation}\label{eq:cs(h+s)}
	\sigma_{el}=\frac{H_s^2e^{2\{L(\alpha_{0s}-1)\}}}{2(\alpha_s'L+{b_s})}
	+\frac{H_h^2e^{2\{L(\alpha_{0h}-1)\}}}{2(\alpha_h'L+{b_h})}
	+2H_sH_he^{L(\alpha_{0s}-1)+L(\alpha_{0h}-1)}
	\frac{\mathfrak{B}\cos\phi_0+\mathfrak{L}\sin\phi_0} {\mathfrak{B}^2 + \mathfrak{L}^2}.
	\end{equation}
	
	In these two equations we used the notations
	
	\begin{equation}\label{eq:Denotation}
	\begin{array}{l}
	L=\ln\left({s/s_{0}}\right),
	\\\phi_0=\frac{\pi}{2}(\alpha_{0s}-\alpha_{0h}),
	\end{array}\qquad
	\begin{array}{l}
	\mathfrak{B}=L\alpha_s' + L\alpha_h'+({b_s}+{b_h}),
	\\\mathfrak{L}=\frac{\pi}{2}(\alpha_s'-\alpha_h'),
	\end{array}
	\nonumber
	\end{equation}
	\begin{equation}
	H_s(\widetilde{Q^2})=\frac{A_s}{{\Bigl(1+\frac{\widetilde{Q^2}}{{Q_s^2}}\Bigr)}^{n_s}}, \quad
	H_h(\widetilde{Q^2})=\frac{A_h\Bigl(\frac{\widetilde{Q^2}}{{Q_h^2}}\Bigr)}{{\Bigl(1+\frac{\widetilde{Q^2}}{{Q_h^2}}\Bigr)}^{n_h+1}}, \nonumber
	\end{equation}
	with
	$$A_{s,h}=-\frac{\sqrt{\pi}}{s_{0}}\widetilde{A_{\ }}_{\!\!s,h}.$$
	
	Finally, we notice that amplitude~(\ref{eq:Amplitude_hs}) can be rewritten in the form
	$$
	A(s,t,Q^2,{M_v}^2)= \widetilde{A_s}e^{-i\frac{\pi}{2}\alpha_s(t)}\left(\frac{s}{s_{0}}\right)^{\alpha_s(t)}
	e^{b_st - n_s\ln{\left(1+\frac{\widetilde{Q^2}}{\widetilde{Q_s^2}}\right)}}
	$$
	\begin{equation}
	+\widetilde{A_h}e^{-i\frac{\pi}{2}\alpha_h(t)}\left(\frac{s}{s_{0}}\right)^{\alpha_h(t)}
	e^{b_ht - (n_h+1)\ln{\left(1+\frac{\widetilde{Q^2}}{\widetilde{Q_h^2}}\right)}
		+\ln{\left(\frac{\widetilde{Q^2}}{\widetilde{Q_h^2}}\right)} },
	\label{eq:Amplitude_hs_modif}
	\end{equation}
	where  the two exponential factors $e^{b_st - n_s\ln{\left(1+\frac{\widetilde{Q^2}}{\widetilde{Q_s^2}}\right)}}$ and $e^{b_ht - (n_h+1)\ln{\left(1+\frac{\widetilde{Q^2}}{\widetilde{Q_h^2}}\right)}
		+\ln{\left(\frac{\widetilde{Q^2}}{\widetilde{Q_h^2}}\right)}}$ can be interpreted as the product of the form factors of upper and lower vertices (see Fig.~\ref{fig:diagram}c). Interestingly, the amplitude~\eqref{eq:Amplitude_hs_modif}  resembles the scattering amplitude of Ref.~\cite{Capua}.

	Let us illustrate
	the important and delicate interplay between the ``soft" and ``hard" components of our unique amplitude. According to the definition~(\ref{two-term-amp}), our amplitude can be
	written as
	\begin{equation}
	A(Q^2,s,t)=A_s(Q^2,s,t)+A_h(Q^2,s,t).
	\label{Ampl-2}
	\end{equation}
	As a consequence, the differential and elastic cross sections contain also an interference term between ``soft" and ``hard" parts, so that they read
	\begin{equation}
	\frac{d\sigma_{el}}{dt}=\frac{d\sigma_{s,el}}{dt}+\frac{d\sigma_{h,el}}{dt}+\frac{d\sigma_{interf,el}}{dt}
	\label{dsigma_2}
	\end{equation}
	and
	\begin{equation}
	\sigma_{el}=\sigma_{s,el}+\sigma_{h,el}+\sigma_{interf,el},
	\label{sigma_2}
	\end{equation}
	according to Eqs.~\eqref{eq:dcsdt(h+s)} and \eqref{eq:cs(h+s)}, respectively.
	
	\begin{figure}
		\centering
		\includegraphics[scale=0.42]{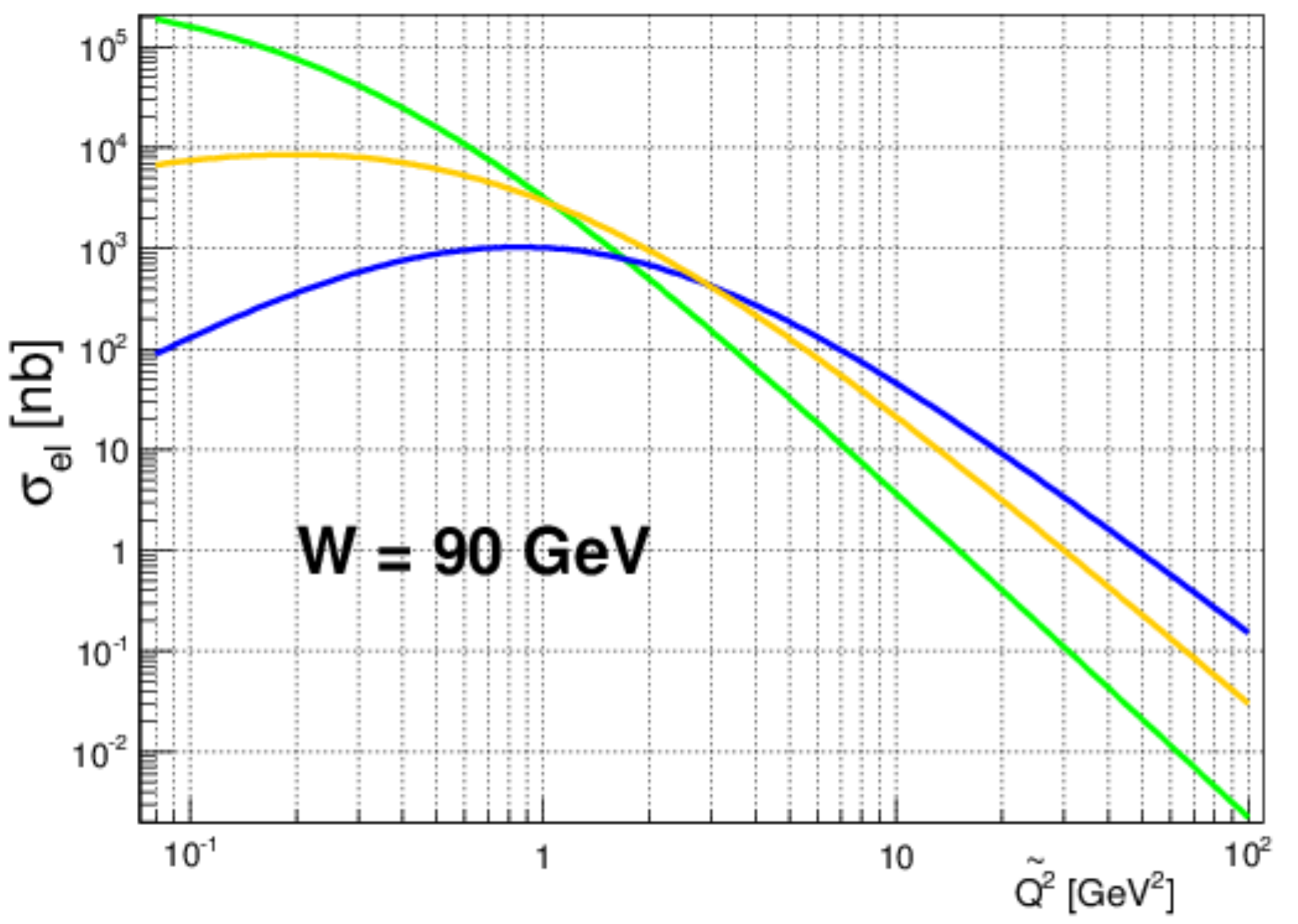}
		\includegraphics[scale=0.43]{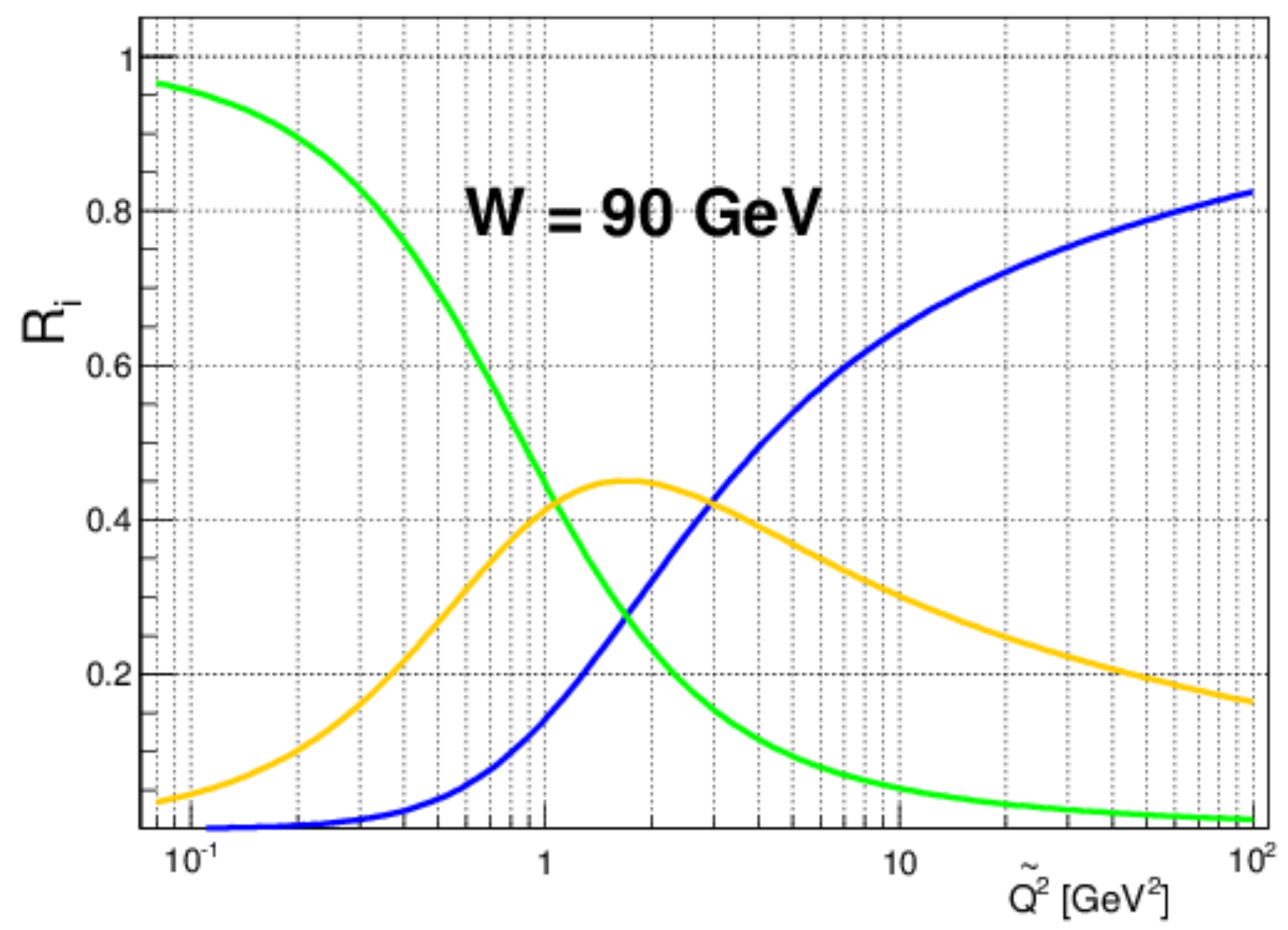}
		\caption{\label{fig:Rsh} Interplay between soft (green line), hard (blue line) and interference (yellow line) components of the cross section $\sigma_{i,el}$ (left plot) and $R_i(\widetilde{Q^2}, t)$ (right plot) as functions of $\widetilde{Q^2}$, for $W=90$ GeV.}
	\end{figure}
	
	Given Eqs.~(\ref{dsigma_2}) and (\ref{sigma_2}), we can define the following ratios for each component:
	
	\begin{equation}
	R_i(\widetilde{Q^2}, W, t)=\frac{ \frac{d\sigma_{i,el}}{dt} }{ \frac{d\sigma_{el}}{dt} }
	\label{ratio_dsigma}
	\end{equation}
	and
	
	\begin{equation}
	R_i(\widetilde{Q^2}, W)=\frac{\sigma_{i,el}}{\sigma_{el}},
	\label{ratio_sigma}
	\end{equation}
	where $i$ stands for $s,\ h,\ interf$.

	
	Fig.~\ref{fig:Rsh} shows the interplay between the components for both $\sigma_{i,el}$ and $R_i(\widetilde{Q^2}, t)$, as functions of $\widetilde {Q^2}$, for $W$ = 90  GeV.
	In Fig.~\ref{fig:Rsh_surf} both plots show that not only $\widetilde{Q^2}$ is the parameter defining softness or hardness of the process, but such is also the  combination of $\widetilde{Q^2}$ and $t$, similar to the variable $z=t-Q^2$ introduced in Ref.~\cite{Capua}.
	On the whole, it can be seen from the plots that the soft component dominates in the region of low $\widetilde{Q^2}$ and $t$, while the hard component dominates in the region of high $\widetilde{Q^2}$ and $t$. 
	
	\begin{figure}[h]
		\centering
		\includegraphics[scale=0.32]{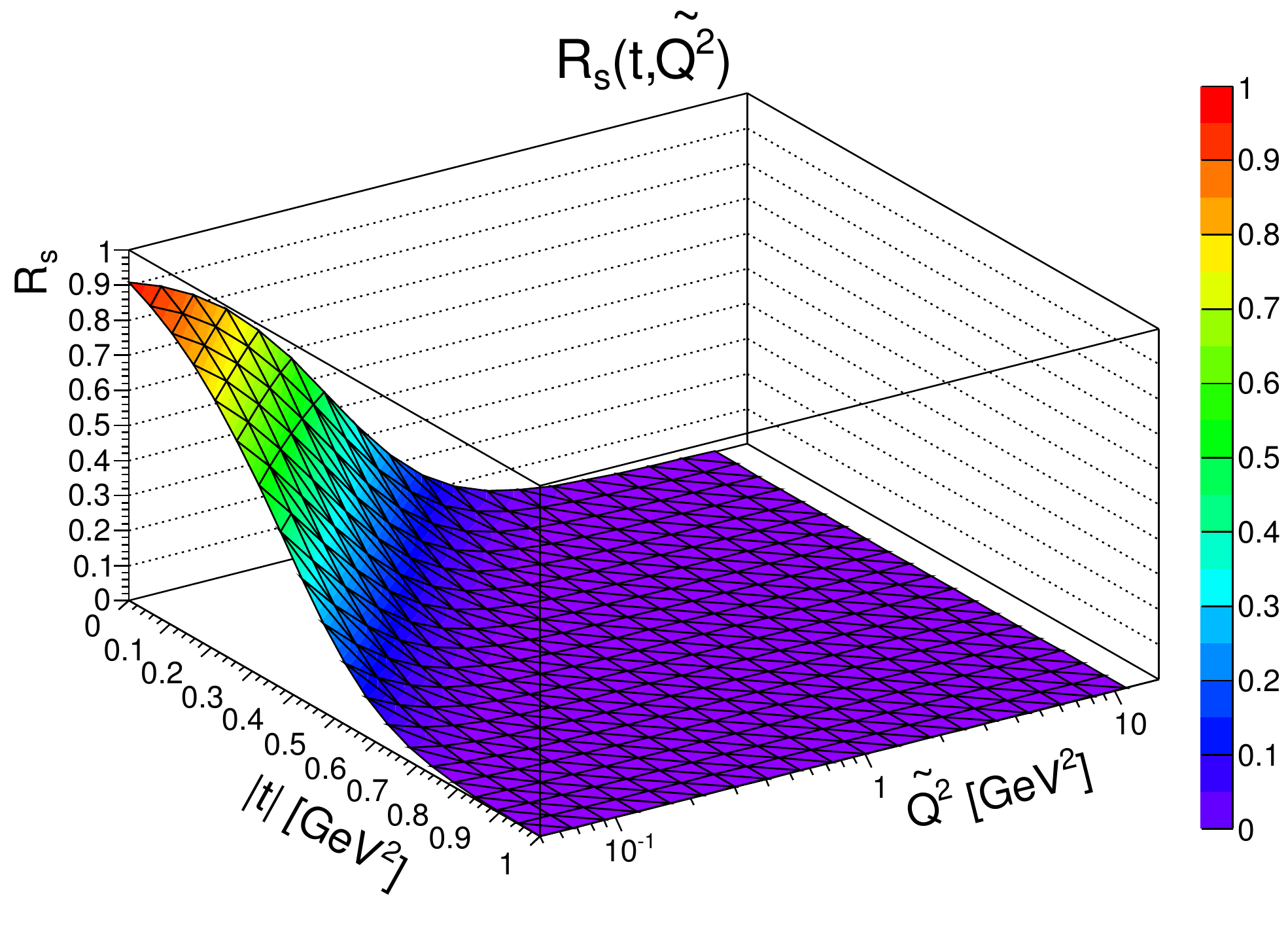}
		\includegraphics[scale=0.28]{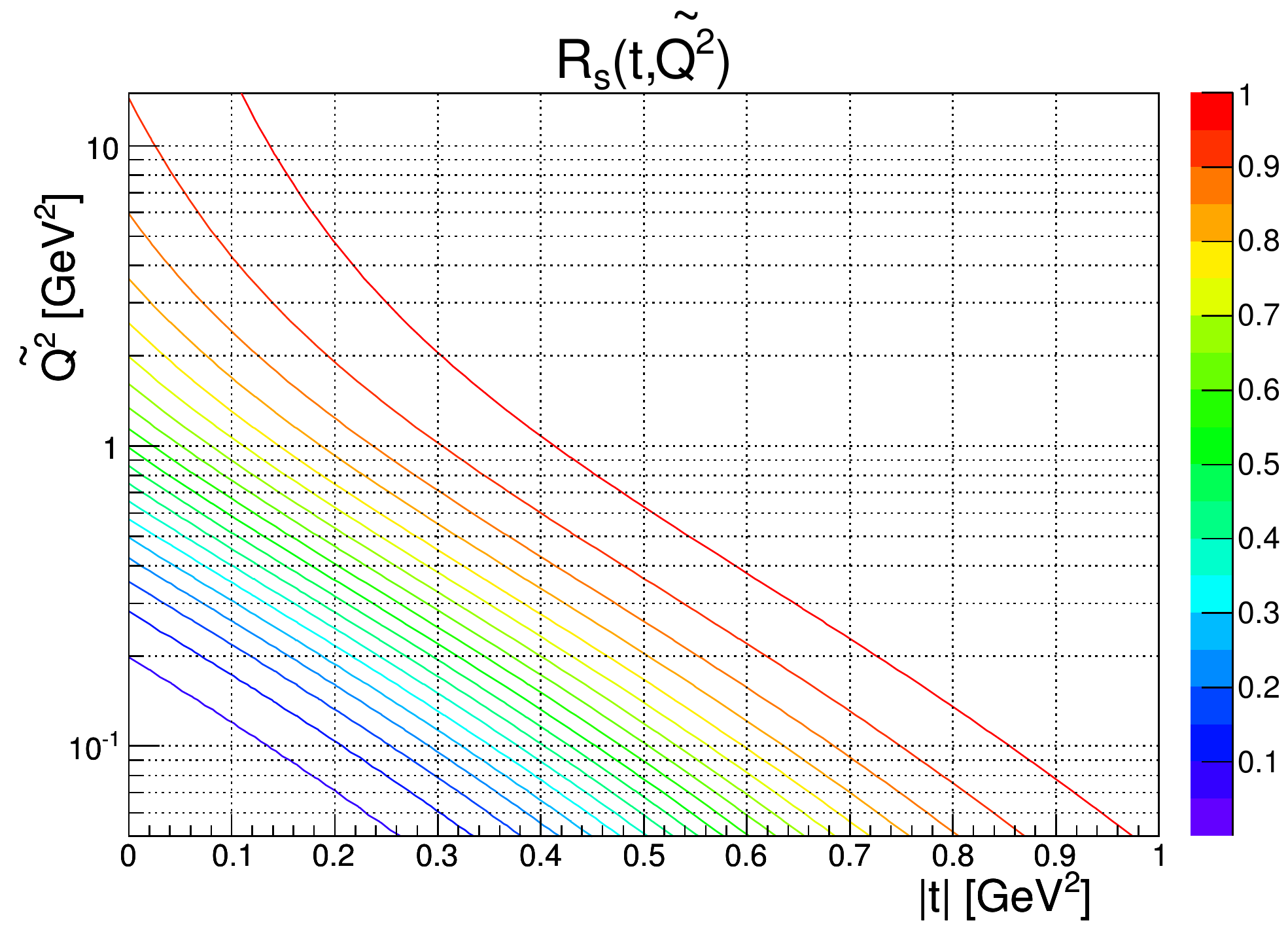}
		\includegraphics[scale=0.32]{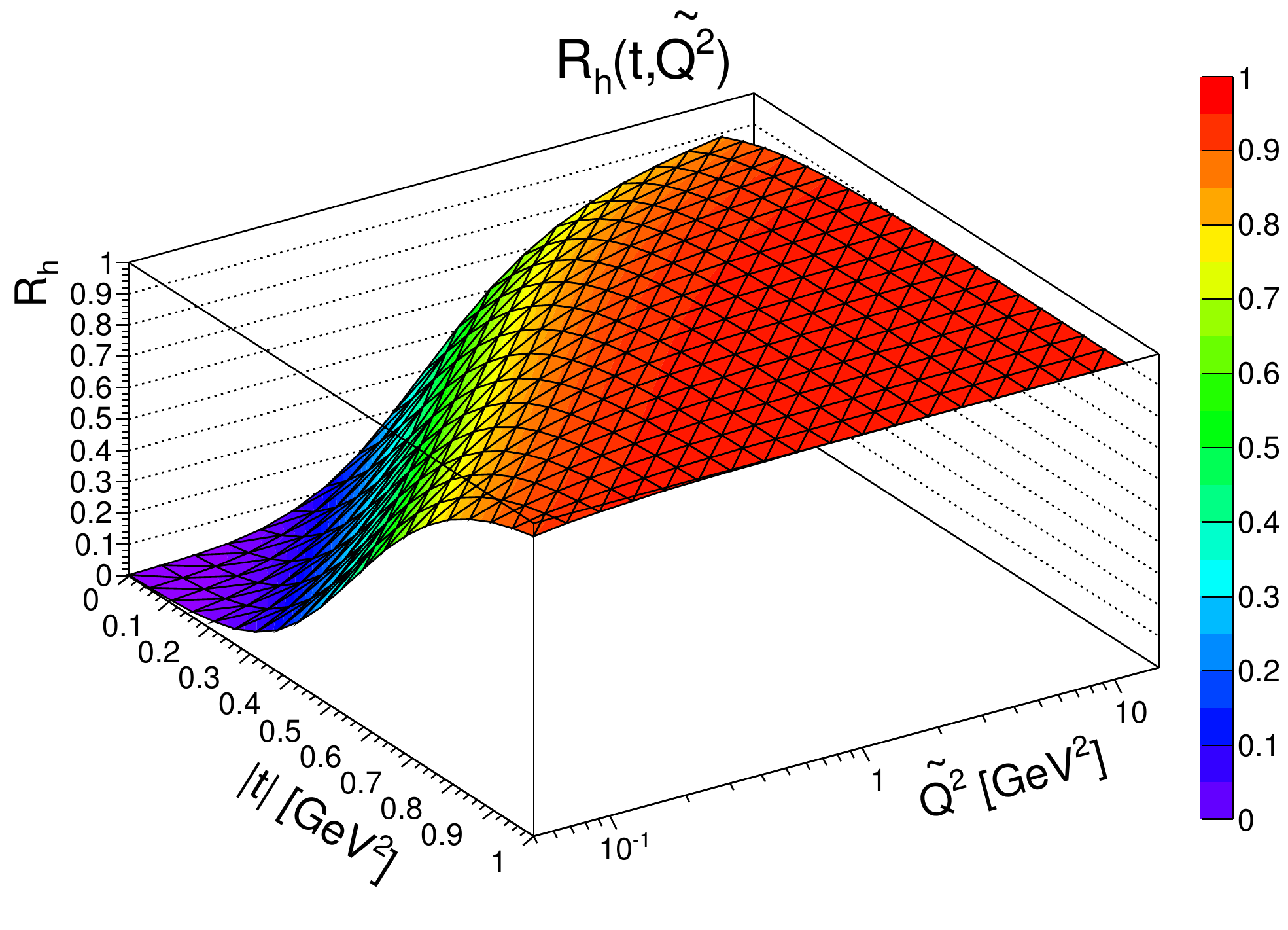}
		\includegraphics[scale=0.28]{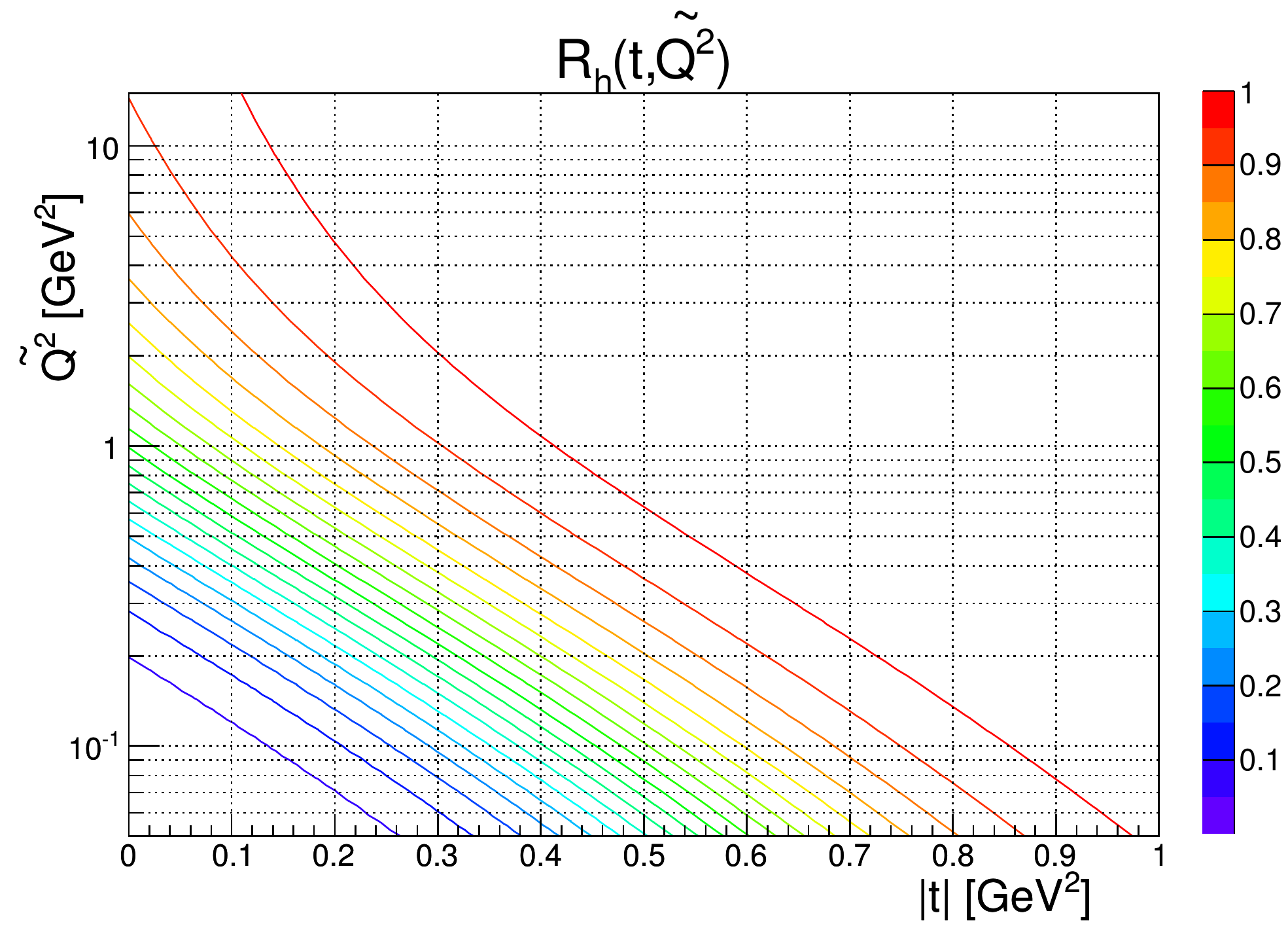}
		\includegraphics[scale=0.32]{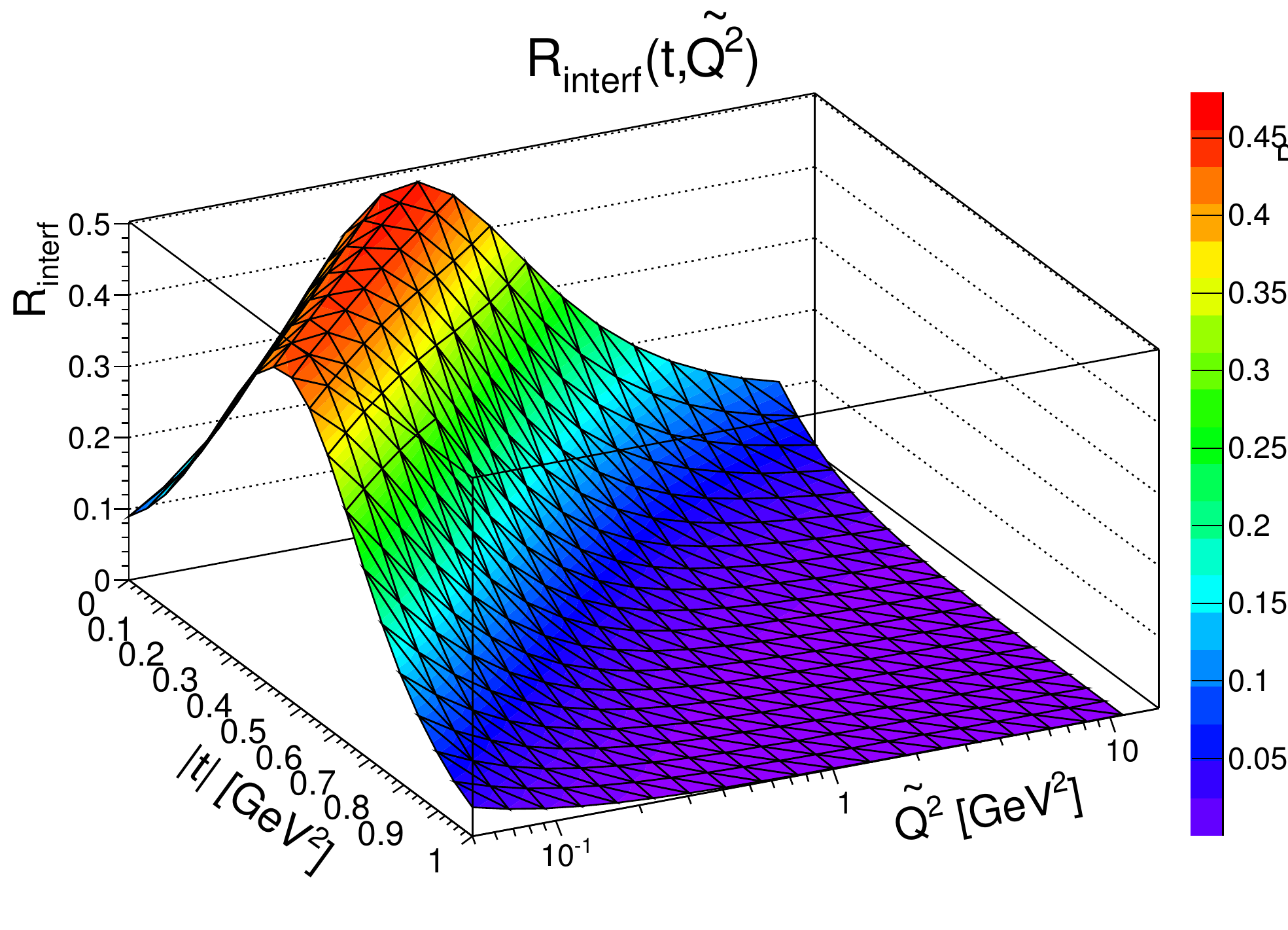}
		\includegraphics[scale=0.28]{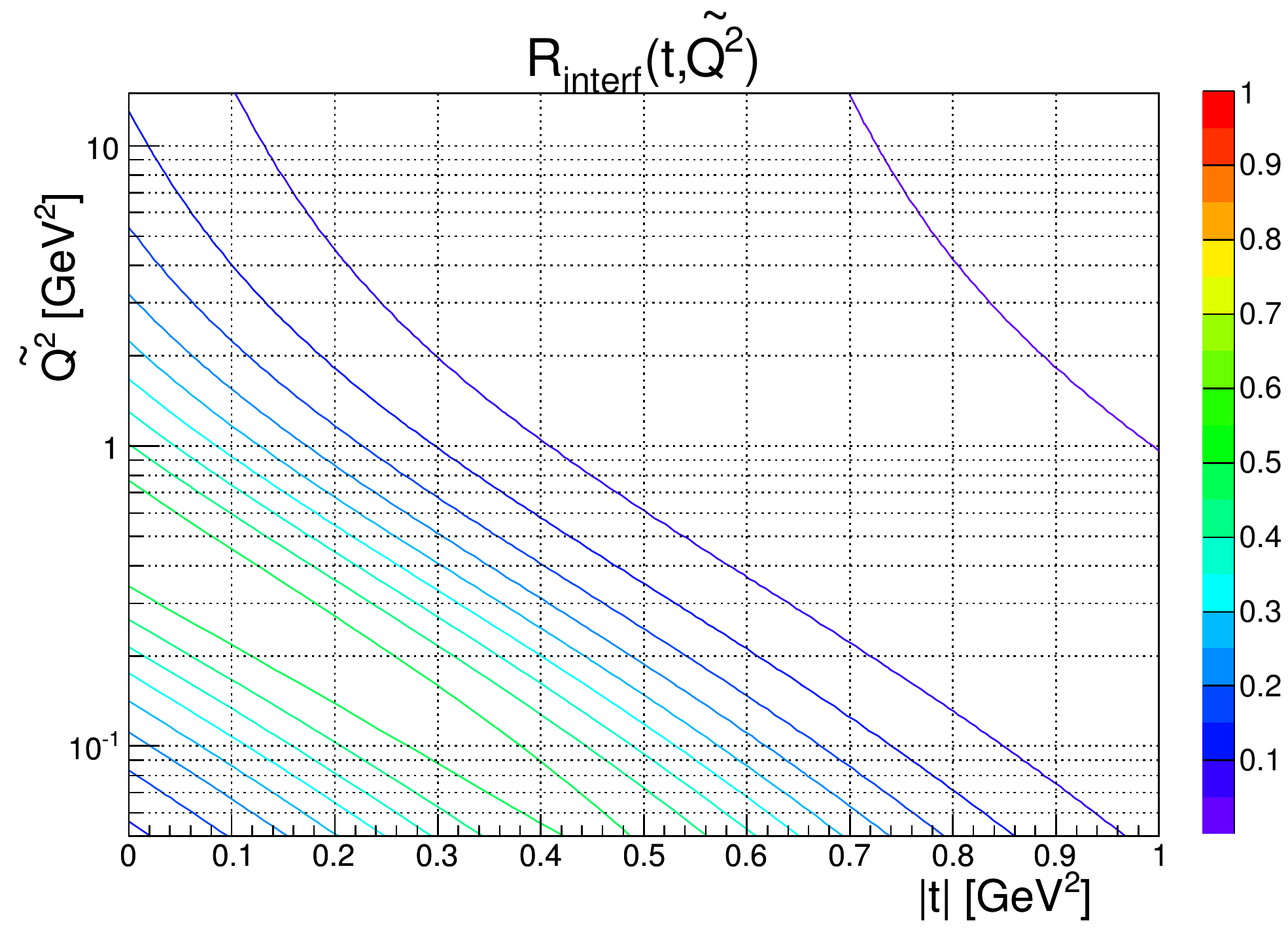}
		\caption{\label{fig:Rsh_surf} Left column: soft (upper surface), hard (middle surface) and interference (bottom surface) components of the ratio $R_i(\widetilde {Q^2}, W, t)$
			are shown  as functions of $\widetilde{Q^2}$ and $t$, for $W=90$~GeV. Right column: some representative curves of the surfaces projected onto the ($t, \widetilde {Q^2}$  ) plane.}
	\end{figure}

	\subsection{Hadron-induced reactions: high-energy $pp$ scattering}\label{sec:pp}
	Hadron-induced reactions differ from those induced by photons at least in two aspects. First, hadrons are on the mass shell and hence the relevant processes are typically ``soft''. Secondly, the mass of incoming hadrons is positive, while the virtual photon has negative squared ``mass". Our attempt to include hadron-hadron scattering into the analysis with our model has the following motivations: a)
	by vector meson dominance (VMD) the photon behaves partly as a meson, therefore meson-baryon (and more generally, hadron-hadron) scattering has much in common with photon-induced reactions. Deviations from VMD may be accounted for the proper $Q^2$ dependence of the amplitude (as we do hope is in our case!); b) of interest is the connection between
	space- and time-like reactions;  c) according to recent claims (see e.g. Ref.~\cite{L,DL_tr}) the highest-energy (LHC) proton-proton scattering data indicate the need for a ``hard" component in the pomeron (to anticipate, our fits do not found support the need of any noticeable ``hard" component in $pp$ scattering).
	
	We did not intend to
	a high-quality fit to the $pp$ data; that would be impossible without the inclusion of subleading contributions and/or the odderon. Instead we normalized the parameters of our leading pomeron term according to recent fits by Donnachie and Landshoff \cite{L} including, apart from a soft term, also a hard one.
	
	The $pp$ scattering amplitude is written in the form similar to the amplitude  (\ref{eq:Amplitude_hs}) for VMP or DVCS, the only difference being that the normalization factor is constant since the $pp$ scattering amplitude does not depend on $Q^2$:
	\begin{equation}\label{eq:Amplitude2_pp}
	A^{pp}(s,t)=
	A^{pp}_s\, e^{-i\frac{\pi}{2}\alpha_s(t)} \left(\frac{s}{s_{0}}\right)^{\alpha_s(t)} e^{b_st}
	+A^{pp}_h\, e^{-i\frac{\pi}{2}\alpha_h(t)} \left(\frac{s}{s_{0}}\right)^{\alpha_h(t)} e^{b_ht}.
	\end{equation}	
	
	We fixed the parameters of pomeron trajectories in accord with those of Refs.~\cite{Lpp,L})
	$$\alpha_{s}(t)=1.084+0.35t,\qquad \alpha_{h}(t)=1.30+0.10t.$$
	
	With these trajectories the total cross section
	\begin{equation}\label{eq:cstot_2}
	\sigma_{tot}=\frac{4\pi}{s} Im\;A(s,{t=0})
	\end{equation}
	was found compatible with the LHC data, as seen in the upper plot of Fig.~\ref{fig:cs_pp}. From the comparison of Eq.~(\ref{eq:cstot_2}) to the LHC data we get
	$$A^{pp}_s=-1.73 \text{\,mb}\cdot\text{GeV}^2,\,\,\quad\quad
	A^{pp}_h=-0.0012 \text{\,mb}\cdot\text{GeV}^2.$$
	The parameter $b_s$ was determined by fitting the differential and integrated elastic cross sections to the measured data. 
	The forward slope $B$, shown in Fig.~\ref{fig:B_pp} was also calculated.
    
    These calculations, fits and figures \ref{fig:cs_pp}, \ref{fig:dcsdt_pp}, \ref{fig:B_pp} are intended to demonstrate that the contribution from the "hard" component of the pomeron is negligible in high-energy elastic $pp$ scattering.  	
	
	\begin{figure}[h]
		\centering
		\includegraphics[trim = 0mm 0mm 12mm 2mm,clip, scale=0.5]{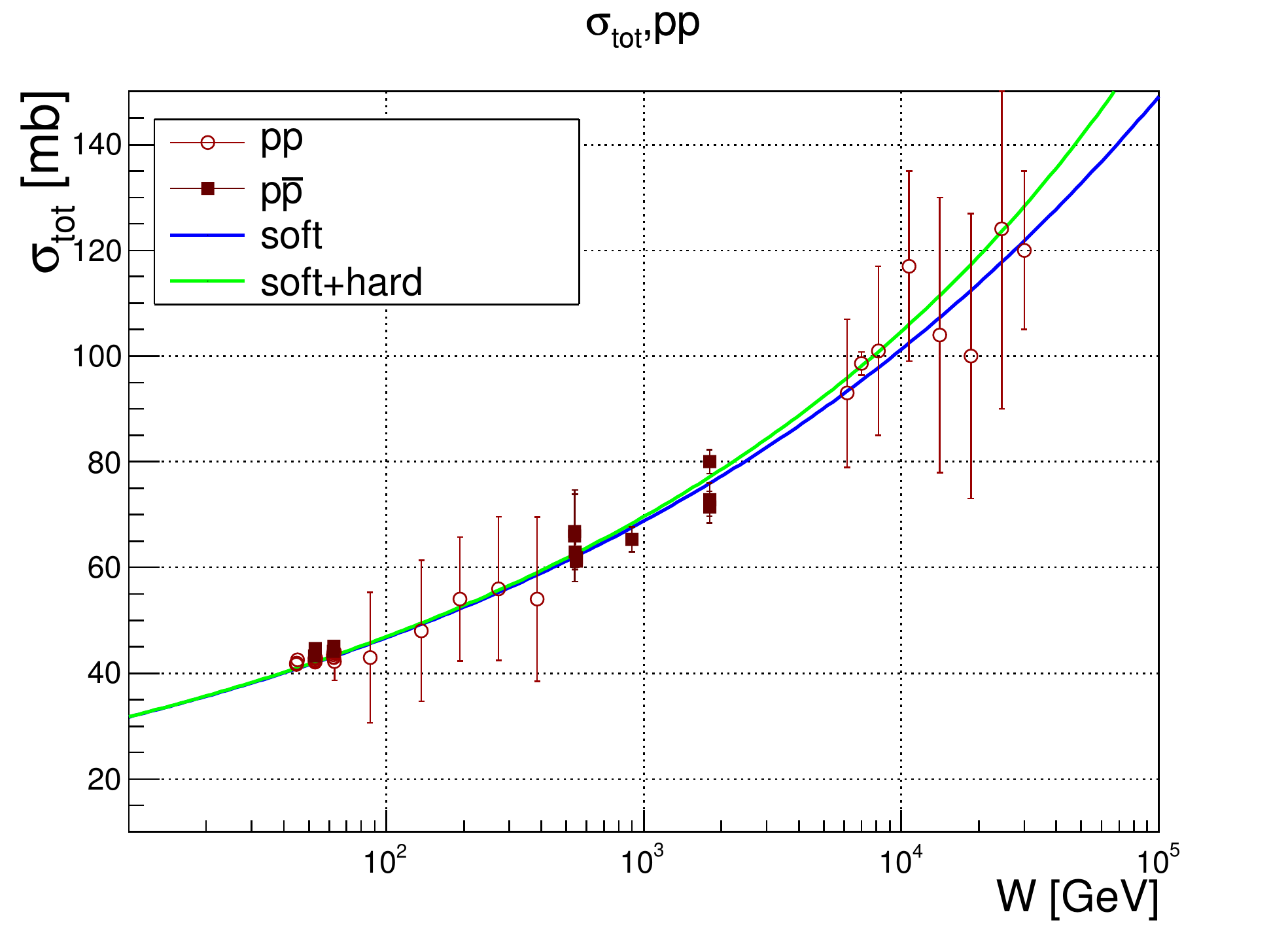}
		\includegraphics[trim = 0mm 0mm 12mm 2mm,clip, scale=0.5]{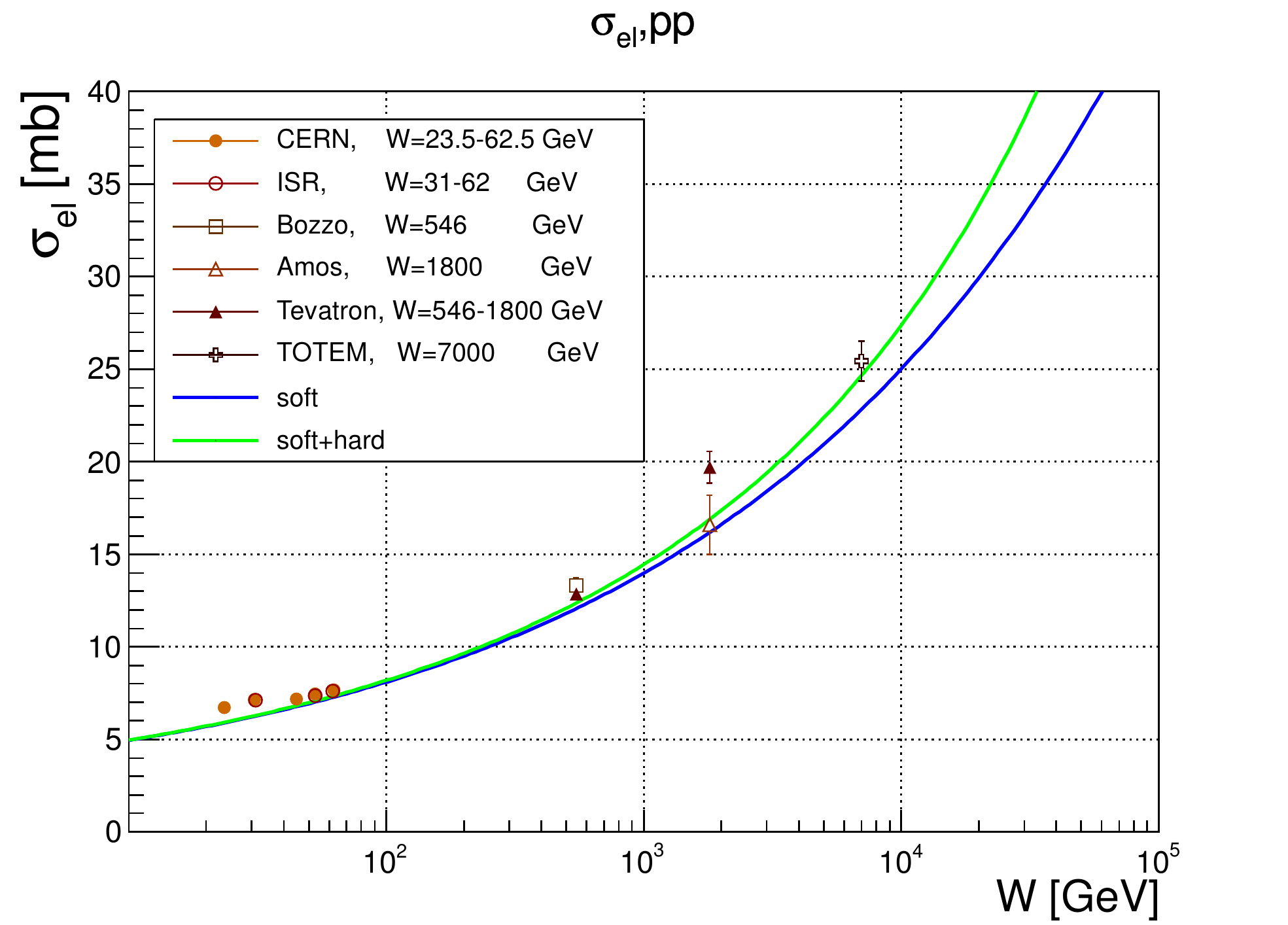}
		\caption{ \label{fig:cs_pp} Calculated total  $\sigma_{tot}(W)$ and the elastic cross sections $\sigma_{el}(W)$  for $pp$ scattering with the data from Refs.~\cite{PDG}.}
	\end{figure}
	
	\begin{figure}[h]
		\centering
		\includegraphics[trim = 0mm 0mm 12mm 2mm,clip, scale=0.5]{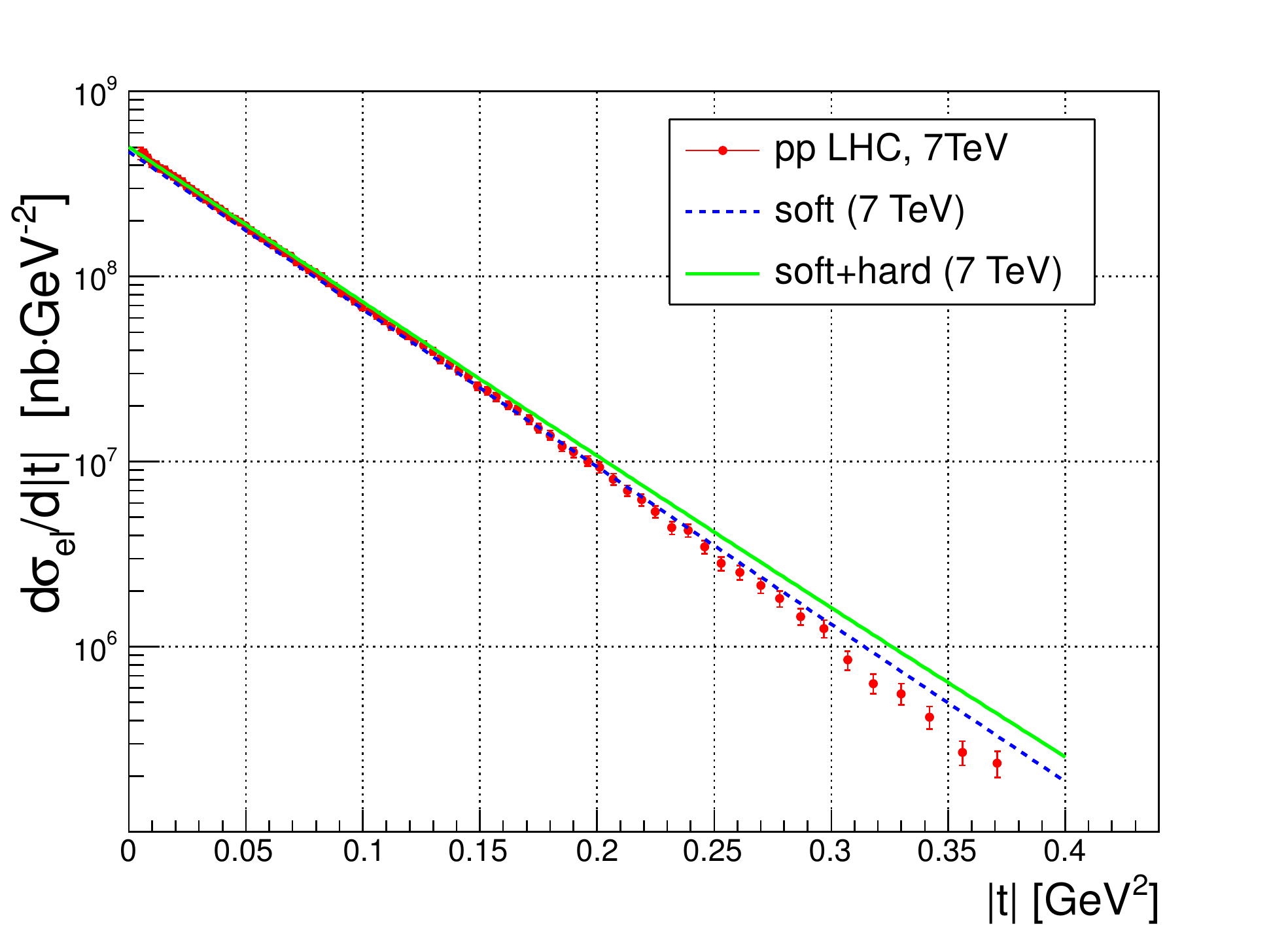}
		\caption{ \label{fig:dcsdt_pp} Differential elastic cross section $d\sigma_{el}/dt$ for $pp$-scattering. The data are from Ref.~\cite{LHCpp}.}
	\end{figure}
	\begin{figure}[H]
		\centering
		\includegraphics[trim = 0mm 0mm 12mm 2mm,clip, scale=0.5]{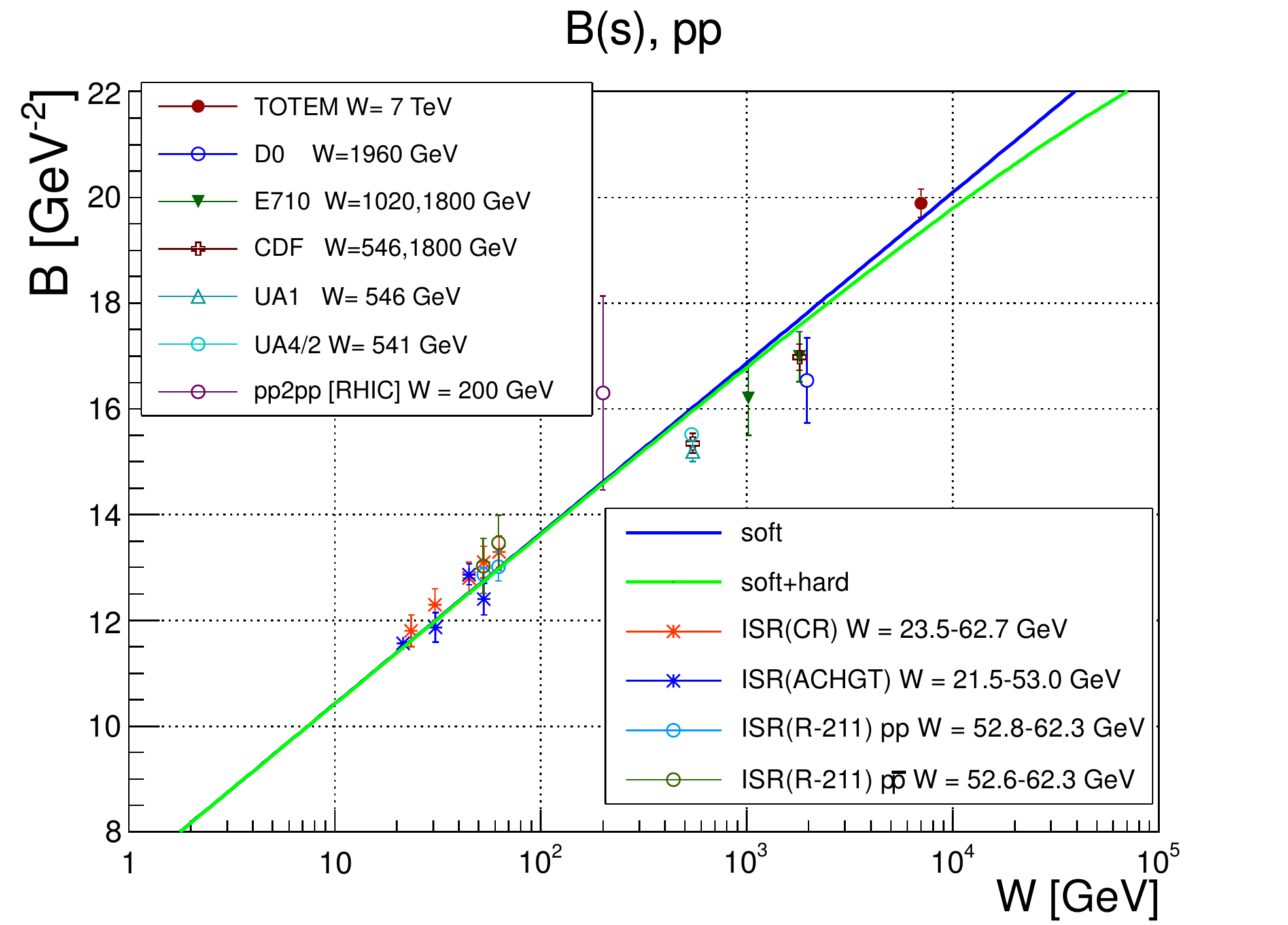}
		\caption{ \label{fig:B_pp}The slop $B$ of the differential elastic cross section $d\sigma_{el}/dt$ for $pp$-scattering as function of $W$. The data are from Refs.~\cite{LHCpp,ISR}.}
	\end{figure}
	
	\section{Ultra-peripheral collisions of hadrons and nuclei}\label{UPC}
	\subsection{Ultra-peripheral collisions}
	
	Following the shut-down of HERA interest in exclusive diffractive vector meson production (VMP) has shifted to the LHC. In ultra-peripheral reactions $h_1h_2\rightarrow h_1Vh_2,$  (where $h_i$ stands for hadrons or nuclei, e.g. Pb, Au, etc.) one of the hadrons (or nuclei) emits quasi-real photons that interact with the other proton/nucleus in a similar way as in $ep$ collisions at HERA. Hence the knowledge of the 
	$\gamma p\rightarrow Vp$ cross section accumulated at HERA is useful at the  LHC. The second ingredient is the photon flux emitted by the  proton (or nucleon). The importance of this class of reactions was recognized in early $70$-ies (two-photon reactions in those times) \cite{Budnev,Terazawa}. 
	In  those papers, in particular, the photon flux was calculated. For a contemporary review on these calculations, see for instance Refs.~\cite{Review1,Review2,Review3}. 
	
	Recent studies of VPM in ultra-peripheral collisions at the LHC
	were reported in Ref.~\cite{TMF1,TMF2}, where references to earlier papers can be also found. 
	In particular, we present predictions for $J/\psi$ and $\psi(2S)$ productions in $pp$ scattering. 
	We also extend the analysis to the lower energies for the photon-proton cross section in order to scrutinize the $\frac{d\sigma^{pp}}{dy}$ cross section behavior in that kinematic regime.
	
	The rapidity distribution of the cross section of vector meson production (VMP) in the reaction $h_1h_2\rightarrow h_1Vh_2,$ as shown in Fig.~\ref{fig:vmp_feynman}, can be written in a factorized form, i.e. it can be presented as a product of the photon flux and photon-proton cross section {\cite{Review1,Review2,Review3,TMF1,TMF2}.
		
		\begin{figure}[H]
			\centering
			\includegraphics[width=1.8in]{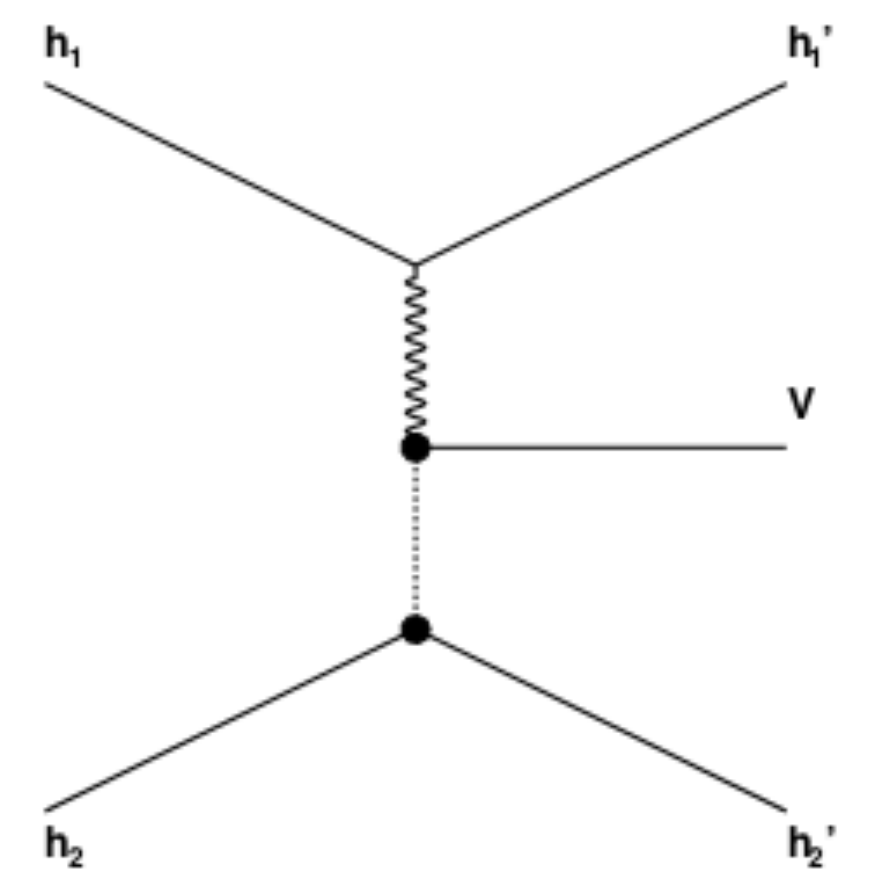}
			\caption{Diagram of vector meson production in a hadronic collision. The wiggle and the dotted lines correspond (interchangeably) pomeron or photon exchange in the $t$-channel.}
			\label{fig:vmp_feynman}
		\end{figure}
		
		The $\gamma p \rightarrow Vp$ cross section ($V$ stands for a vector meson) depends on three variables: the total energy $W$ of the $\gamma p$ system, the squared momentum transfer $t$ and $\widetilde Q^2=Q^2+M_V^2$, where $Q^2=-q^2$ is the photon virtuality. Since, in ultraperipheral\footnote {In ultraperipheral collisions the impact parameter $b\gg R_1+R_2,$ i.e. the closest distance between the centers of the colliding particles/nuclei, $R_{1,2}$ being their radii.} collisions  photons are nearly real ($Q^2\approx 0$), the vector meson mass $M_V^2$ remains the only measure of ``hardness''. 
		The $t$-dependence (the shape of the diffraction cone) is known to be nearly exponential.
		It can be either integrated, or kept explicit. 
		The integrated $\sigma_{\gamma p\rightarrow Vp}(\tilde Q^2, W)$ and differential    $\frac{d\sigma(t)}{dt}$ cross sections are well known from HERA measurements.

		As mentioned, the differential cross section as function of rapidity can be factorized:\footnote{More precisely, the cross section can be presented as the sum of two factorized terms, depending on the photon or pomeron emitted by the relevant proton.}
		\begin{equation}\label{eq:1}
		\frac{d\sigma}{dy}^{h_1h_2\rightarrow h_1Vh_2}
		=r(y)E_{\gamma_+}\frac{dN_{\gamma_+h_1}}{dE_{\gamma_+}}\sigma^{\gamma h_2\rightarrow Vh_2}(E_{\gamma_+})+
		r(y)E_{\gamma_-}\frac{dN_{\gamma_-h_2}}{dE_{\gamma_-}}\sigma^{\gamma h_1\rightarrow Vh_1}(E_{\gamma_-}).
		\end{equation}
		Here $\frac{dN_{\gamma h}}{dE_{\gamma}}
		=\frac{\alpha_{em}}{2\pi E_{\gamma}}\left[1+(1-\frac{2E_{\gamma}}{W_{pp}})^2\right]
		\left(\ln\Omega-\frac{11}{6}+\frac{3}{\Omega}-\frac{3}{2\Omega^2}+\frac{1}{3\Omega^3}\right)$
		is the {``equivalent''} photon flux~\cite{Review1,Review2,Review3}, 
		$\sigma^{\gamma h_i\rightarrow Vh_i}(E_{\gamma})$ is the total (i.e. integrated over $t$) exclusive VMP cross section (the same as at HERA \cite{ActaPol,FFJS}),
		{$r(y)$ is the rapidity gap survival correction,}
		and $E_{\gamma}=W^2_{\gamma p}/(2W_{pp})$ is the photon energy, with
		$E_{\gamma\,\mathrm{min}}=M_V^2/(4\gamma_Lm_p),$ where $\gamma_L=W_{pp}/(2m_p)$
		is the Lorentz factor (Lorentz boost of a single beam).
		Furthermore,
		$\Omega=1+Q_0^2/Q_\mathrm{min}^2,$ $Q_\mathrm{min}^2=\left(E_{\gamma}/\gamma_L\right)^2,$ $Q_0^2=0.71$GeV$^2,$ and $y=\ln(2E_{\gamma}/m_V)$. The signs $+$ or $-$ near $E_\gamma$ and $N_\gamma$ in Eq.~(\ref{eq:1}) correspond to the particular proton, to which the photon flux is attached.
		
		For definiteness we assume that: a) the colliding particles are protons;
		b) the produced vector meson $V$ is $J/\psi$ (or $\psi(2S)$), and c) the collision energy $W_{pp}=7\,$TeV.
		
		%
		
		\subsection{The $\gamma p \to Vp$ cross section}
		Below we present theoretical predictions for $J/\psi$ and $\psi(2S)$ production in $\gamma p$ scattering.
		In doing so, we use the so-called Reggeometric model \cite{ActaPol}, a two-component (``soft'' and ``hard'') pomeron model \cite{FFJS} and a model \cite{Martynov} including also the low-energy region.
		In the Reggeometric model we use   
		\begin{equation}
		\sigma_{\gamma p \to J/\psi}=A_0^2\,\frac{(W_{\gamma p}/W_0)^{4(\alpha_0-1)}}{(1+\widetilde Q^2/Q^2_0)^{2n}\left[4\alpha'\ln(W_{\gamma p}/W_0)+4\left(\frac{a}{\widetilde Q^2}+\frac{b}{2m_\mathrm{p}^2}\right)\right]}\ ,
		\end{equation}
		where $\widetilde Q^2=Q^2+m_V^2,$ 
		%
		
		The above models, apart from $W$ and $t$, contain also dependence on the virtuality $Q^2$ and the mass of the vector meson $M_V$, relevant in extensions to the $\psi(2S)$ production cross section. 
		As shown in Ref.~\cite{FFJS}, to obtain the $\psi(2S)$ cross section one needs also an appropriate normalization factor, which is expected to be close to $f_{\psi(2S)}=\frac{m_{\psi(2S)}\Gamma(\psi(2S)\to e^+e^-)}{m_{J/\psi}\Gamma(J/\psi\to e^+e^-)}=0.5$. According to a fit of $\frac{\sigma^{\gamma p\to p+\psi(2S)}(W)}{\sigma^{\gamma p\to p+J/\psi}(W)}$ to the data \cite{psiHERAData1,psiHERAData2,psiHERAData3} with a two-component pomeron model, the value  $f_{\psi(2S)}=0.4$ is reasonable.
		Thus, if the formula for the cross section $\sigma(W,Q^2,\,m_{J/\psi})$ describes $\gamma p\to J/\psi+p$ production, then $f_{\psi(2S)}\sigma(W,Q^2,\,m_{\psi(2S)})$ should describe $\gamma p\to \psi(2S)+p$ production as well.

		The above models, fitted to the HERA electron-proton VMP data can be applied also to VMP in hadron-hadron scattering.
		The LHCb Collaboration has recently measured ultraperipheral $J/\psi$ and $\psi(2S)$ photoproduction cross sections in $pp$-scattering (at $7\,$TeV) \cite{LHCb2}.  

		\subsection{Rapidity distributions}\label{sec:Martynov}
		
		\begin{figure}[b]
			\centering
			\includegraphics[trim = 0mm 1mm 0mm 0mm,clip,width=3.1in]{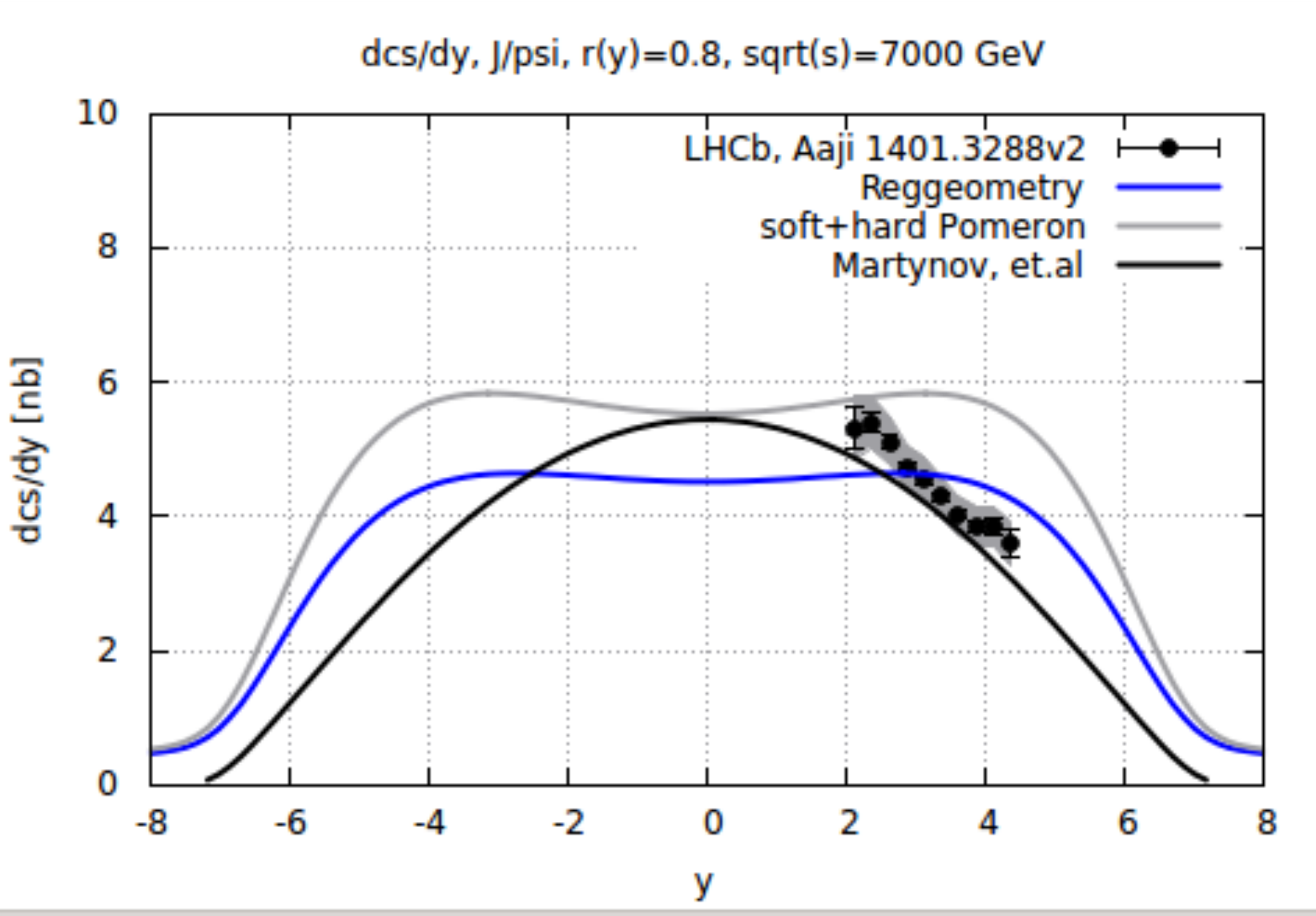}
			\includegraphics[trim = 1mm 1mm 0mm 0mm,clip,width=3.1in]{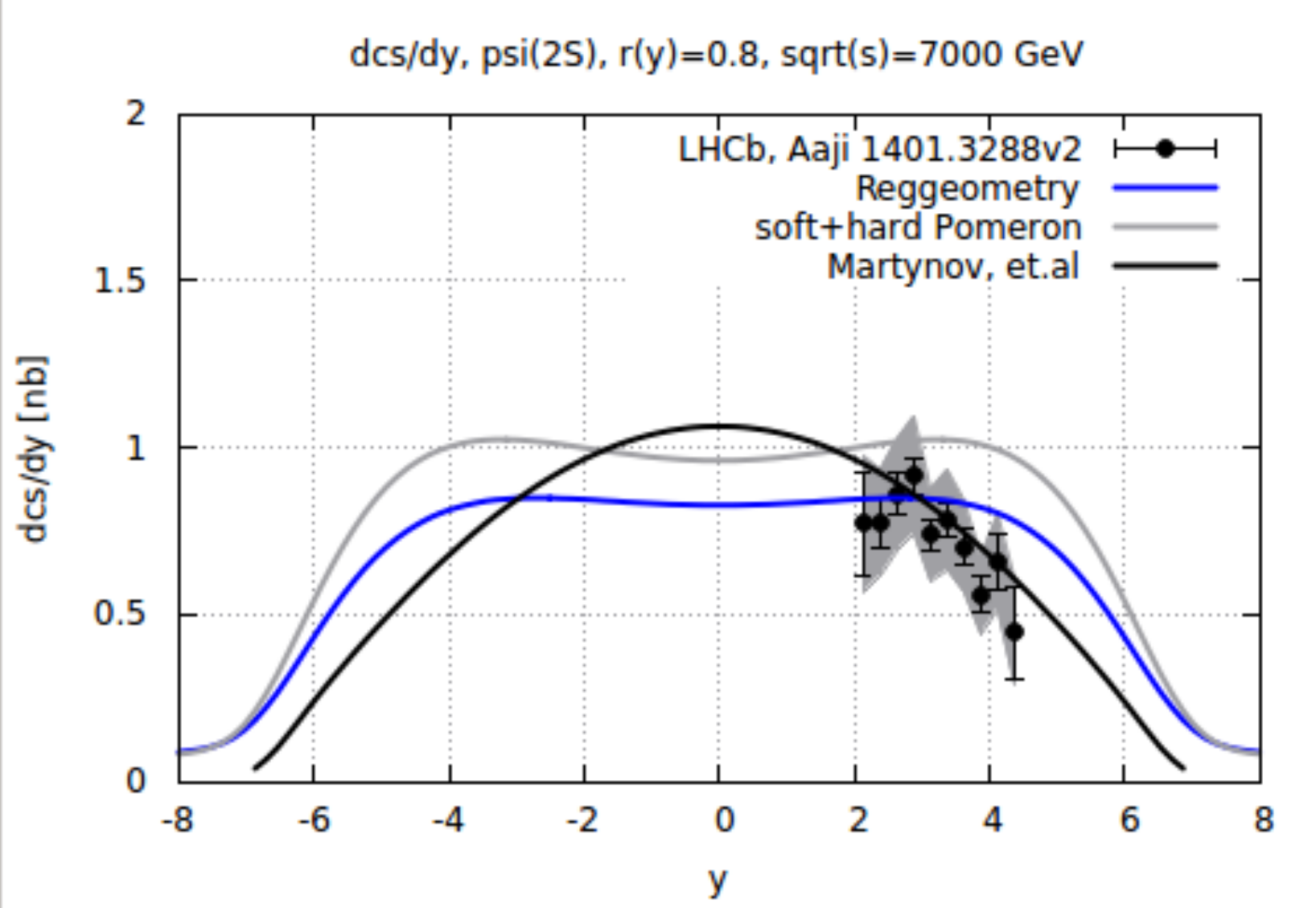}
			\caption{Comparison of the LHCb \cite{LHCb2} data on (a)\,$J/\psi$ and (b)\,$\psi(2S)$ photoproduction cross section as a function of rapidity, $y$, with the two-component pomeron model \cite{FFJS}, the Reggeometric model \cite{ActaPol} and that of Ref.~\cite{Martynov}.
				An absorption correction $r(y)=0.8$ was applied to theoretical predictions.}\label{fig:dcsdy}
		\end{figure}
		
		To calculate the rapidity distribution $\frac{d\sigma}{dy}^{pp\to pVp}(y)$ we use Eq.~(\ref{eq:1}), with an appropriate $\gamma p$ cross section $\sigma^{\gamma p\rightarrow Vp}(W_{\gamma p})$. 
		In~Fig.~\ref{fig:dcsdy} we show the LHCb \cite{LHCb2} data together with the predictions for the $J/\psi$ and the $\psi(2S)$ differential rapidity cross sections obtained from the Regge model \cite{ActaPol}, the two-component pomeron model \cite{FFJS} and that of Ref.~\cite{Martynov}. The rapidity gap survival factor $r(y)=0.8$\, was used.
		
		The energy of the $\gamma p$ system $W_{\gamma p}$ is related to rapidity $y$ via $W^{\pm}_{\gamma p}=\sqrt{M_{J/\psi} W_{pp}\, e^{\pm y}}$ (the choice of the sign depends on the  propagation direction of $\gamma$). 
		Hence, the differential rapidity cross section $\frac{d\sigma}{dy}$,  in the range $y\in[2,\, 4.5]$  at energy $W_{pp}=7\,$TeV, needs the knowledge of the integrated cross section  $\sigma(W_{\gamma p})$ in the range $W_{\gamma p}\in [15.5,\;54]\cup[400,1397]\,$GeV.
		Fig.~\ref{fig:dcsdy}(a) and \ref{fig:dcsdy}(b) show how sensitive the differential $\frac{d\sigma}{dy}$ cross section predictions are to the choice of $\sigma(W_{\gamma p})$ cross sections.   
		
		The curve of the model \cite{Martynov} for $J/\psi$ (and its extension for $\psi(2S)$) gives better description of the differential rapidity cross section than the Reggeometric and the 
		two-component pomeron models, but it seems to underestimate the $J/\psi$ data (see Fig.~\ref{fig:dcsdy}(a)). This may result  from underestimation of the $\gamma p$ cross sections by the model \cite{Martynov} at higher energies. To properly describe the rapidity distribution of VMP cross section in $pp$ scattering, we need to correctly describe the $\gamma p$ VMP cross section in the whole energy region. Each $\gamma p$ energy range corresponds to its particular rapidity range.

		\section*{Conclusions}
The total, elastic and inelastic cross sections at the LHC did not reveal surprises; the rate of their rise (not predicted by theory) follows extrapolations \cite{PDG} of phenomenological models, typically $\ln^2 s$ or, equivalently that of Donnachie and Landshoff's  supercritical pomeron, with $\alpha(0)\approx 0.08$ \cite{DLl}. Forward physics at the LHC is dominated by  
pomeron exchange, the role of secondary ({\it e.g.} $f$ or $\rho$) exchanges is negligible, their relative contribution there \cite{JLL } being smaller than the experimental uncertainties. The odderon is not "seen" in fits to total cross sections.  Although the common belief is that cross sections will continue rising indefinitely,  alternatives, e.g. tending to a constant, oscillations etc. are not excluded by theory. Fits to the LHC total cross sections by the Brazilian group \cite{Paulo} however hints to a slower rise in the ATLAS data.  

Contrary to total cross sections, the data on the forward slope $B(s,0)$ and the phase (ratio of the real to imaginary part $\rho(s,0)$ offers surprises, triggering theoretical work. The forward slope $B(s,0)$, typically logarithmic in Regge-pole models, was found by the TOTEM collaboration to accelerate from $\ln s$ to $\ln^2 s$ at highest LHC energy needing theoretical explanation and understanding. Another news from TOTEM is the surprisingly low value $\rho(13,0)$. Both are discuss in detail in Sec. 4.1 of the present review. The low value of the ratio $\rho(13,0)$ is almost a direct evidence for the odderon, predicted many years ago and discussed in numerous papers, see e.g. \cite{Jenk2, JLL} and references therein. Recent fits to the TOTEM data with its low $\rho$ value cannot prove or disprove the existence of the odderon until larger values of $t$, namely those at the dip will be shown to work. 

There is little doubt about the existence of the odderon, just because nothing forbids its existence. Its parameters are not predicted from theory. By a plausible estimate \cite{PEPAN}, based on the string model the odderon's slope is roughly $\alpha'_O\approx 2\alpha'_P/3$. The odderon could be detected directly by measuring $pp$ and $\bar pp$ differential cross sections at the same energy, {\it e.g.} by rescaling the LHC down to the closed Tevatron energy, 
$\sim 2$ TeV$^2$. 

The dip-bump structure is a crucial test for most of the models. Usually it is generated by 
unitarity correction (rescattering), for example in the Regge-eikonal formalism. The point is that these models predict multiple dips and bumps, while experiments, including those most recent at $13$ TeV$^2$ do not exhibit them: there is only one, typicall of the dipole pomeron, presented in Sec. 4.1.		

In that section another structure in the differential cross section, namely the "break" near $t=-0.1$ GeV$^2$ and it connection with the proton "atmosphere" is scrutinized. That phenomenon was first observed at the ISR. Its universality is of great interest.

The knowledge of elastic scattering is crucial for the understanding of diffraction dissociation. By Regge factorization, many features of elastic scattering may be translated to the much more complicated inelastic processes, namely single, double and central diffraction. Models and predictions for these reactions presented in Sects. 6 and 7 may be helpful in planning, guiding experimental studies of  SD, DD and CED at the LHC. 

The pomeron is the central object in forward physics at the LHC. As repeatedly stressed in this paper, in the LHC energy region one for the first time has a chance to identify the pomeron, uncontaminated by secondary exchanges. Perturbative quantum chromodynamics (pQCD) predicted that the intercept of the (bare) pomeron is much higher than its "canonical" value $0.08$. In Sec. 8 we explore a ("reggeometric") model of the pomeron with two components, whose weight depends on the "hardness" of the reaction (soft in $pp$). We concluded that the "hard" component is negligible at LHC. 

Finally we note that we ignore the so-called rapidity gap survival corrections that brought much confusion in studies of diffraction dissociation. In our opinion, the confusion comes from the mixture of the space-time treatment of inelastic processes with the analytic $S$ matrix theory, part of which are Regge-poles, operational only for asymptotic states. A reasonable Rege-pole model compatible with unitarity should not contain "rapidity gap survival corrections", otherwise it should be improved within its own formalism. In other words, the size of these corrections reflect the "goodness" of the model.

\section*{Acnowledgement}
During the work on this paper we profited from numerous enlightening
discussions with Professor O. Nachtmann. Fruitful collaboration on various parts of this review with Oleg Kuprash, Magno Machado, Risto Orava and Chung-I Tan is gratefully acknowledged.

L. J. was supported by the Ukrainian Nat. Academy's project "Structure and dynamics of statical and quantum-mechanical systems".  

We thank the Referee of this paper for his inspiring criticism and useful remarks.


\begin{thebibliography}{99}
\bibitem{Collins} P. D. B. Collins, {\it An introduction to Regge Theory and High Energy Physics}, Cambridge University Press (1977).

\bibitem{DLl} S. Donnachie and P.V. Landshoff, \textit{Phys. Lett. B}, {\bf 727}, 500 (2013), and earlier references therein.
\bibitem{Petrov} V.A. Petrov, A. Prokudin, \textit{Phys. Rev. D} {\bf 87}, 036003 (2013), arXiv: 1212.1924.
\bibitem{Paulo} D.A. Fagundes, M.J. Menon, P.V.R.G. Silva,\textit{ Nucl. Phys. A} {\bf966}, 185 (2017), arXiv:1703.07486.
\bibitem{Pancheri} D.A. Fagundes, G. Pancheri, A.Grau, S. Pacetti, Y.N. Srivastava, \textit{Phys. Rev. D} {\bf 88}, 094019 (2013).
\bibitem{Khoze1} V. A. Khoze, A. D. Martin, and M. G. Ryskin,  \textit{Eur. Phys. J. C}, {\bf 74}, 2756 (2014), arXiv:1312.3851; 

\bibitem{Khoze2} V. A. Khoze, A. D. Martin, and M. G. Ryskin, \textit{Int. J. Mod. Phys. A} {\bf30}, 1542004 (2015), arXiv:1402.2778; 

\bibitem{Khoze3} V.A. Khoze, A.D. Martin, M.G. Ryskin, (2017), arXiv:1712.00325.

\bibitem{Alkin} A. Alkin, O. Kovalenko, E. Martynov, \textit{EPL} {\bf 102}, 31001 (2012), arXiv:1304.0850.
\bibitem{Godizov}A.A. Godizov, \textit{Phys. Rev. D}  {\bf96}, 034023 (2017), arXiv:1705.09126.

\bibitem{BFKL1} V.S. Fadin, E.A. Kuraev, L.N. Lipatov, \textit{Phys. Lett.} {\bf 60} 50 (1975); 
\bibitem{BFKL2} I.I. Balitsky, L.N. Lipatov, \textit{Yad. Fizika,} {\bf 28}, 822 (1978).
\bibitem{TOTEM_tot}	G. Antchev {\it et al.} (TOTEM Collab.),(2017), arXiv:1712.06153.
\bibitem{TOTEM_rho} G. Antchev {\it et al.} (TOTEM Collab.), (2017), https://cds.cern.ch/record/2298154.
\bibitem{MN} E.S. Martynov, E. B. Nicolescu, \textit{Phys. Lett. B}
\textbf{778}, 414 (2018), arXiv:1711.03288.
\bibitem{COMPETE} J.R. Cudell et al (COMPETE Collaboration),\textit{ Phys. Rev. D} {\bf65}, 074024 (2002). 
\bibitem{TT0} S.M.~Troshin and N.E.~Tyurin, \textit{Mod. Phys. Lett. A} {\bf 32}, 1750168 (2017), hep-ph/1708.00302. 
\bibitem{TT2} S.M.~Troshin and N.E.~Tyurin, (2017), arXiv:1704.00443.
\bibitem{TT01} S.M.~Troshin and N.E.~Tyurin, \textit{Eur. Phys. J. A} {\bf 53} (2017), hep-ph/1701.01815; 
\bibitem{TT02}S.M.~Troshin and N.E.~Tyurin, (2016), hep-ph/1602.08972; 
\bibitem{TT03}S.M.~Troshin and N.E.~Tyurin, (2016), hep-ph/1601.00483.

\bibitem{RPM} L.~Jenkovszky, I.~Szanyi, {\it Phys. Part. Nuclei Lett.}, {\bf 14}  (2017), 687 arXiv:1701.01269.
\bibitem{JSZ2} L.~Jenkovszky, I.~Szanyi, {\it Mod. Phys. Lett. A} {\bf 32}, 1750116 (2017),  arXiv:1705.04880.
\bibitem{JSZT} L.~Jenkovszky, I.~Szanyi and C.-I Tan. {\it Shape of Proton and the Pion Cloud} to be published.
\bibitem{JL} L.~Jenkovszky and A.~Lengyel, \textit{Acta Phys. Pol. B} {\bf 46}, 863 (2015), arXiv:1410.4106.

\bibitem{Burq} J.P.~Burq {\it et al.} \textit{CERN/EP preprint 82-195, Part. Phys. B} {\bf 217}, 285 (1983).

\bibitem{JLL} L.L.~Jenkovszky, A.I.~Lengyel, D.I.~Lontkovsky, \textit{Int'l J. Mod. Phys. A}, {\bf 26}, 4755 (2011), arXiv/1105.1202.
\bibitem{Deile} M. Deile, \textit{Soft (and Hard) QCD Processes in TOTEM}, WE-Heraeus Physics School "QCD - Old Challenges and New Opportunities" (Bad Honnef, 2017).
\bibitem{totem82} G. Antchev {\it et al.} (TOTEM Collab.), \textit{Eur. Phys. J. C} {\bf 76}, 661 (2016), arXiv:1610.00603.
\bibitem{Chew} G.F.P. Chew, {\it Theory of Strong Interactions},
Benjamin, New Yourk, 1961.

\bibitem{ChF} G.F.P. Chew and S.C. Frautschi, Phys. Rev. Lett. {\bf 7} (1961) 394. 




\bibitem{DHSch} R. Dolen, D. Horn and C. Schmit, \textit{Phys. Rev.} {\bf 166} (1968) 1768. 
\bibitem{Veneziano} G. Veneziano, \textit{Nuovo Cim.} {\bf 57A} (1968) 190.

\bibitem{BP}V.~Barone and E.~Predazzi, \textit{High-energy particle diffraction}, Springer-Verlag Berlin Heidelberg NewYork, ISBN 3540421076, (2002).
\bibitem{DDLN}S.~Donnachie, H.G.~Dosch, O.~Nachtmann, and P.~Landshoff, \textit{Pomeron physics and QCD}, \textit{Camb. Monogr. Part. Phys. Nucl. Phys. Cosmol.} {\bf 19}, 1 (2002).

\bibitem{Regge1} Regge, T. (1959),\textit{ Nuovo Cimento} \textbf{14}, 951. 
\bibitem{Regge2} Regge, T. (1960), \textit{Nuovo Cimento} \textbf{18}, 947.
\bibitem{Barut} A.O.~Barut and D.E.~Zwanziger, {\it Phys. Rev.} {\bf 127}, 974 (1962).
\bibitem{Jenk2} L. Jenkovszky, \textit{Rivista Nuovo Cim.}, {\bf 10} (1987), p. N12.

\bibitem{Fiore2}R. Fiore, L. Jenkovszky, V. Magas, F. Paccanoni, A. Papa,
\textit{Eur.Phys.J. A} \textbf{10} (2001) 217-221, hep-ph/0011035.
\bibitem{Goulianos} K. Goulianos, \textit{Physics Reports}, {\bf 1017} (1983) 170. 
\bibitem{Pomeranchuk} V.N. Gribov, I.Ya. Pomeranchuk, \textit{Phys. Rev. Letters}, {\bf 9} (1962) 238.
\bibitem{Fort} L.L. Jenkovszky, \textit{Fortschritte d. Physik}, {\bf 34} (1986) 791-816.


\bibitem{DAMA}A.I. Bugrij {\it et al.} \textit{Fortschritte d. Physik}, {\bf 21}
(1973) 427-506.
\bibitem{Complete} R. Fiore, A. Flachi, L.L. Jenkovszky, A.I. Lengyel, V.K. Magas {\it A Kinematically Complete Analysis of the CLAS data on the Proton Structure Function $F_2$ in a Regge-Dual model}, Phys.Rev. D69 (2004) 014004; arXiv:0308178.
\bibitem{DJS} P. Desgrolard, Laszlo L. Jenkovszky, B.V. Struminsky, \textit{Eur.Phys.J. C} \textbf{11} (1999) 145-151.  

\bibitem{AB} W. Broniowski, E. R. Arriola, \textit{Acta Phys. Pol. B} {\bf 10}, 1203, (2017).
\bibitem{Brazil1} A.K. Kohara, E. Ferreira, T. Kodama and M. Rangler, (2017), arXiv:1710.09862. 
\bibitem{Brazil2} A.K. Kohara, E. Ferreira, T. Kodama and M. Rangler, (2017) arXiv:1709.05713.
\bibitem{KKL} J.~Kontros, K.~Kontros, and A.~Lengyel,(2001), hep-ph/0104133.
\bibitem{KKL1} J.~Kontros, K.~Kontros, and A.~Lengyel, (2000), hep-ph/0006141.
\bibitem{PEPAN}  A.N.~Wall, L.L.~Jenkovszky, and B.V.~Struminsky, \textit{Sov. J. Particles and Nuclei}, {\bf 19}, 180 (1988).
\bibitem{Land} S.~Donnachie, G.~Dosch, P.~Landshoff and O.~Nachtmann, {\it Pomeron physics and QCD} (Cambridge University Press, 2002). 
\bibitem{BJSz} N.~Bence, L.~Jenkovszky, I.~Szanyi. {\it Approaching the asymptotics at the LHC}, arXiv:1711.06380.
\bibitem{PDG} C. Patrignani {\it et al.} (Particle Data Group), \textit{Chin. Phys. C}, {\bf 40}, 100001 (2016).

\bibitem{data} http://durpdg.dur.ac.uk/review/pp2/exphtml/CERN-ISR.shtml
\bibitem{atlas7} G. Aad {\it et al.} (ATLAS Collab.), \textit{Nucl. Phys. B} {\bf 889}, 486 (2014). 
\bibitem{atlas8} M. Aaboud {\it et al.} (ATLAS Collab.) \textit{Phys. Lett. B} {\bf 761}, 158 (2016). 
\bibitem{totem7} G. Antchev {\it et al.} (TOTEM Collab.), \textit{Europhys. Lett.} {\bf 101}, 21004 (2013). 
\bibitem{totem81} G. Antchev {\it et al.} (TOTEM Collab.), \textit{Phys. Rev. Lett.} {\bf 111}, 012001 (2013).
\bibitem{totem83} G. Antchev {\it et al.} (TOTEM Collab.), \textit{Nucl. Phys. B} {\bf 899}, 527 (2015), arXiv:1503.08111.
\bibitem{Auger} P. Abreu et al. (Pierre Auger Collab.) \textit{Phys. Rev. Lett.} {\bf 109}, 062002 (2012), arXiv:1208.1520.

\bibitem{C-I1} C.-I Tan and D.M. Tow, {\it Phys. Letters} {\bf 53B}, 452 (1975).
\bibitem{C-I2} U.~Sukhatme, Chung-I~Tan, and Tran Thanh Van, {\it Z. Phys. C, Particles and Fields}, {\bf 1}, 95 (1979). 
\bibitem{Bronzan} J.B.~Bronzan, Fine structure of the Pomeron, in {\it Symposium on the Pomeron, ANL report} ANL/HEP 7327, p.34, (Argonne National Laboratory, 1973).
\bibitem{totem8.3} TOTEM Collab. (G. Antchev {\it et al.}), {\it Phys. Rev. Lett.} {\bf 111}, 012001 (2013)
\bibitem{Bar} B.~Barbiellini {\it et. al.}, {\it Phys. Lett. B} {\bf 39}, 663 (1972).
\bibitem{LNC} G. Cohen-Tannoudji {\it et al.}, {\it Nuov. Cim.} {\bf 5},  957 (1972).
\bibitem{Brazilb} D.A.~Fagundes, L.~Jenkovszky, E.Q.~Miranda, G.~Pancheri, P.V.R.G.~Silva, { \it Int. J. Mod. Phys. A} {\bf 31}, 1645022 (2016), arXiv:1509.02197.
\bibitem{ISR} http://durpdg.dur.ac.uk/review/pp2/exphtml/CERN-ISR.shtml
\bibitem{Goulianos_renorm} K. Goulianos, Phys. Rev. D 80, 111901 (2009).


\bibitem{JLa1}G.A. Jaroszkiewicz and P.V. Landshoff, \textit{Phys. Rev. D} {\bf 10}
(1974) 170. 
\bibitem{JLa2}A. Donnachie and  P.V. Landshoff, \textit{Nucl. Phys.} {\bf 244} (1984) 322.
\bibitem{DL}A.~Donnachie and P.V.~Landshoff, \textit{Successful description of exclusive vector meson electroproduction}, (2008), arXiv:0803.0686 [hep-ph].
\bibitem{Paccanoni}R. Fiore, Laszlo L. Jenkovszky, V. Magas, F. Paccanoni, A. Papa, \textit{Phys.Part.Nucl.} {\bf 31} (2000) 46; hep-ph/9911503.


\bibitem{Albrow1}M. Albrow, V. Khoze, Ch. Royon, \textit{Special Issue Int. J. Mod. Phys. A} {\bf 29} (2014) 28. 
\bibitem{COMPASS}A. Austregesilo, for the COMPASS Coll., \textit{Proc. 15th EDS Conference}, Sept 2013, arXiv:1310.3190.
\bibitem{CDF}M. Albrow, for the CDF Coll., \textit{Int. J. Mod.Phys.A} {\bf 29} (2014) 28, 1446009, arXiv:1409.0462.
\bibitem{STAR}L. Adamczyk, W. Guryn and J. Turnau, \textit{Int. J. Mod.Phys.A} {\bf 29} (2014) 28, 1446010, arXiv:1410.5752.
\bibitem{ALICE}R. Schicker et al.,\textit{ AIP Conference Proceedings} {\bf 1819}, 040003 (2017), https://doi.org/10.1063/1.4977133, arXiv:1612.06379. 
\bibitem{ATLAS} E. Bols, master thesis,\textit{ http://discoverycenter.nbi.ku.dk/} \newline 
teaching/thesis\_page/MasterEmilBolsFinal.pdf.
\bibitem{CMS}The CMS Collaboration, (2017), arXiv: 1706.08310.
\bibitem{LHCb}R. McNulty, for the LHCb Collaboration, arXiv:1711.06668.
\bibitem{Jenk1} L. Jenkovszky, O. Kuprash, J.W. Lamsa, V.K. Magas, and R.Orava, \textit{Phys.Rev.D} {\bf 83} (2011) 056014, arXiv:1011.0664. 

\bibitem{Fiore1} R. Fiore, A. Flachi, L. Jenkovszky, A. Lengyel and  V. Magas, \textit{Phys. Rev. D} {\bf 69} (2004), 014004, hep-ph/0308178. 
\bibitem{Fiore3}R. Fiore, L. Jenkovszky, R. Schicker, \textit{Eur.Phys.J. C} \textbf{76} (2016) no.1, 38, arXiv:1512.04977. 
\bibitem{Fiore4} R. Fiore, L. Jenkovszky and R. Schicker, {\it Exclusive diffractive resonance production in proton-proton collisions at high energies}, arXive:1711.08353, submitted to EPJ C. 
\bibitem{Acta}S.~Fazio, R.~Fiore, L.L.~Jenkovszky, A.~Lavorini, and A.~Salii,
\textit{Acta. Phys. Pol. B} {\bf 44} (2013) 1333, [arXiv:1304.1891].
\bibitem{Capua}
M.~Capua, S.~Fazio, R.~Fiore,  L.L.~Jenkovszky, and F.~Paccanoni, \textit{Phys. Lett. B} {\bf 645} (2007) 161, [hep-ph/0605319].
\bibitem{FazioPhysRev}
S.~Fazio, R.~Fiore, L.L.~Jenkovszky, and A.~Lavorini, \textit{Phys. Rev. D} {\bf 85} (2012) 054009, [arXiv:1109.6374].
\bibitem{Francesco}
R.~Fiore, L.L.~Jenkovszky, and F.~Paccanoni,
\textit{Eur. Phys. J. C} {\bf 10} (1999) 461, [hep-ph/9812458].
\bibitem{Confer1} S.~Fazio, R.~Fiore, L.L.~Jenkovszky, A.~Lavorini, and A.~Saliy, ``Reggeometry of Deeply Virtual Compton Scattering and Exclusive Diffractive Vector Meson Production at HERA'', In: Proceedings of the 14th Workshop on Elastic and Diffractive Scattering (EDS Blois Workshop), Frontiers of QCD: From Puzzles to Discoveries, Quy Nhon, Vietnam, December 15 - 21, 2011, www.slac.stanford.edu/C111215/.
\bibitem{Confer2} L.L.~Jenkovszky, A.~Salii, and V.~Batozskaya, ``Reggeometry of deeply virtual compton scattering and exclusive vector meson production at HERA'', in Proceedings of Quarks 2012, Yaroslavl, Russia, June 4 -10, 2012.
\bibitem{Confer3} L.L.~Jenkovszky and A.~Salii, ``Reggeometry of lepton- and hadron-induced reactions'', the 4th International Conference, Current Problems in Nuclear Physics and Atomic Energy, Kyiv, Ukraine, September 3 - 7, 2012.
\bibitem{Confer4} R.~Fiore, L.L.~Jenkovszky, A.~Lavorini, and A.~Salii, ``Reggeometry of lepton- and hadron-induced reactions'', in the Proceedings of ``Diffraction 2012: International Workshop on Diffraction in High Energy and Nuclear Physics'', Puerto del Carmen, Spain, September 10 - 15, 2012; AIP Conf. Proc. 1523, pp. 83-86.
\bibitem{L}
P.V.~Landshoff,
\textit{Acta Phys. Polon. B} {\bf 40} (2009) 1967, arXiv:0903.1523 [hep-ph].
\bibitem{FFJS} S. Fazio, R. Fiore, L. Jenkovszky, A. Salii, \textit{Phys. Rev. D} {\bf 90}, (2014) 016007, arXiv:1312.5683.
\bibitem{DL_tr}
A.~Donnachie and P.V.~Landshoff, \textit{Elastic Scattering at the LHC},arXiv:1112.2485 [hep-ph].
\bibitem{Lpp}
P.V.~Landshoff, \textit{How well can we predict the total cross section at the LHC?},
arXiv:0811.0260v1 [hep-ph].
\bibitem{LHCpp}
G.~Antchev et al. [TOTEM Collaboration], \textit{Measurement of proton-proton elastic scattering and total cross-section at $\sqrt s$ = 7 TeV}, CERN-PH-EP-2012-239 (2012).
\bibitem{Budnev}V.~E.~Balakin, V.~M.~Budnev and I.~F.~Ginzburg, \textit{Pisma Zh. Eksp. Teor. Fiz. } {\bf 11}, 559 (1970).
\bibitem{Terazawa}S.~J.~Brodsky, T.~Kinoshita and H.~Terazawa, \textit{Phys. Rev. Lett.}  {\bf 25}, 972 (1970).
\bibitem{Review1} G.~Baur {\it et al.}, \textit{Rev. Mod. Phys.} {\bf 50} 26 (1978);
\bibitem{Review2} G.~Baur {\it et al.}, \textit{Phys. Rept.} {\bf 364}, 359 (2002), hep-ph/0112211; 
\bibitem{Review3} K.~Hencken {\it et al.}, \textit{Phys. Rept. }{\bf 458}, 1 (2008), arXiv:0706.3356.
\bibitem{TMF1} R.~Fiore {\it et al.}, \textit{Theor. and Mathematical Physics}, {\bf 182} (2015) 141–149, arXiv:1408.0530; 
\bibitem{TMF2} R.~Fiore {\it et al.}, {\it ``Vector meson production at the LHC''}, in the Proceedings of "Diffraction-2014" held in Primosten, arXiv:1506.01990.
\bibitem{ActaPol} S.~Fazio {\it et al.}, \textit{Acta Phys. Polon. B} {\bf 44}, 1333 (2013), arXiv:1304.1891. 
\bibitem{Martynov} E.~Martynov, E.~Predazzi and A.~Prokudin, \textit{Phys. Rev. D} {\bf 67}, 074023 (2003), hep-ph/0207272. 
\bibitem{psiHERAData1}
ZEUS Collaboration, 
{\it ``Measurement of the cross-section ratio $\mathit{\sigma_{\psi(2S)}/\sigma_{J/\psi(1S)}}$ in deep-inelastic exclusive $ep$ scattering at HERA''} (prelim.) (2014).
%
\bibitem{psiHERAData2} H1 Collaboration, C.~Adloff {\it et al.} 
\textit{Phys. Lett. B} {\bf 421} (1998) 385
[hep-ex/9711012].

\bibitem{psiHERAData3} H1 Collaboration, C.~Adloff {\it et al.}, 
\textit{Eur. Phys. J. C} {\bf 10} (1999) 373
[hep-ex/9903008].
\bibitem{LHCb2} LHCb Collaboration, R. Aaij {\it et al.},\textit{ J. Phys. G} {\bf 41}, 055002 (2014), arXiv:1401.3288; {\it ibid.}  {\bf 40}, 045001 (2013), arXiv:1301.7084.

\end{thebibliography}
\end{document}